\documentclass{cernyrep}

\usepackage{epsfig}
\usepackage{lscape}
\usepackage{amsmath,amssymb,epsf,color,colordvi,shadow,pifont}

\begin{document}
\title{Cosmology and Gravitation: the grand scheme for High-Energy Physics}

\author{P. Bin\'etruy}

\institute{Universit\'e Paris Diderot, Paris, France}

\maketitle 

 \begin{abstract}
These lectures describe how the Standard Model of cosmology ($\Lambda$CDM) 
has developped, based on 
observational facts but also on ideas formed in the context of the theory of
fundamental interactions, both gravitational and non-gravitational, the latter 
being described by the Standard Model of high energy physics. It focuses on  
the latest developments, in particular 
the precise knowledge of the early Universe provided by the observation of the 
Cosmic Microwave Background  and the discovery of the present acceleration of
the expansion of the Universe. While insisting on the successes of the Standard 
Model of cosmology, we will stress that it rests on three pillars which involve
many open questions: the theory of inflation, the nature of dark matter and 
of dark energy. We will devote one chapter to each of these issues, describing 
in particular how this impacts our views on the theory of fundamental 
interactions.   
{\em More technical parts are given in italics. They may be skipped 
altogether.}
\end{abstract}

\def\noi{\noindent}
\def\sq{\hbox {\rlap{$\sqcap$}$\sqcup$}}
\def\1{{\rm 1\mskip-4.5mu l} }

\section{A not so brief history of modern cosmology} 
\label{chap:1}
Cosmology has been an enquiry of the human kind probably since the dawn of 
humanity. Modern cosmology was born in the early XX$^{{\rm th}}$ century with 
the bold move of Einstein and contemporaries to apply the equations of general 
relativity, the theory of gravity, to the whole Universe. This has led to many 
successes and/or suprises, the most notable of which being presumably the
discovery of extra-galactic objects which recede from our own Galaxy, i.e. the 
discovery of the expansion of the Universe~\cite{Le27,Hu29}. This led to the 
development of the 
Big Bang theory, with the early Universe being a hot and dense medium (a 
prediction confirmed by the discovery of the cosmic microwave background by 
Penzias and Wilson in 1965~\cite{PW65}), and thus 
a laboratory for studying elementary particles. A picture thus emerged in the 
1970s, not only based on the theory of gravity, but also on non-gravitational 
interactions described by the Standard Model of high energy physics, which was
being finalized at the same time (its experimental confirmation would take 
another 40 years and 
has culminated in the discovery of the Higgs particle in 2012).

A first success of the particle physics approach to cosmology has been the 
understanding of the abundancy of light elements in the 80s. This was 
the first quantitative success of cosmology. Meanwhile, the development
of gauge symmetries and the understanding of the r\^ole of spontaneous
symmetry breaking in fundamental interactions led the community to focus its 
attention on phase transitions in the early Universe, in particular associated 
with the quark-gluon transition, the breaking of the electroweak symmetry or 
even of the grand unified symmetry. It is in this context that, in the early 
80s, the theory of inflation was proposed~\cite{St80,Gu81} to solve some of the 
mysteries of the standard Big Bang theory.

The theory of inflation included a model for the genesis of density 
fluctuations responsible for the formation of large scale structures, such as 
galaxies or clusters of galaxies: the quantum fluctuations during the 
exponential (de Sitter) expansion. But this implied the presence of fluctuations
in the otherwise homogeneous and istropic Cosmic Microwave Background. Such 
fluctuations were observed by the COBE satellite, at the level of one 
part in 100 000. 
Generic models of inflation predicted also in a very elegant manner 
that space (not spacetime!) is flat, any spatial curvature being erased by the 
exponential expansion. According to Einstein's equations, this implied that
the average energy density in the Universe had the critical value 
$\rho_c \sim 10^{-26} {\rm kg}/{\rm m}^3$. 

This was a prediction not supported by observation. It was known since the 
1930s that there was a significant amont of non-luminous --or dark-- matter in
the Universe: in 1933, Fritz Zwicky, by studying the velocity distribution
of galaxies in the Coma cluster, had identified that there was $400$ 
times more mass than expected from their luminosity. This had been confirmed 
by studying subsequently the rotation curves of many other galaxies.
But the total of luminous and dark matter could not account for more than $30\%$
of the critical energy density (other components like radiation are 
subdominant at present times). Models of open inflation were even constructed 
to reconcile inflation with observation.

The clue came in 1999~\cite{HZS98a,SCP99} when it was observed that the 
expansion of the Universe is presently accelerating. Since matter or radiation 
tend to decelerate the expansion, one has to resort to a new form of energy,
named dark energy, to understand this acceleration. Was this the  
component which would provide the missing $70\%$ to account for a total
energy density $\rho_c$ and thus a spatially flat Universe?
The answer came from a more precise study of the fluctuations in the CMB
through the space mission WMAP (and more recently Planck): they conclude indeed
that these fluctuations are consistent with spatial flatness.

The latest cosmology results from the Planck mission, released this year, have 
confirmed the predictions of the simplest models of inflation, a rather 
remarkable feat since they are associated with dynamics active in the first 
fractions of seconds after the big bang, and they allow tu fully understand 
the imprints observed 350 000 years after the big bang.

We thus have at our disposal a Standard Model of cosmology which is sometimes 
compared with the Standard Model of high energy physics: in both cases, no
major  experimental/observational data seems to be in conflict with the Model.
There is however one big difference. The Standard Model of cosmology rests on 
three ``pillars'' --inflation, dark matter, dark energy-- which are very poorly 
known: we have at present no microscopic theory of inflation, and we ignore the 
exact nature of dark matter or dark energy. 

For example, there are convincing arguments that dark matter is made of 
weakly interacting massive particles of a new type, and a large experimental 
programme has been set up to identify them. Their discovery would be of utmost 
importance because this would be the first sign of physics beyond the Standard 
Model of particle physics. But it remains a possibility --though not a favored 
one-- to explain the 
observed facts through a 
modification of gravity at different scales (from galaxies to clusters and 
cosmological scales). Finally, axion dark matter would be a minimal 
extension fof the Standard Model, accounting for dark matter.

The discovery of the Higgs has provided us with the first example of a
fundamental scalar field (at least fundamental at the scale where we observe 
it). this a welcome for cosmology since microscopic models of accleration of
the expansion of the Universe --whether inflation or dark energy-- make 
heavy use of such fields. They have the double advantage of being non-vanishing 
without breaking the symmetries of spacetime (like Lorentz symmetry) and of
having the potential of providing an unclustered  background. they also appear 
naturally in the context of extensions of the Standard Model, like 
supersymmetry or extra dimensions. 

Scalar fields thus provide valuable toy models of inflation or dark energy. 
However, such toys models are difficult to implement into realistic high 
energy physics models because of the constraints existing on physics beyond 
the Standard Models of particle physics and cosmology.

In what follows, we will review the successes of the Standard Model of 
cosmology, focussing on the most recent results. And we will consider the three 
pillars of this Model ---inflation, dark matter, dark energy-- and identify the 
most pressing questions concerning these three concepts.

{\em In what follows some more technical discussions are given in italics. 
They can 
be skipped altogether. Background material is also provided in Appendices, most 
of it being intended for the reader who wants to reproduce some of the 
more advanced results.}

We start in this first Chapter with an introduction, partly historical, to the 
main concepts of cosmology.

\subsection{Gravity governs the evolution of the Universe} \label{sect:1-1}

\begin{figure}[h]
\begin{center}
\includegraphics[scale=0.45]{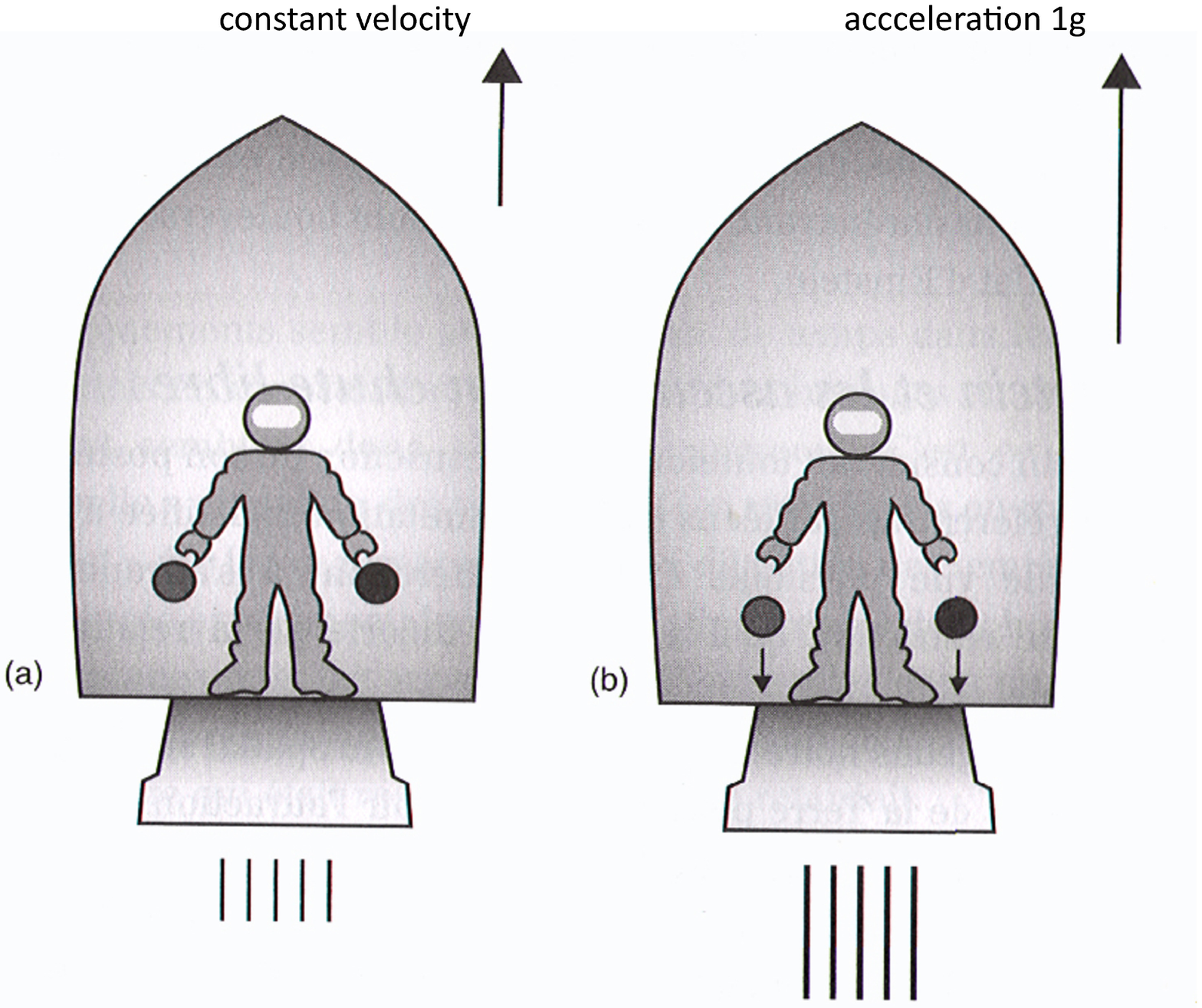}
\end{center}
\caption{(a) When the astronaut in a rocket in uniform motion (constant 
velocity) drops the balls, they float around. (b) If the rocket accelerates by 
$1$ g, the balls fall to the ground and the astronaut has no way to identify 
whether this due to acceleration or to a gravitational field (i.e. the Earth 
attraction).}
\label{fig_rocket1}
\end{figure}

\begin{figure}[h]
\begin{center}
\includegraphics[scale=0.45]{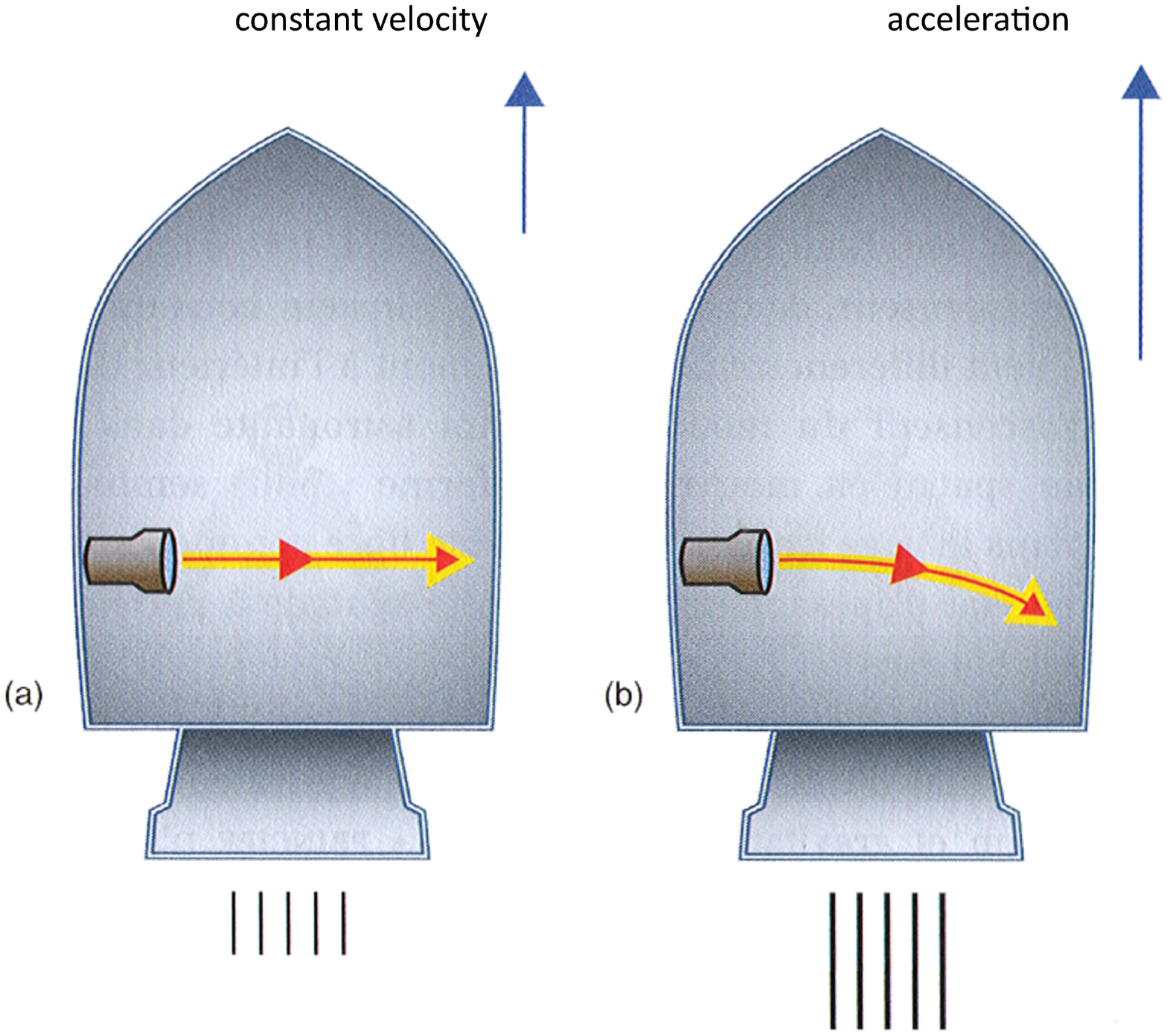}
\end{center}
\caption{If light propagates along straight lines in the rocket in uniform 
motion (a), it must deviate from a straight line when the rocket accelerates
(b).
Thus it must propagate along curved lines in gravitational fields. }
\label{fig_rocket2}
\end{figure}

The evolution of the universe at large is governed by gravity, and thus 
described by Einstein's equations. We recall that general relativity
is based on the 
assumption made by Einstein that observations made in an accelerating reference
are indistinguishable from those made in a gravitational field (as
illustrated on a simple example in Fig.~\ref{fig_rocket1}).
This has several consequences, the most notable of which being that the curves 
followed by light (null geodesics) are not straight lines (see Fig. 
\ref{fig_rocket2}). 
Einstein's equations are 
highly non-linear second order differential equations for the metric 
$g_{\mu\nu}$\footnote{We recall that the invariant infinitesimal spacetime 
interval reads
\begin{equation}
\label{interval}
ds^2 = g_{\mu\nu} dx^\mu dx^\nu \ .
\end{equation}
The metric signature we adopt throughout is 
Einstein's choice: 
$(+,-,-,-)$.}. They read\footnote{{\em These equations can be obtained  from 
the Einstein-Hilbert action:
\begin{equation}
\label{2-Einstein-Hilbert}
{\cal S} = {1 \over 8 \pi G_{_N}} \int d^4x \sqrt{-g} \left[-{1 \over 2} R - 
\lambda\right] + {\cal S}_m(\psi, g_{\mu\nu})\ ,
\end{equation}
where the generic fields $\psi$ contribute to the energy-momentum: $T_{\mu\nu} =
(2/\sqrt{-g})(\delta {\cal S}_m /\delta g^{\mu\nu})$}}: 
\begin{equation}
\label{2-Einstein}
G_{\mu\nu}  \equiv R_{\mu\nu} - {1 \over 2} g_{\mu\nu} R 
=  8 \pi G_{_N} T_{\mu\nu} + \lambda g_{\mu\nu} \ .
\end{equation}
where $R_{\mu\nu}$ is the Ricci tensor,
$R$ the associated curvature scalar ({\em see Appendix~\ref{app:B}}) 
and $T_{\mu\nu}$ the 
energy-momentum tensor; finally $\lambda$ is the cosmological constant
which has the dimension of an inverse length squared.

Thus Einstein's equations relate the geometry of spacetime (the left-hand 
side of (\ref{2-Einstein})) with its matter field content (the right-hand 
side). 

Einstein equations are field equations. Of which field? This is better 
understood in the weak gravitational field limit, that is in the limit of an 
almost flat spacetime. In this case, the metric is approximated by
\begin{equation}
\label{weakfield}
g_{\mu\nu} \sim \eta_{\mu\nu} + h_{\mu\nu}(x) \ , 
\end{equation}
where $\eta_{\mu\nu}$ is the 
Minkowski metric and $h_{\mu\nu}(x)$ is interpreted as the spin-2 graviton field.
In particular, we note that  
\begin{equation}
\label{Newton}
g_{00} = 1 + 2 \Phi \ ,
\end{equation}
where $\Phi$ is the Newtonian potential which satisfies the Poisson equation
$\Delta \Phi = 4 \pi G_{_N} \rho$.

Einstein used his equations not just to describe a given 
gravitational system like a planet or a star but the evolution of the whole
universe. In his days, this was a bold move: it 
should be remembered how little of the universe was known at the time these 
equations were written.
``In 1917, the world was supposed to consist of our galaxy and presumably a 
void beyond. The Andromeda nebula had not yet been certified to lie beyond 
the Milky Way.''[Pais~\cite{Pais} p. 286] Indeed, it is in this context that 
Einstein introduced the cosmological constant in order to have a static 
solution (until it was observed by Hubble that the universe is expanding)
for the universe.

More precisely~\cite{Pais}, Einstein first noticed that a slight modification 
of the Poisson equation, namely
\begin{equation}
\label{7-1a} 
\Delta \Phi - \lambda \Phi = 4\pi\  G_{_N} \rho
\end{equation}
allowed a solution with a constant density $\rho$ ($\Phi = -4 \pi G_{_N} \rho
/ \lambda$) and thus a static Newtonian universe. In the context of general 
relativity, he found a static solution of (\ref{2-Einstein}) under the 
condition that
\begin{equation}
\label{7-1b}
\lambda = {1 \over r^2} = 4 \pi G_{_N} \rho \ ,
\end{equation}
where $r$ is the spatial curvature {\em (see Exercise 1-1)}. 
It was soon shown that this Einstein universe is unstable to small 
perturbations. 

\vskip .5cm
{\em \underline{Exercise 1-1}~: Consider the following metric
\begin{equation}
\label{7-ex1}
g_{00} = 1 \ , \quad g_{ij} = - \delta_{ij} + {x_i x_j \over x^2 - r^2} \ ,
\ x^2 = \sum_{i=1}^3 x_i^2  \ .
\end{equation}

a) Show that it is a solution of Einstein's equations 
(\ref{2-Einstein}) in the case 
of non-relativistic matter with a constant energy 
density $\rho$ satisfying  the condition (\ref{7-1b}). 

b) Prove that, in the Newtonian limit, one recovers (\ref{7-1a}).

\vskip .3cm 
Hints: a) $\Gamma^i{}_{jk} = r^{-2} \left[ x_i \delta_{jk} - x_i x_j x_k/
(x^2-r^2) \right]$ which gives $R_{ij} = - 2 g_{ij}/r^2$.   

b) In the Newtonian limit, $G_{00} \sim \Delta g_{00}$ with $g_{00}$ given by 
(\ref{weakfield}).}

\subsection{An expanding Universe} \label{sect:1-2}

Other solutions to the Einstein equations were soon discovered.
The first exact non-trivial one was found in late 1915 by Schwarzschild,
who was then fighting in the German army, 
within a month of the publication of Einstein's theory and presented on his 
behalf by Einstein at the Prussian Academy in the first days of 1916
\cite{Sc16}, just 
before Schwarzschild death from a illness contracted at the front.
It describes static isotropic regions of empty spacetime ($\lambda = 0$), 
such as the ones 
encountered in the exterior of a static star of mass $M$ and radius $R$:
\begin{equation}
\label{2-Schwarz}
ds^2 = \left( 1 - {2G_{_N} M \over r} \right) dt^2 - 
\left( 1 - {2G_{_N} M \over r} \right)^{-1} dr^2 - r^2 d\theta^2 - r^2 \sin^2 
\theta d\phi^2 \ .
\end{equation}
In 1917, de Sitter proposed a time-dependent vacuum solution in the case where 
$\lambda \not = 0$~\cite{dS17a,dS17b}:
\begin{equation}
\label{deSitter}
ds^2 = dt^2 - e^{2Ht} d{\bf x}^2 \ , \quad H^2 \equiv \lambda^2/3 \ .
\end{equation}
But, since there exists time-dependent solutions, why should the Universe be 
static? This led to the so-called 
``Great Debate'' between Harlow Shapley and Heber D. Curtis in 
1920\footnote{see http://apod.nasa.gov/diamond\_jubilee/debate20.html}.
H. Shapley was supporting the view that the Universe is 
composed of only one big Galaxy, the spiral nebulae being just nearby gas clouds
(he was also arguing --rightfully-- that our Sun is far from the center of this 
big Galaxy). H.D. Curtis, on the other hand considered, that the Universe is 
made 
of many galaxies like our own, and that some of these galaxies had already been 
identified, in the form of the spiral nebulae (and he was supporting the view 
that our Sun is  close to the centre of our relatively small Galaxy).

In 1925, Edwin Hubble studies, with the 100 inch Hooker telescope of Mount 
Wilson, the Cepheids which are variable stars in the  Andromeda nebula M31. He
shows that the distance is even greater than the size proposed by Shapley for
our Milky Way: M31 is a galaxy of its own, the Andromeda galaxy (at a distance 
of $3.10^6$ light-years from us)~\cite{Hu26}. 

\begin{figure}[h]
\begin{center}
\includegraphics[scale=0.45]{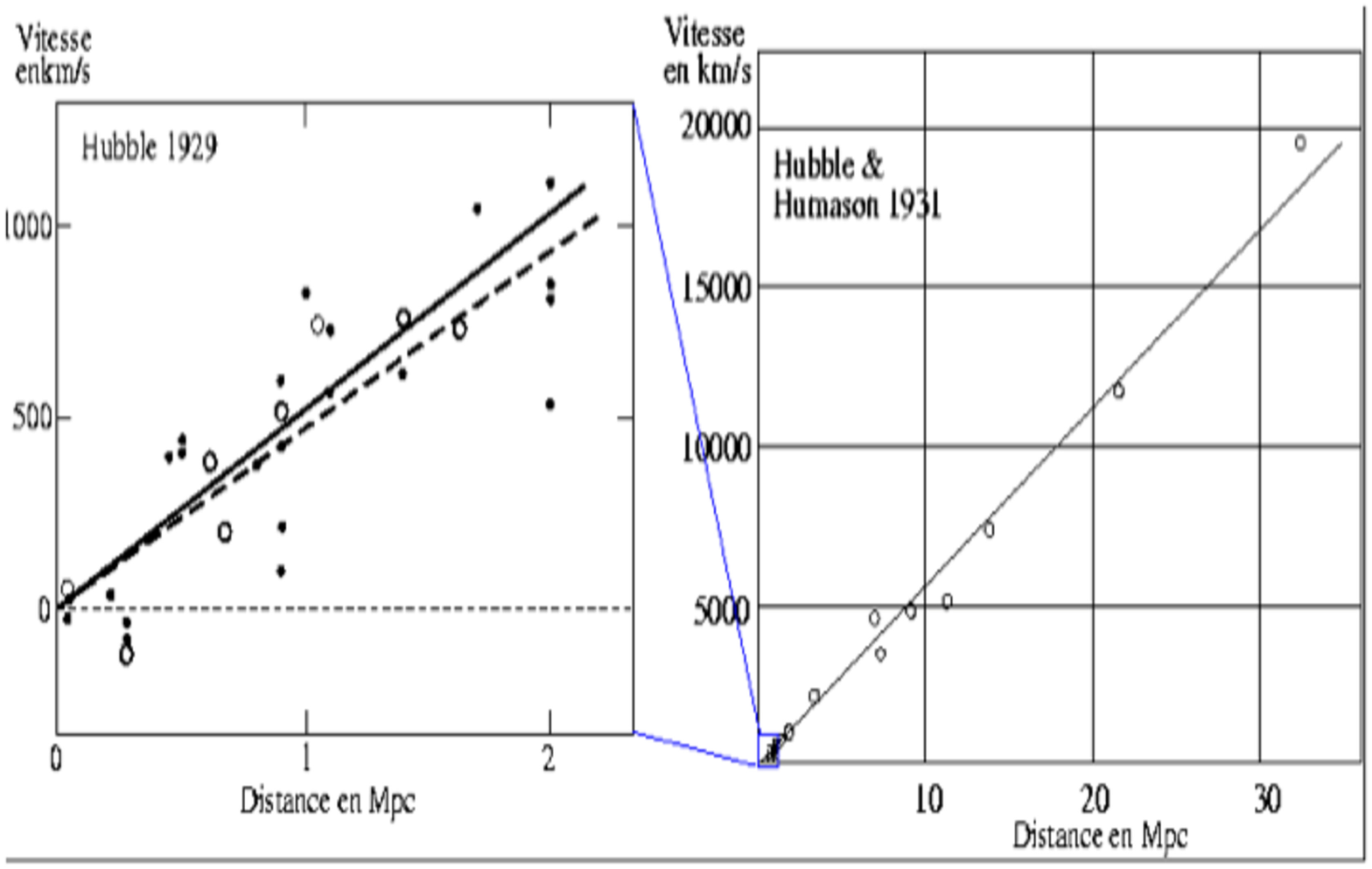}
\end{center}
\vskip -.5cm
\caption{Hubble plot (radial velocity vs distance) of extragalactic ``nebulae''
obtained by Hubble in his original 
article of 1929~\cite{Hu29} and by Hubble and Humason two years later 
\cite{HH31}.}
\label{Hubble_29-31}
\end{figure}

In 1929, Hubble~\cite{Hu29} discovers, by combining spectroscopic measurements 
with measures of distance, that galaxies at a distance $d$ from us 
recede at a velocity $v$ following the law:
\begin{equation}
\label{Hubble}
v \sim H_0 d \ ,
\end{equation}
the constant of proportionality being henceforth called the Hubble 
constant (see Fig.~\ref{Hubble_29-31})\footnote{Hubble's result  
was actually anticipated by 
G. Lema\^itre who 
published the law (\ref{Hubble}) two years earlier~\cite{Le27}.}.
As a consequence of the Hubble law, the Universe is expanding! 

The velocities $v$ are measured by Hubble through the Doppler shift of
spectroscopic lines of the light emitted by the galaxy: if $\lambda_{{\rm emit}}$
is the wavelength of some spectrocopic line of the light emitted by a galaxy 
(receding from us at velocity $v$) and  
$\lambda_{{\rm obs}}$ of the corresponding line in the light observed by us, 
then 
\begin{equation}
\label{Doppler}
1+z \equiv {\lambda_{{\rm obs}} \over \lambda_{{\rm emit}}} = {1 + v/c \over 
\sqrt{1-v^2/c^2}} \sim 1 + v/c \ .
\end{equation}
Hence $v \sim zc$. 

Let us note that stars within galaxies, such are our Sun, are not individually 
subject to the expansion: they have fallen into the gravitational potential
ofthe galaxy and are thus not receding from one another ({\em see 
Exercise 1-2}). This is why one had 
first to identify extragalactic objects before discovering the expansion: only
galaxies, or clusters of galaxies, recede from one another.

\vskip .5cm
{\em \underline{Exercise 1-2}~: The purpose of this exercise is to show 
that, 
in the case where a fluctuation of density appears (such as when galaxies 
form), the massive objects (stars) decouple from the general expansion to fall 
into the local gravitational potential~\cite{Pe67,We87}. 

In a matter-dominated universe of uniform density 
$\rho$, a perturbation appears in the form of a sphere with uniform excess
density $\Delta \rho$.  Assuming that the gravitational field inside the 
sphere is described 
by the Robertson-Walker metric with positive curvature constant $\Delta k>0$,
the evolution is governed by the equations
\begin{eqnarray}
{{\dot a}^2   \over a^2 } + {\Delta k \over a^2}  &=& 
{1 \over 3} (\rho + \Delta \rho) \ , \label{2-ex1} \\ 
a^3 (\rho + \Delta \rho) &=& C \ , \label{2-ex2}
\end{eqnarray}
where $C$ is a constant and we neglect the cosmological constant.

a) Show that the solution is given parametrically by
\begin{eqnarray}
a(t) &=& {C \over 6 \Delta k} (1 - \cos \eta) \ , \label{2-ex3} \\
t &=& t_0 + {C \over 6 \Delta k^{3/2}} (\eta - \sin \eta) \ . \label{2-ex4}
\end{eqnarray}
Note that this does not assume that $\Delta \rho$ is small.

b) Show that, as $t \rightarrow t_0$, one has $a(t) = (C/3)^{1/3}\left[3 
(t-t_0)/2\right]^{2/3}\sim (t-t_0)^{2/3}$, as in
the rest of the matter-dominated universe. Verify that, whereas $\rho =
(4/3) (t-t_0)^{-2}(\sim 
a^{-3})$, $\Delta \rho =(9/5) \Delta k a^{-2} \sim (t-t_0)^{-4/3}$.

c) How long does it take before the system starts to collapse to a bound 
system or to a singularity?

\vskip .3cm 
Hints: a) $a {\dot a}^2 + a \Delta k = C/3$

b) Make an expansion in $\eta$ to the subleading order ($\eta$ is small 
when $t \rightarrow t_0$).

c) From (\ref{2-ex3}) the expansion stops and the collapse starts at $\eta 
= \pi$ or $t-t_0 = \pi C /(6 \Delta k^{3/2})$. }

\vskip .5cm
\underline{Exercise 1-3}~: Identify the redefinition of coordinates $\bar t = 
\bar t(t,r)$, $\bar r = \bar r (t,r)$ which allows to write the de Sitter 
metric (\ref{deSitter})  
\begin{equation}
\label{deSitterflat}
ds^2 = dt^2 - e^{2Ht} (dr^2 + r^2 d\Omega^2) \ ,
\end{equation}
into the following form:
\begin{equation}
\label{deSitterstatic}
ds^2 = \left(1- {\bar r^2 \over R_H^2} \right) d\bar t^2 -  
\left(1- {\bar r^2 \over R_H^2} \right)^{-1} d{\bar r}^2 - \bar r^2 d\Omega^2 
\ , \quad R_H \equiv 
H^{-1} \ .
\end{equation}
The first equation is known as the flat form of the de Sitter metric (see 
(\ref{2-2a}) below), the second one is the static form (compare with the 
Schwarzschild metric (\ref{2-Schwarz})).

\vskip .3cm 
Hints:  $e^{Ht} =  e^{H \bar t}\sqrt{1-H^2\bar r^2}$ and $r = \bar r e^{-Ht}$.

\subsection{Friedmann-Lema\^itre-Robertson-Walker universe} \label{sect:1-3}

As one gets to larger and larger distances, the Universe becomes more 
homogeneous and isotropic. Under the assumption that it reaches homogeneity 
and isotropy on scales
of order $100$ Mpc ($1$ pc $= 3.262$ light-year $= 3.086 \times 10^{16}$ 
m) and larger, one may first try to find a
homogeneous and isotropic metric as a solution of Einstein's equations. 
The most 
general ansatz is, up to coordinate redefinitions, the Robertson-Walker metric:
\begin{eqnarray}
ds^2 &=& c^2 dt^2 -  a^2(t) \ \gamma_{ij} dx^i dx^j, \label{2-2a} \\
\gamma_{ij} dx^i dx^j &=& {dr^2 \over 1-kr^2} + r^2 \left( 
d\theta^2 + \sin^2 \theta d\phi^2 \right) , \label{2-2b}
\end{eqnarray}
where $a(t)$ is the cosmic scale factor, which is time-dependent in an 
expanding or contracting universe. Such a universe is called a 
Friedmann-Lema\^{\i}tre 
universe. The constant $k$ which appears in the spatial metric $\gamma_{ij}$ 
can take the values $\pm 1$ or $0$: the value $0$ corresponds to flat space, 
i.e. usual 
Minkowski spacetime; the value $+1$ to closed space ($r^2 <1$) and the value 
$-1$ to open space. Note that $r$ is dimensionless whereas $a$ has the 
dimension of a length. From now on, we set $c=1$, except otherwise stated.

For the energy-momentum tensor that appears on the right-hand side of 
Einstein's equations, we follow our assumption of homogeneity 
and isotropy and assimilate the content of the Universe to a perfect fluid:
\begin{equation}
\label{2-4}
T_{\mu\nu} = -  p g_{\mu\nu} + (p+\rho ) U_\mu U_\nu \quad ,
\end{equation}
where  $U^\mu$ is the velocity 4-vector ($U^t=1, U^i = 0$). It follows from  
(\ref{2-4}) that $T_{tt} = \rho$ and $T_{ij}=a^2 p \gamma_{ij}$. 
The pressure $p$ and energy density $\rho$ usually satisfy the equation 
of state:
\begin{equation}
\label{2-5}
p = w \rho \quad .
\end{equation}
The constant $w$, called the equation of state parameter, takes the value 
$w \sim 0$ for non-relativistic matter 
(negligible pressure) and $w=1/3$ for relativistic matter (radiation).
In all generality, the perfect fluid consists of several components with 
different values of $w$.

One obtains from the $(0,0)$ and $(i,j)$ components of the Einstein  
equations (\ref{2-Einstein}) ({\em see Exercise B-1 of Appendix~\ref{app:B}}):
\begin{eqnarray}
3  \left( {{\dot a}^2   \over a^2 } + {k \over a^2} \right) &=& 
8\pi G_{_N} \rho + \lambda, \label{2-6}\\
 {\dot a}^2 + 2 a {\ddot a} + k &=& - 8 \pi G_{_N} a^2 p  + a^2 \lambda,
\label{2-7}
\end{eqnarray}
where  we  use standard notations: $\dot a$ is the first time derivative of
the cosmic scale factor, $\ddot a$  the second time derivative.

The first of the preceding equations can be written as  the Friedmann
equation, which gives an expression for the Hubble parameter 
$H \equiv \dot a /a$ 
measuring the rate of the expansion of the Universe:
\begin{equation}
\label{2-8}
H^2 \equiv {{\dot a}^2   \over a^2 } = {1\over 3} \left( \lambda
+ 8 \pi G_{_N} \rho \right) - {k \over a^2} \ .
\end{equation}
The cosmological constant appears  as a constant contribution to 
the Hubble parameter.  We note that, setting $k=0$ and $\rho = 0$, one recovers 
de Sitter solution (\ref{deSitter}): $\dot a^2/a^2 = \lambda/3$ i.e. $a(t) \sim
e^{Ht}$. For the time being, we will set $\lambda$ to zero and return to it in 
subsequent chapters.

Next, we note that, assuming $k = 0$, we have at present time:
\begin{equation}
\label{4-3}
\rho = {3 H_0^2 \over  8 \pi G_{_N}} \equiv \rho_c \sim 10^{-26} 
{\rm kg}/{\rm m}^3 \ ,
\end{equation}
where $H_0$ is the Hubble constant, i.e. the present value of the Hubble 
parameter. This corresponds to approximately one galaxy per Mpc$^3$ or $5$ 
protons per m$^3$. In fundamental units where $\hbar = c = 1$, this is of the 
order of $\left( 10^{-3} \hbox{eV} \right)^4$. We easily deduce from 
(\ref{2-8}) that space is open (resp. closed)
if at present time $\rho < \rho_c$ (resp. $\rho < \rho_c$). Hence the name 
critical density for $\rho_c$.

Friedmann equation (\ref{2-8}) can be understood on very simple grounds: 
since the universe at large scale is homogeneous and
isotropic, there is no specific location and motion in the universe 
should not allow to identify any such location. This implies that the most    
general motion has the form ${\bf v}(t) = H(t) {\bf x}$ where ${\bf x}$ and
${\bf v}$ denote the position and the velocity and $H(t)$ is an arbitrary 
function of time. Since ${\bf v} = \dot {\bf x}$, one obtains ${\bf x} =
a(t) {\bf r}$, where ${\bf r}$ is a constant for a given body (called the 
comoving coordinate) and $a(t)$ is related to $H(t)$ through $H = \dot a / a$. 
Now, consider a particle of mass $m$ located at position ${\bf x}$: the
sum of its kinetic and gravitational potential energy is constant. Denoting by
$\rho$ the energy density of the (homogeneous) universe, we have
\begin{equation}
\label{2-9}
{1 \over 2} m {\bf v}^2 - {4 \pi \over 3} G_{_N} m \rho  \left| {\bf x} 
\right|^2 = \hbox{cst} \ .
\end{equation}
Writing this constant $-km{\bf r}^2/2$, we obtain from (\ref{2-9})
\begin{equation}
\label{2-10}
{{\dot a}^2   \over a^2 } = {8 \pi \over 3}  G_{_N} \rho  - {k \over a^2} \ , 
\end{equation}
which is nothing but Friedmann equation (\ref{2-8}) (with vanishing 
cosmological constant).

Friedmann equation should be supplemented by the conservation of the 
energy-momentum tensor which simply yields:
\begin{equation}
\label{2-11}
\dot \rho = - 3 H (p + \rho) \quad .
\end{equation}
Hence a component with equation of state (\ref{2-5}) has its energy density
scaling as $\rho \sim a(t)^{-3(1+w)}$. Thus non-relativistic matter
(often referred to as matter) energy density scales as $a^{-3}$. In other 
words, the energy density of matter 
evolves in such a way that $\rho a^3$ remains 
constant. Radiation scales as $a^{-4}$ and a component with equation of 
state $p=-\rho$ ($w=-1$) has constant  energy density\footnote{The latter case 
corresponds to a cosmological constant as can be seen from
(\ref{2-6}-\ref{2-7}) where the cosmological constant can be replaced by a 
component with $\rho_\Lambda = - p_\Lambda = \lambda/(8\pi G_{_N})$.}.

We note for future use that, if a component  with equation of state 
(\ref{2-5}) dominates the energy density of the universe (as well as the 
curvature term $-k/a^2$), then (\ref{2-10}) has a scaling solution 
\begin{equation}
\label{2-11a}
a(t) \sim t^\nu \ , \quad \hbox{with} \ \nu = {2 \over 3(1+w)} \ .
\end{equation} 
For example, in a matter-dominated universe, $a(t) \sim t^{2/3}$, and in a 
radiation-dominated universe, $a(t) \sim t^{1/2}$.

Differentiating the Friedmann equation with respect to time, and using the
energy-momentum conservation (\ref{2-11}), one easily obtains
\begin{equation}
\label{2-12}
\ddot a = - {4 \pi G_{_N} \over 3} a (3p + \rho) + a{\lambda \over 3} \ .
\end{equation}
This allows to recover (\ref{2-7}) from Friedmann equation and energy-momentum 
conservation.

\subsection{Redshift} \label{sect:1-4}

In an expanding or contracting universe, the Doppler frequency shift undergone 
by the light emitted from a distant source 
gives a direct information on the time 
dependence of the cosmic scale factor $a(t)$. To obtain the explicit relation, 
we consider a photon propagating in a fixed direction ($\theta$ and $\phi$ 
fixed). Its equation of motion is given as in special relativity by setting 
$ds^2=0$ in (\ref{2-2a}):
\begin{equation}
\label{2-13}     
c^2 dt^2 = a^2(t) {dr^2 \over 1-kr^2} \quad .
\end{equation}
Thus, if a photon (an electromagnetic wave) leaves at time $t$ a galaxy 
located at distance $r$ from  us, it will reach us at time $t_0$ such that
\begin{equation}
\label{2-14}
\int_{t}^{t_0} {cdt \over a(t)} = \int_0^{r} {dr \over \sqrt{1-kr^2}}
\end{equation}
The electromagnetic wave is emitted with the same amplitude at a time
$t+T$ where the period $T$ is related to the wavelength of the 
emitted wave $\lambda$ by  the relation $\lambda = cT$. It is thus 
received with the same amplitude at the time $t_0+T_0$ given by
\begin{equation}
\label{2-15}
\int_{t+T}^{t_0+T_0} {cdt \over a(t)} 
= \int_0^{r} {dr \over \sqrt{1-kr^2}} \quad ,
\end{equation}
the wavelength of the received wave being simply $\lambda_0 = cT_0$. Since
$T_0,T\ll t_0,t$, we obtain from comparing (\ref{2-14}) and (\ref{2-15})
\begin{equation}
\label{2-16}
{cT_0 \over a_0} = {cT \over a(t)} \qquad \hbox{i.e.} \qquad
{\lambda_0 \over \lambda} = {a_0 \over a(t)} \quad ,
\end{equation}
where $a_0$ is the present value of the cosmic scale factor.

Defining the redshift parameter $z$ as the fractional increase in wavelength
$z= (\lambda_0-\lambda)/\lambda$, we have
\begin{equation}
\label{2-17}
1+ z = {a_0 \over a(t)}  \quad .
\end{equation}
One may thus replace time by redshift since time decreases monotonically 
as redshift increases.

\subsection{The universe today: energy budget} \label{4s1}

The Friedmann equation 
\begin{equation}
\label{4-1}
H^2 \equiv {{\dot a}^2   \over a^2 } = {1\over 3} \left( \lambda
+ 8 \pi G_{_N} \rho \right) - {k \over a^2} \ .
\end{equation} 
allows to define the Hubble constant 
$H_0$, i.e. the present value of the Hubble parameter, which sets the scale 
of our Universe at present time. Because of the troubled history of the 
measurement of the Hubble constant, it has become customary to express it 
in units of $100 \ \hbox{km}.\hbox{s}^{-1}.\hbox{Mpc}^{-1}$ which gives its 
order of magnitude. Present measurements give 
$$h_0 \equiv {H_0 \over 100 \ \hbox{km}.\hbox{s}^{-1}.\hbox{Mpc}^{-1}} 
= 0.7\pm0.1 \quad .$$
The corresponding length and time scales are:
\begin{eqnarray}
\ell_{H_0} & \equiv & {c \over H_0} = 3000 \ h_0^{-1} \ \hbox{Mpc} = 9.25 
\times 10^{25} \ h_0^{-1} \ \hbox{m}, 
\label{4-2a}\\
t_{H_0} & \equiv & {1 \over H_0} = 3.1 \times 10^{17}\ h_0^{-1} \ \hbox{s} = 
9.8 \ h_0^{-1} \ \hbox{Gyr}. \label{4-2b}
\end{eqnarray}

It has become customary to normalize the different forms of energy density in 
the present Universe in terms of the critical density $\rho_c = 3H_0^2/
(8\pi G_{_N})$ defined in (\ref{4-3}). Separating the energy 
density $\rho_{_{M0}}$ presently stored in non-relativistic matter 
(baryons, neutrinos, dark matter,...) from the density $\rho_{_{R0}}$ 
presently stored in radiation (photons, relativistic neutrino if any), 
one defines:
\begin{equation}
\label{4-4}
\Omega_{_M} \equiv {\rho_{_{M0}} \over \rho_c}, \; \; \; \; 
\Omega_{_R} \equiv {\rho_{_{R0}} \over \rho_c}, \; \; \; \; 
\Omega_{\Lambda} \equiv {\lambda \over 3 H_0^2}, \; \;
\Omega_k \equiv -{k \over  a_0^2 H_0^2} \ .
\end{equation}
The last term comes from the spatial curvature and is not strictly speaking
a contribution to the energy density. One may add other components: we will 
refrain to do so in this Chapter and defer this to the last one.
 
Then the Friedmann equation taken at time $t_0$ simply reads 
\begin{equation}
\label{4-5}
\Omega_{_M} + \Omega_{_R} + \Omega_{\Lambda} = 1 - \Omega_k  \ .
\end{equation}
Since matter dominates over radiation in the present Universe, we may neglect 
$\Omega_{_R}$ in the preceding equation. As we will see in the next Chapters, 
present observational data tend to favor the following set of values: 
$\Omega_k\sim 0$ (see Section~\ref{4s3}) and $\Omega_{_M} \sim 0.3$, 
$\Omega_\Lambda \sim 0.7$ (see Section~\ref{sect:1obs-1} and Fig. 
\ref{fig4-2}). The matter density is not consistent with the observed density 
of luminous matter ($\Omega_{{\rm luminous}} \sim 0.04$) and thus calls for a 
nonluminous form of matter, dark matter which will be studied in Section 
\ref{chap:4}.

Using the dependence of the 
different components with the scale factor $a(t)= a_0/(1+z)$,
one may then rewrite the Friedmann equation at any time as:
\begin{eqnarray}
\label{4-6}
H^2(t)  &=& H_0^2 \left[ \Omega_\Lambda 
+ \Omega_{_M} \left({a_0 \over a(t)}\right)^3
+ \Omega_{_R} \left({a_0 \over a(t)}\right)^4
+ \Omega_k  \left({a_0 \over a(t)}\right)^2 \right] \ , \\
\hbox{or} \ H^2(z) &=& H_0^2 \left[ \Omega_{_M} (1+z)^3 +  \Omega_{_R} (1+z)^4
+ \Omega_{k} (1+z)^2 + \Omega_\Lambda \right] \ . \label{4-31}
\end{eqnarray}
where $a_0$ is the present value of the cosmic scale factor and all time 
dependences (or alternatively redshift dependence) have been written 
explicitly.
We note that, even if $\Omega_{_R}$ is negligible in (\ref{4-5}), 
this is not so in the 
early Universe because the radiation term increases faster than the matter 
term in (\ref{4-6}) as one gets back in time (i.e. as $a(t)$ decreases).
If we add an extra component $X$ with equation of state $p_{_X} = w_{_X} 
\rho_{_X}$, it contributes an extra term  
$\Omega_{_X} \left(a_0 / a(t)\right)^{3(1+w_{_X})}$
where $\Omega_{_X} = \rho_{_X}/\rho_c$.

An important information about the evolution of the universe at a given time
is whether its expansion is accelerating or decelerating.
The acceleration of our universe  is usually measured by the 
deceleration parameter $q$ which is defined as:
\begin{equation}
\label{4-7}  
q \equiv - {\ddot a a \over \dot a^2}  \quad .
\end{equation}
Using (\ref{2-12}) of Section~\ref{chap:2} and separating again matter and 
radiation, we may write it at present time $t_0$ as:
\begin{equation}
\label{4-8}  
q_0 =  - {1 \over H_0^2} \left( {\ddot a  \over a} \right)_{t=t_0}
= {1 \over 2} \Omega_{_M} + \Omega_{_R} -  \Omega_{\Lambda} \quad .
\end{equation}
Once again, the radiation term $\Omega_{_R}$ can be neglected in this relation.
We see that in order to have an acceleration of the expansion ($q_0 <0$), we 
need the cosmological constant to dominate over the other terms.

We can also write the deceleration parameter in (\ref{4-7}) in terms of 
redshift as in  (\ref{4-31})
\begin{equation}
\label{4-33}
q = {H_0^2 \over 2H(z)^2} \left[  \Omega_{_M} (1+z)^3 
+  2 \Omega_{_R} (1+z)^4 - 2 \Omega_\Lambda \right] \quad .
\end{equation}
This shows that the universe starts accelerating at redshift values
$1 + z \sim \left(2 \Omega_\Lambda/\Omega_{_M} \right)^{1/3}$ (neglecting 
$\Omega_{_R}$), that is typically redshifts of order $1$.

The measurements of the Hubble constant and of the deceleration parameter 
at present time
allow to obtain the behaviour of the cosmic scale factor in the last 
stages of the evolution of the universe:
\begin{equation}
\label{4-9}
a(t) = a_0 \left[ 1 + {t-t_0 \over t_{H_0}} - {q_0 \over 2} 
{(t-t_0)^2 \over t_{H_0}^2} + \cdots \right] \ .
\end{equation}

\subsection{The early universe}
\label{sect:1-5}

Table~\ref{tab5-1} summarizes the history of the universe in the context of 
the big bang model with an inflationary epoch. We are referring to the 
different stages using time, redshift or temperature. The last two can be 
related using the conservation of entropy.

Indeed, one can show, using the second law of thermodynamics ($TdS=dE+pdV$) 
that the entropy per unit volume is simply the quantity 
\begin{equation}
\label{5-00a}
s \equiv {S \over V} = {\rho + p \over T}
\end{equation}
and that the entropy in a covolume $sa^3$ remains constant. The entropy 
density is dominated by relativistic particles and reads 
\begin{eqnarray}
\label{5-00b}
s &=& {2 \over 3} g_s(T) a_{_{BB}}  T^3 \quad ,  \\
g_s(T) &=& \sum_{{\rm bosons}\ i} g_i \left({T_i \over T}\right)^3 +
{7 \over 8}\sum_{{\rm fermions}\ i} g_i \left({T_i \over T}\right)^3 \quad 
\label{5-00c},
\end{eqnarray}
where the sum extends {\em only} to the species in thermal equilibrium
and $a_{_{BB}} \equiv \pi^2 k^4 / (15 c^3 \hbar^3) = 7.56 \times 10^{-16} 
\ {\rm J} \cdot{\rm m}^{-3}\cdot{\rm K}^{-4}$ is the blackbody constant.

We deduce from the constancy of $sa^3$  that  $g_s \left(aT\right)^3$
remains constant. Hence the temperature $T$ of the universe behaves as $a^{-1}$
whenever $g_s$ remains constant. We conclude that as we go back in time ($a$ 
decreases), temperature increases, as well as energy density. The early universe
is hot and dense. It might even reach a stage where our equations no longer 
apply because it becomes infinitely hot and dense: this is the initial 
singularity, sarcastically (but successfully) called big bang by Fred Hoyle, on 
a BBC radio show in 1949. 

We note that some caution has to be paid whenever  some species drop out of
thermal equilibrium. Indeed, a given species drops out of equilibrium when 
its interaction rate  $\Gamma$ drops below the expansion rate $H$. For 
example  
neutrinos decouple  at temperatures below 1 MeV. Their temperature continues 
to decrease as $a^{-1}$ and thus remains equal to $T$. However when  $kT$ 
drops below $2 m_e$, electrons annihilate against positrons with no 
possibility of being regenerated and the entropy of the electron-positron pairs
is transferred to the photons. Since  $\left. g_s\right|_{\gamma,e^\pm }
= 2+4\cdot 7/8=11/2$ and  $\left. g_s\right|_{\gamma} = 2$, the 
temperature of the photons becomes multiplied by a factor $(11/4)^{1/3}$. Since
the neutrinos have already decoupled, they are not affected by this entropy
release and their temperature remains untouched. Thus we have
\begin{equation}
\label{5-00d}
{T \over T_\nu} = \left( {11 \over 4} \right)^{1/3} \sim 1.40 \quad .
\end{equation}
We can then compute the value of $g_s$ for temperatures much smaller than 
$m_e$:
$g_s = 2+(7/8)6(4/11)=3.91$. We deduce that, at present time ($T_0 =
2.725$ K), $s_0/k= 2890 \
{\rm cm}^{-3}$. 

\vskip .5cm
\begin{table}
\caption{The different stages of the cosmological evolution in the standard 
scenario, given in terms of time $t$ since the big bang singularity, the 
energy $kT$ of the background  photons and the redshift $z$. The double line 
following nucleosynthesis indicates the part of the evolution which has been 
tested through observation. The values 
$(h_0=0.7,\Omega_{_M}=0.3, \Omega_\Lambda =0.7)$ are adopted to compute 
explicit values.}
\begin{center}
\begin{tabular}{|c|c|c|l|}
\hline
&&&\\
$t$ & $k T_\gamma$ (eV) & $z$ &  \\
&&& \\
\hline
&&& \\
$t_0 \sim 15$ Gyr & $2.35 \times 10^{-4}$ & $0$ & now \\
&&& \\
\hline
&&& \\
$\sim$ Gyr & $\sim 10^{-3}$ & $\sim 10$ & formation of galaxies \\
&&& \\
\hline
&&& \\
$t_{_{\rm rec}} \sim 4\times 10^5$ yr & $0.26$ & $1100$ & recombination \\
&&& \\
\hline
&&& \\
$t_{_{\rm eq}} \sim 4\times 10^4$ yr & $0.83$ & $3500$ & matter-radiation 
equality \\
&&& \\
\hline
&&& \\
$3$ min & $6 \times 10^4$ & $2\times 10^8$ & nucleosynthesis\\
&&& \\
\hline \hline
&&& \\
$1$ s & $10^6$ & $3\times 10^{9}$ & $e^+ e^-$ annihilation \\
&&& \\
\hline&&& \\
$4\times 10^{-6}$ s & $4\times 10^8$ & $10^{12}$ & QCD phase transition \\
&&& \\
\hline
&&& \\
$<4\times 10^{-6}$ s & $> 10^9$ & & baryogenesis \\
&&& \\
\hline
&&& \\
& & & inflation \\
&&& \\
\hline
&&& \\
$t=0$ & & $\infty$ & big-bang \\
&&& \\
\hline
\end{tabular}
\end{center}

\label{tab5-1}
\end{table}

As we have seen in  (\ref{2-11a}),
it follows from the Friedmann equation  that, if the Universe is 
dominated by a component of equation of state $p=w\rho$, 
then the cosmic scale factor $a(t)$
varies with time as $t^{2/[3(1+w)]}$. We start at time $t_0$ with the 
energy budget: $\Omega_{_M}=0.3, \Omega_\Lambda =0.7, \Omega_k \sim 0$ 
(see next Chapters). Radiation consists of photons and relativistic neutrinos. 
Since generically
\begin{equation}
\label{5-00e}
\rho_{_R} = {1 \over 2} g(T) a_{_{BB}} T^4 \ ,
\end{equation}
where $g(T)$ is the effective number of degrees of freedom
\begin{equation}
\label{5-00f} 
g(T) = \sum_{{\rm bosons}\ i} g_i \left({T_i \over T}\right)^4 +
{7 \over 8}\sum_{{\rm fermions}\ i} g_i \left({T_i \over T}\right)^4 
\end{equation}
(we have taken into account the possibility that the species $i$ may have 
a thermal distribution at a temperature $T_i$ different from the temperature 
$T$ of the photons), we have
\begin{equation}
\label{5-0a}
\rho_{_R}(t) = \rho_\gamma(t) \left[ 1 + {7 \over 8} 
\left({4 \over 11}\right)^{4/3}
 \ N_\nu^{{\rm rel}}(t) \right] 
\quad ,
\end{equation}
where $N_\nu^{{\rm rel}}(t)$ is the number of relativistic neutrinos at time 
$t$. At present time $t_0$, we have $\Omega_\gamma = \rho_\gamma(t_0)/
\rho_c = 2.48 \times 10^{-5} \ h^{-2}$ and the mass limits on neutrinos imply 
$N_\nu^{{\rm rel}}(t_0) \le 1$. In any case, $\Omega_{_R} \ll \Omega_{_M}$.

For redshifts larger than 1, the cosmological constant becomes subdominant and 
the universe is matter-dominated ($a(t) \sim t^{2/3}$). In the early phase of 
this matter-dominated epoch, the Universe is a ionized plasma with electrons
and protons: it is opaque to photons. But, at a time $t_{_{{\rm rec}}}$, electrons 
recombine with the protons to form atoms of hydrogen and, 
because hydrogen is neutral, this induces the decoupling of matter and photon: 
from then on ($t_{_{{\rm rec}}} <t<t_0$), the universe is transparent\footnote{After 
recombination, the intergalactic medium remains neutral during a period
often called the dark ages, until the first stars ignite and the first quasars 
are formed. The ultraviolet photons produced by these sources progressively 
then re-ionize the universe. This period, 
called the re-ionization period, may be long since only small volumes 
around the first galaxies start to be ionized until these volumes coalesce to 
re-ionize the full intergalactic medium. But, in any case, the 
universe is then sufficiently dilute to prevent recoupling.}.
This is the important recombination stage. After decoupling the energy density 
$\rho_\gamma \sim T^4$ of the primordial photons is redshifted according to the 
law
\begin{equation}
\label{5-0c}
{T(t) \over T_0} = {a_0 \over a(t)} = 1 + z \quad .
\end{equation}
One observes presently this cosmic microwave background (CMB) as a radiation 
with a black-body spectrum at temperature $T_0 = 2.725$ K or energy 
$kT_0 = 2.35 \times 10^{-4}$ eV. 

Since the binding energy of the ground state of atomic hydrogen is $E_b = 13.6$
eV, one may expect that the energy $k T_{_{{\rm rec}}}$ is of the same order. 
It is substantially smaller because of the smallness of the ratio of baryons 
to photons $\eta = n_b/n_\gamma \sim 5 \times 10^{-10}$. {\em Indeed, according
to the Saha equation, the fraction $x$ of ionized atoms is given by
\begin{equation}
\label{5-0d} 
{n_p n_e \over n_H n_\gamma} = {x^2 \over (1-x)} \eta = {4.05 c^3 \over \pi^2}
\left( {m_e \over 2\pi k T} \right)^{3/2} e^{-E_b/kT} \quad . 
\end{equation}
Hence, because $\eta \ll 1$, the ionized fraction $x$ becomes negligible 
only for energies much smaller than $E_b$.} A careful treatment gives 
$k T_{_{{\rm rec}}}\sim 0.26$ eV.

As we proceed back in time, radiation energy 
density increases more rapidly (as $a(t)^{-4}$) than matter ($a(t)^{-3}$) 
(as $a(t)$ decreases). At time $t_{{\rm eq}}$, there is 
equality. This corresponds to  
\begin{equation}
\label{5-0b}
{1 \over 1 + z_{{\rm eq}}} = {a(t_{{\rm eq}}) \over a_0} 
= {1.68 \ \Omega_\gamma \over \Omega_{_M}} = {4.17 \times 
10^{-5} \over \Omega_{_M} h_0^2} \quad ,
\end{equation}
where we have assumed 3 relativistic neutrinos at this time.


As we go further back in time, we presumably reach a period where matter
overcame antimatter. It is indeed a great puzzle of our Universe to observe so 
little antimatter, when our microscopic theories treat matter and antimatter on 
equal footing. More quantitatively, one has to explain the following very small 
number:
\begin{equation}
\label{asymmetry}
\eta \equiv {n_B \over n_\gamma} = {n_b - n_{\bar b} \over n_\gamma} \sim 
6 \times 10^{-10} \ ,
\end{equation}
where $n_B=n_b - n_{\bar b}$ (resp. $n_\gamma$) is the baryon (resp. photon)
number density, based on baryon $b$ and antibaryon $\bar b$ counts. The actual 
number comes from the latest Planck data~\cite{Planck13_16}.

Sakharov~\cite{Sa67} gave in 1967 the necessary ingredients to 
generate an asymmetry between matter and antimatter:
\begin{itemize}
\item a process that destroys baryon number,
\item a violation of the symmetry between matter and antimatter (the so-called 
charge conjugation), as well as a violation of the time reversal symmetry,
\item an absence of thermal equilibrium.
\end{itemize}  
The Standard Model ensures the second set of conditions (CP violation which was 
discovered by Cronin and Fitch~\cite{CF64} is accounted for by the phase of the 
CKM matrix). The expanding early universe provides the third condition. It 
remains to find a process that destroys baryon number. Different roads were 
followed: non-perturbative processes (sphaleron) at the electroweak phase 
transition; proton decay in the context of grand unified theories; decay of 
heavy neutrinos which leads to lepton number violation, and consequently to 
baryon number violation (leptogenesis).

\section{The days where cosmology became a quantitative science: cosmic 
microwave background} 
\label{chap:2}

We have recalled briefly in Section~\ref{sect:1-5} the history of the Universe
(see Table~\ref{tab5-1}). The very early universe is a ionized plasma, and thus 
is opaque to light. But we have seen that, soon after matter-radiation 
equality, electrons recombined with the protons to form neutral atoms of 
hydrogen, which induces the decoupling of matter and photon. From this epoch 
on, the universe becomes transparent to light. The primordial gas of photons 
produced at this epoch cools down as the universe expands (following 
\ref{5-0c}) and forms nowadays the cosmic microwave background (CMB). 

Bell Labs radio astronomers Arno Penzias and Robert Wilson were using a large 
horn antenna in 1964 and 1965 to map signals from the Milky Way, when they 
accidently discovered the CMB~\cite{PW65}. 
The discovery of this radiation was a major confirmation of the hot big bang 
model: its homogeneity 
and isotropy was a signature of its cosmological origin. However, the degree 
of homogeneity and isotropy of this radiation was difficult to reconcile with 
the history of the Universe as understood from the standard big bang theory: 
radiation coming from regions of the sky which were not supposed to be causally 
connected in the past had exactly the same properties.

It was for such reasons that the scenario of inflation was
proposed~\cite{St80,Gu81}: an exponential expansion of the Universe right 
after the big bang. In the solution proposed by A. Guth~\cite{Gu81}, the set up 
was the spontaneous breaking of the grand unified theory: the corresponding 
phase transition was providing the vacuum energy necessary to initiate such 
an exponential expansion. This scenario proved to be difficult to realize and it
was followed by many variants: new inflation~\cite{Li82}, chaotic inflation
\cite{Li83}, ...    

\begin{figure}[t]
\begin{center}
\includegraphics[scale=0.45]{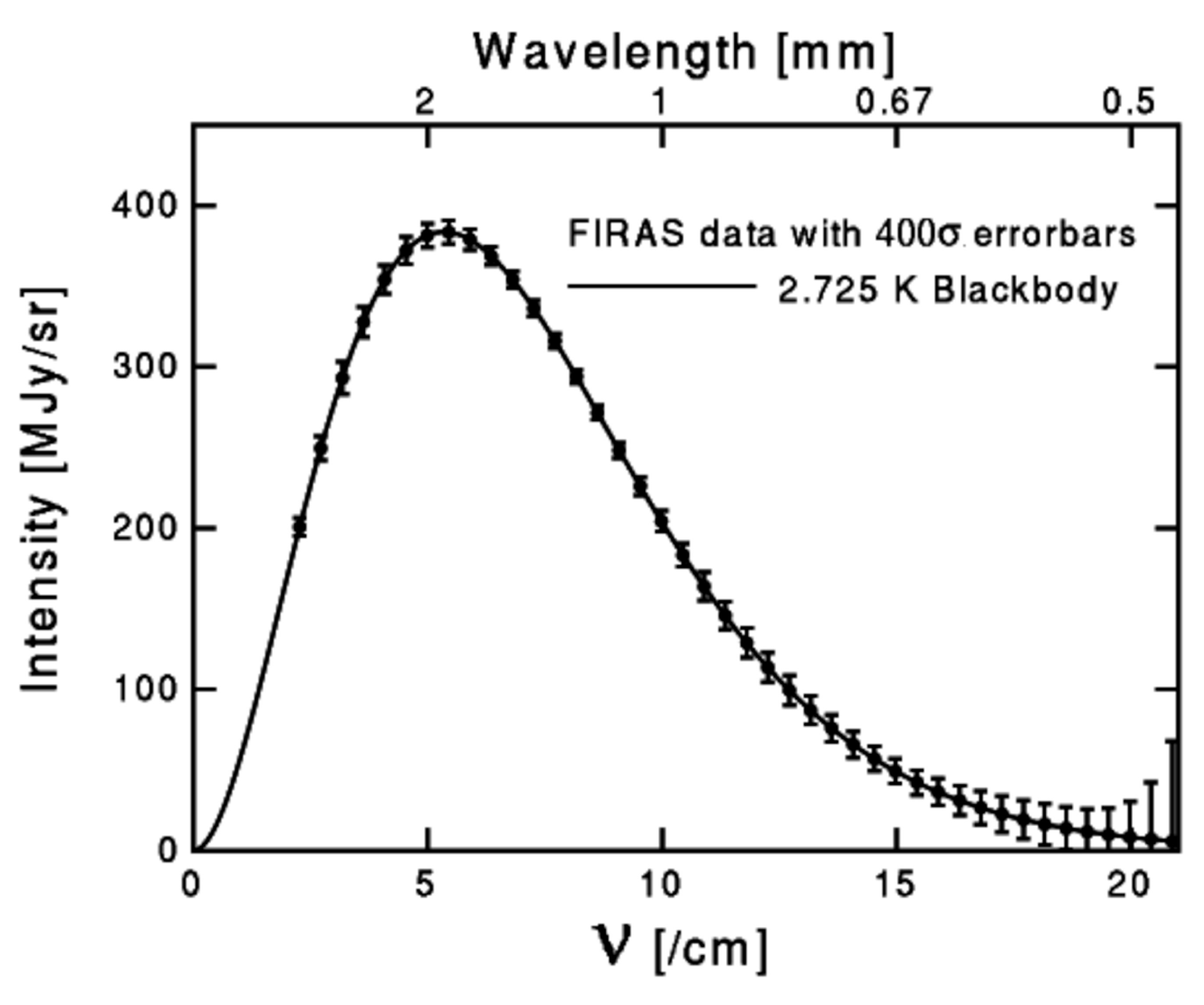}
\end{center}
\caption{Blackbody spectrum of the CMB as observed by the instrument FIRAS 
onboard the COBE satellite~\cite{MA90}} 
\label{FIRAS}
\end{figure}

We have said that, for $t < t_{_{{\rm rec}}}$, i.e. $z > 1100$, the Universe is a 
ionized plasma, opaque to electromagnetic radiation. This means that, when we 
observe the early Universe, we hit a ``wall'' at the corresponding redshift:
the earlier Universe appears to our observation as a blackbody, and should 
thus radiate according to the predictions of Planck. This is certainly the 
largest blackbody that one could think of. This blackbody spectrum of the 
CMB was indeed observed~\cite{MA90} by the FIRAS instrument onboard the COsmic 
Background Explorer (COBE) satellite launched in 1989 by NASA (see Fig. 
\ref{FIRAS}).

More precisely, the CMB has the nearly perfect thermal spectrum of a black body 
at temperature 
$T_\gamma = 2.725$ K (corresponding to a number density $n_\gamma = 411$
cm$^{-3}$):
\begin{equation}
\label{4-38l}
d\rho_\gamma = 2 hf {1 \over e^{hf/kT_\gamma}-1}{4\pi f^2 df \over c^3}
\end{equation}
(the first factor accounts for the two polarizations) or
\begin{equation}
\label{4-38m}
{d\rho_\gamma \over d\ln f} = 3.8 \times 10^{-15} \hbox{J/m}^3 \left(
{f \over f_\gamma}\right)^4 {e-1 \over e^{f/f_\gamma}-1} \ ,
\end{equation}
where $f_\gamma = kT_\gamma/h = 5.7 \times 10^{10}$ Hz.      

\begin{figure}[h]
\begin{center}
\includegraphics[scale=0.45,angle=90]{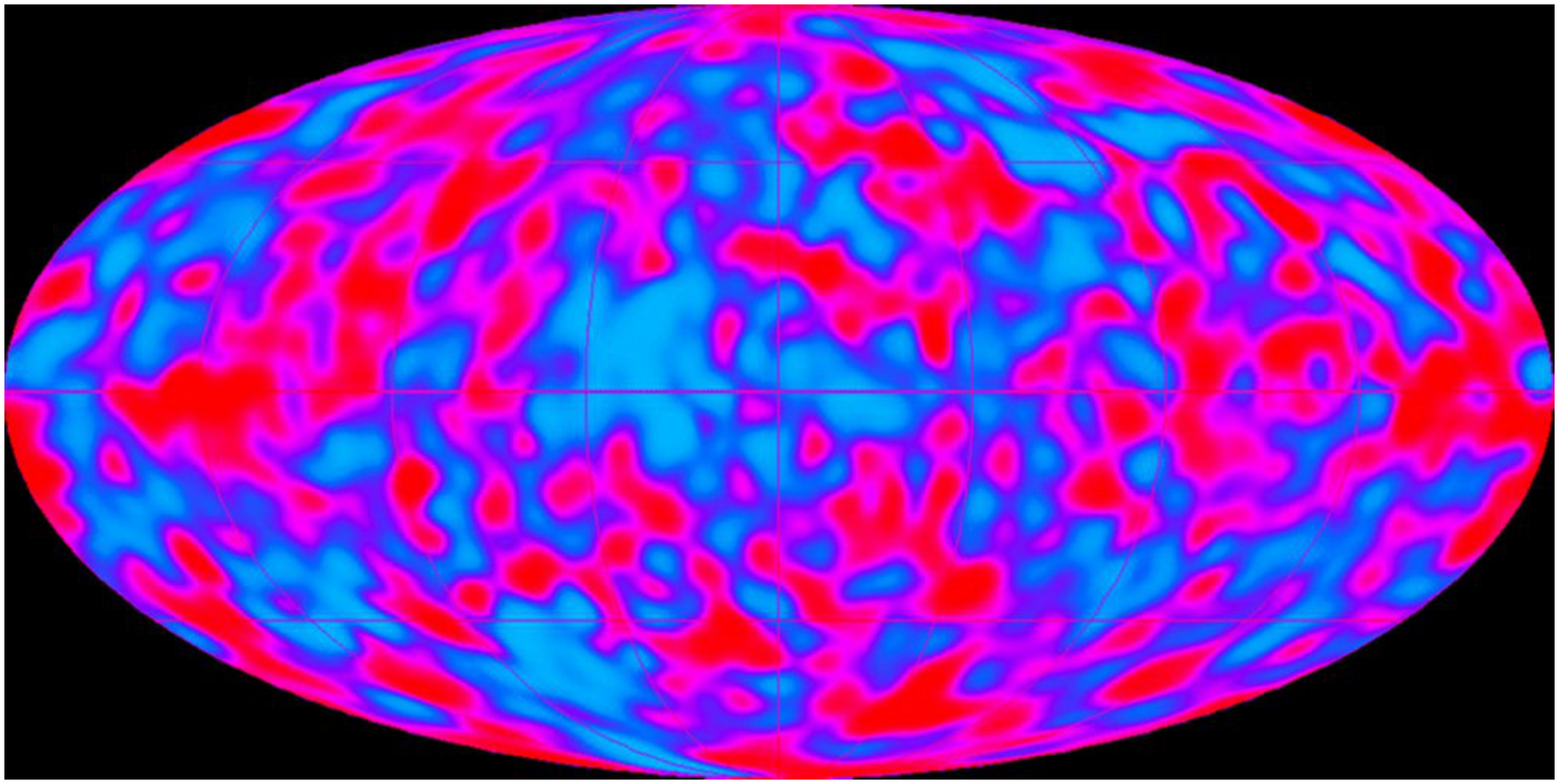}
\end{center}
\caption{Temperature fluctuations in the CMB observed by COBE~\cite{Sm92} 
(the galaxy is removed). The galaxy as well as the dipole 
component associated to the Doppler effect due to the motion of the Earth 
have been removed. Blue areas are colder than average, red areas are warmer 
but no $7^\circ$ region varies from the mean by more than $200$ $\mu$K 
($\Delta T/T \sim 8.10^{-5}$).}
\label{COBE_CMB}
\end{figure}

Another expectation for COBE was the presence of fluctuations in the CMB.
Indeed, if inflation was to explain the puzzle of isotropy and homogeneity of
the CMB through the whole sky, one expected that quantum fluctuations produced 
during the inflation phase would show up to some degree as tiny fluctuations of 
temperature in the CMB. Such fluctuations were discovered by the instrument DMR
(Differential Microwave Radiometers) onboard COBE, at the level of one part to 
$10^5$~\cite{Sm92} (see Fig.~\ref{COBE_CMB}).
   
It is primarily homogeneous and isotropic
but includes fluctuations at a level of $10^{-5}$, which are of much interest 
since they are .

Since the days of COBE, there has been an extensive study of the CMB 
fluctuations
to identify the imprints of the recombination and earlier epochs. This uses the 
results of ground, balloon or space missions (WMAP in the US and most recently 
Planck in Europe). We will review this in some details in the next Section.

\subsection{CMB} \label{4s3}

Before discussing the spectrum of CMB fluctuations, we introduce the important 
notion of a particle horizon in cosmology.

Because of the speed of light, a photon which is emitted at the big bang 
($t=0$) will have travelled a finite distance at time $t$. The proper
distance (\ref{4-30}) measured at time $t$ is simply given by the integral:
\begin{eqnarray}
\label{4-38a}
d_{ph}(t) &=& a(t) \int_0^t {cdt' \over a(t')}  \\
&=& {\ell_{H_0}\over 1+z} \int_z^\infty {dz \over \left[\Omega_{_M} (1+z)^3 
+  \Omega_{_R} (1+z)^4 + \Omega_{k} (1+z)^2 + \Omega_\Lambda \right]^{1/2}} 
\quad ,\nonumber
\end{eqnarray}
where, in the second line, we have used (\ref{4-32}).
This is the maximal distance that a photon (or any particle) could have 
travelled at time $t$ since the big bang. In other words, it is possible to 
receive signals at a time $t$ only from comoving particles within a sphere 
of radius $d_h(t)$. This distance is known as the particle 
horizon at time $t$. 

A quantity of relevance for our discussion of CMB fluctuations is the horizon 
at the time of the recombination i.e. $z_{_{{\rm rec}}} \sim 1100$. 
We  note that
the integral on the second line of (\ref{4-38a}) is dominated by the lowest 
values of $z$: $z\sim z_{_{{\rm rec}}}$ where the universe is still matter
dominated. Hence
\begin{equation}
\label{4-38b}   
d_{ph}(t_{_{{\rm rec}}}) \sim {2 \ell_{H_0} \over \Omega_{_M}^{1/2} 
z_{_{{\rm rec}}}^{3/2}} \sim 0.3 \ \hbox{Mpc}\ .
\end{equation}
One may introduce also the Hubble radius
\begin{equation}
\label{4-38ab}
R_H(t) = H^{-1} (z) \ ,
\end{equation}
which will play an important role in the following discussion. This scale 
characterizes the curvature of spacetime at the time $t$ ({\em see for example 
Exercise B.1.b}). We note that the particle horizon is simply twice the Hubble 
radius at recombination, as can be checked from (\ref{4-31}):
\begin{equation}
\label{4-38c}   
R_H(t_{_{{\rm rec}}}) \sim { \ell_{H_0} \over \Omega_{_M}^{1/2} 
z_{_{{\rm rec}}}^{3/2}} \ .
\end{equation}
This radius is seen from an observer at present time under an angle
\begin{equation}
\label{4-38d}
\theta_H(t_{_{{\rm rec}}}) = {R_H(t_{_{{\rm rec}}}) \over 
d_A(t_{_{{\rm rec}}})} \ ,
\end{equation}
where the angular distance has been defined in (\ref{4-38}). We can compute
analytically this angular distance under the assumption that the universe is 
matter dominated ({\em see Exercise C-1}). Using (\ref{D-4ex3}), we have
\begin{equation}
\label{4-38e}
d_A(t_{_{{\rm rec}}}) = {a_0 r \over 1+z_{_{{\rm rec}}}} \sim {2\ell_{H_0}
\over \Omega_{_M} z_{_{{\rm rec}}}} \  .
\end{equation}
Thus, since, in our approximation, the total energy density $\Omega_{_T}$ is
given by $\Omega_{_M}$,
\begin{equation}
\label{4-38f}
\theta_H(t_{_{{\rm rec}}}) \sim \Omega_{_T}^{1/2}/(2 z_{_{{\rm rec}}}^{1/2}) 
\sim 0.015 \ \hbox{rad} \ \Omega_{_T}^{1/2}  \sim 1^\circ \ \Omega_{_T}^{1/2} 
\ .
\end{equation}
We have written in the latter equation $\Omega_{_T}$ instead of $\Omega_{_M}$ 
because numerical computations show that, in case where $\Omega_\Lambda$ is
non-negligible, the angle depends on $\Omega_{_M} + \Omega_\Lambda = 
\Omega_{_T}$. 

We can now discuss the evolution of photon temperature fluctuations. For 
simplicity, we will assume a flat primordial spectrum of fluctuations: this 
leads to predictions in good agreement with experiment; moreover, as we will 
see in the next Section, it is naturally explained in the context of inflation 
scenarios.

Before decoupling, the photons are tightly coupled with the baryons through
Thomson scattering. In a 
gravitational potential well, gravity tends to pull this  baryon-photon fluid 
down the well whereas radiation pressure tends to push it out. Thus, the fluid 
undergoes a series of acoustic oscillations.  These oscillations can obviously 
only proceed if they are compatible with causality i.e. if the corresponding 
wavelength is smaller than the horizon scale or the Hubble radius: 
$\lambda =2\pi a(t) /k < R_H(t)$ 
or 
\begin{equation}
\label{4-38g}
k > 2 \pi   {a(t) \over R_H(t)} \sim t^{-1/3} \ . 
\end{equation}

\begin{figure}
\begin{center}
\includegraphics[scale=0.6]{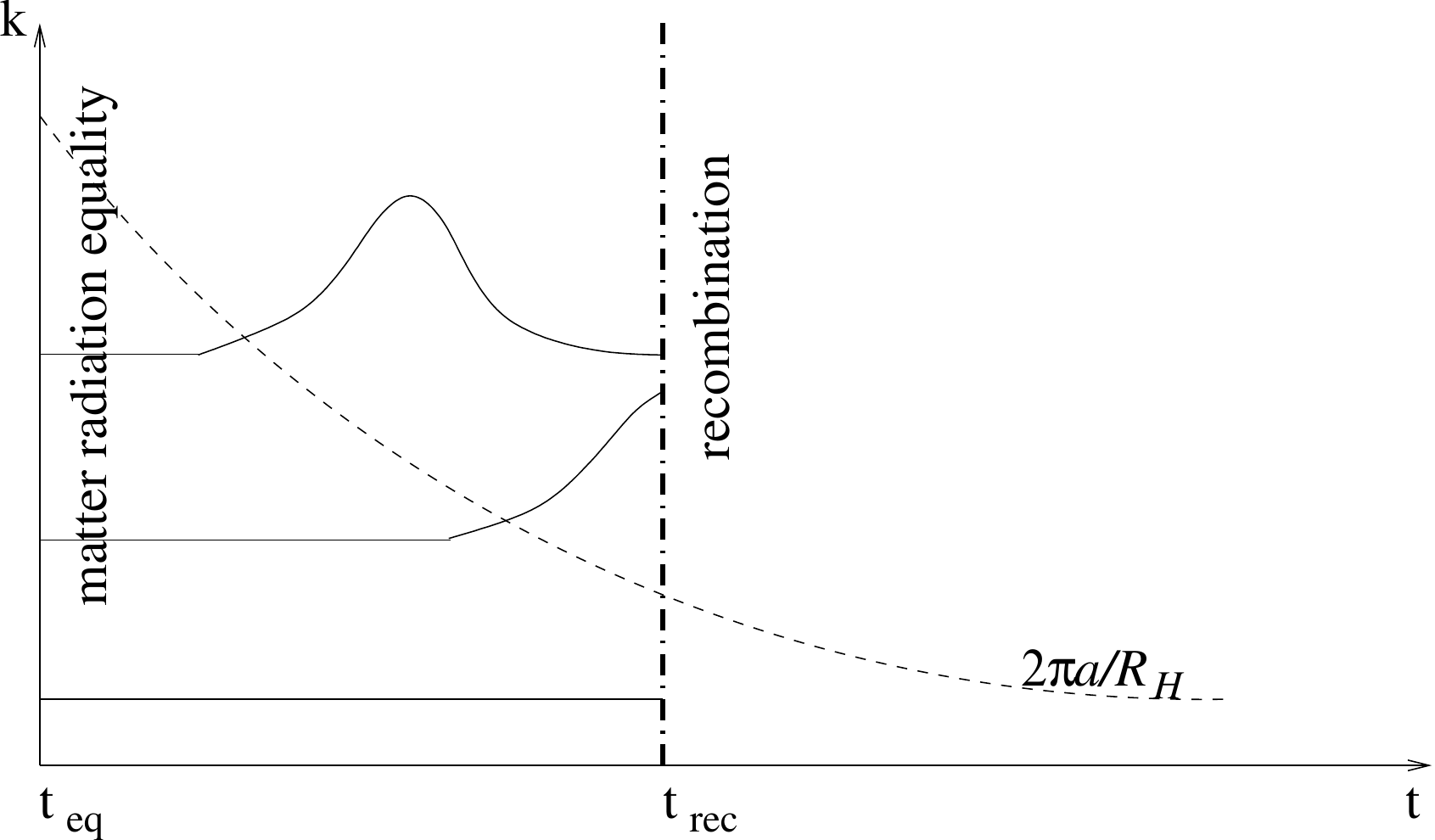}
\end{center}
\caption{Evolution of the photon temperature fluctuations before the 
recombination. This diagram illustrates that oscillations start once the 
corresponding Fourier mode enters the Hubble radius (these oscillations are 
fluctuations in temperature, along a vertical axis orthogonal to the two axes
that are drawn on the figure).}
\label{fig4-3}
\end{figure}

Starting with a flat 
primordial spectrum, we see that the first oscillation peak corresponds to  
$\lambda \sim R_H(t_{_{{\rm rec}}})$, followed by other compression peaks at
$R_H(t_{_{{\rm rec}}})/n$ (see Fig.~\ref{fig4-3}). They correspond to an 
angular scale on the sky:
\begin{equation} 
\label{4-38h}
\theta_n \sim {R_H(t_{_{{\rm rec}}}) \over d_A(t_{_{{\rm rec}}})} {1\over n} = 
{\theta_H(t_{_{{\rm rec}}}) \over n}\ .
\end{equation}
Since photons decouple at $t_{_{{\rm rec}}}$, we observe the same spectrum 
presently (up to a redshift in the photon temperature)\footnote{{\em A more 
careful analysis indicates the presence of Doppler effects besides the 
gravitational effects that we have taken into account here. Such Doppler 
effects turn out to be non-leading for odd values of $n$.}}.

Experiments usually measure the temperature difference of photons received 
by two antennas separated by an angle $\theta$, averaged over a large fraction 
of the sky. Defining the correlation function
\begin{equation}
\label{4-38i}
C(\theta) = \left \langle {\Delta T \over T_0}({\bf n}_1) {\Delta T \over T_0}
({\bf n}_2) \right \rangle
\end{equation}
averaged over all ${\bf n}_1$ and  ${\bf n}_2$ satisfying the condition
 ${\bf n}_1 \cdot {\bf n}_2 = \cos \theta$, we have indeed
\begin{equation}
\label{4-38j}
\left \langle \left( {T ({\bf n}_1) -  T ({\bf n}_2) \over T_0} \right)^2 
\right \rangle = 2 \left(C(0) - C(\theta) \right).
\end{equation} 
We may decompose $C(\theta)$ over Legendre polynomials:
\begin{equation}
\label{4-38k}
C(\theta) = {1 \over 4 \pi} \sum_l^\infty (2l+1) C_l P_l(\cos \theta) \ .
\end{equation}
The monopole ($l=0$), related to the overall temperature $T_0$, and the dipole 
($l=1$), due to the Solar system peculiar velocity, bring no information on 
the primordial fluctuations. A given coefficient $C_l$ characterizes the 
contribution of the multipole component $l$ to the correlation function. 
If $\theta \ll 1$, the main contribution to $C_l$ corresponds to an 
angular 
scale\footnote{The $C_l$ are related to the coefficients $a_{lm}$ in the 
expansion of $\Delta T/T$ in terms of the spherical harmonics $Y_{lm}$: 
$C_l = \langle \left| a_{lm}\right|^2 \rangle_m$. The relation between the 
value of $l$ and the angle comes from the observation that $Y_{lm}$ has 
$(l-m)$ zeros for $-1<\cos \theta<1$ and  Re($Y_{lm}$) $m$ zeros for 
$0<\phi<2\pi$.} $\theta \sim \pi/l \sim 200^\circ / l$. The previous 
discussion (see (\ref{4-38f}) and (\ref{4-38h})) implies that we expect the 
first acoustic peak at a value $l \sim 200\Omega_{_T}^{-1/2}$.

\begin{figure}
\begin{center}
\includegraphics[scale=0.85]{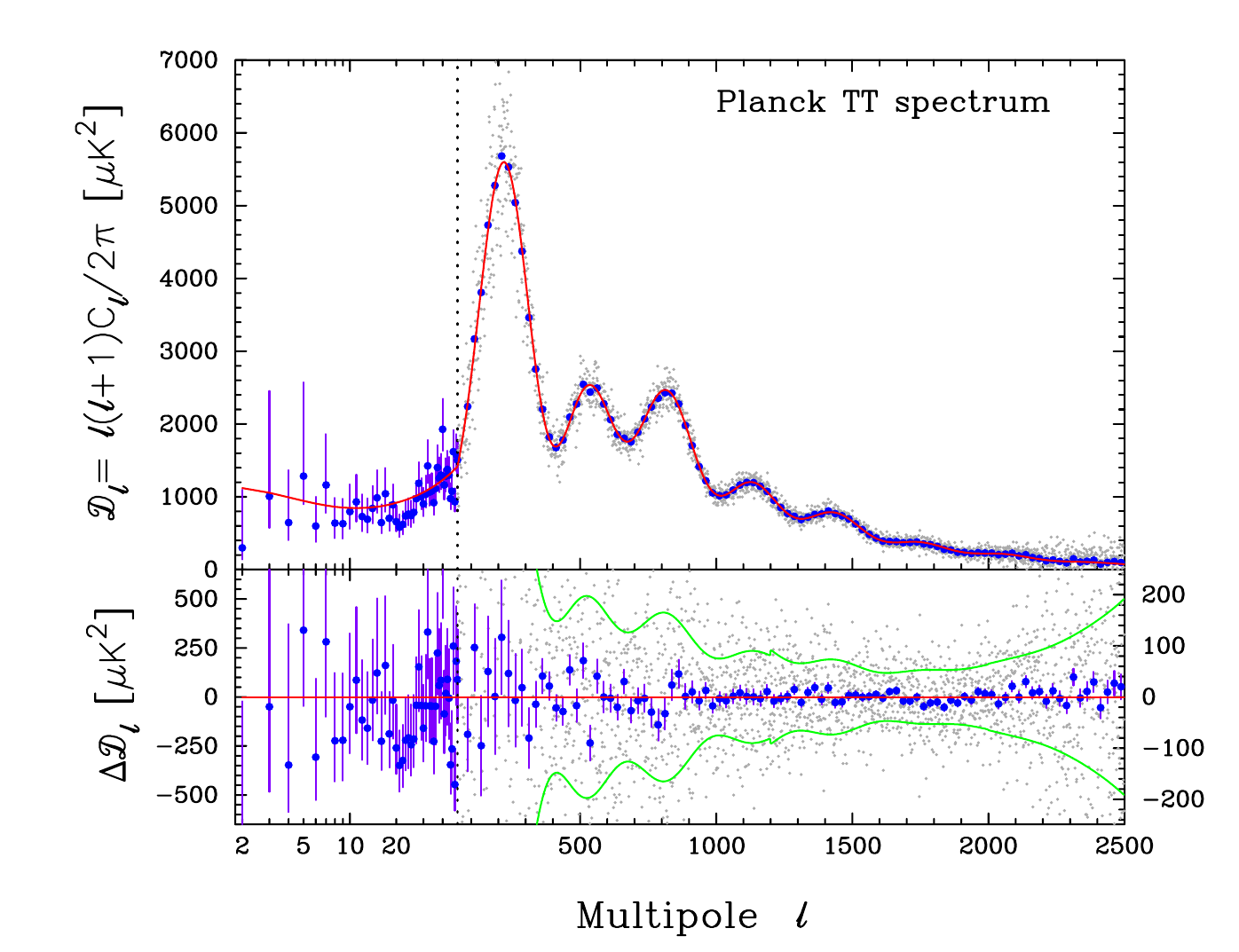}
\end{center}
\caption{This figure compares the temperature spectrum for the best fit 
$\Lambda$CDM  model, in red, with the temperature angular power spectrum 
observed by the Planck collaboration (in blue, averaged over bins of width 
$\Delta l \sim 31$ with 1 $\sigma$ errors).  The gray dots are the unbinned data.
In the lower panel, the green lines show the $\pm 1 \sigma$ errors on the 
individual power spectrum estimates.\cite{Planck13_16}
\label{fig4-4}}
\end{figure}

The power spectrum obtained by the Planck experiment is shown in Fig. 
\ref{fig4-4}. One finds the first acoustic peak at $l\sim 200$, which 
constrains the $\Lambda$CDM model used to perform the fit to
$\Omega_{_T} = \Omega_{_M} + \Omega_\Lambda \sim 1$. Many other constraints 
may be inferred from a detailed study of the power spectrum 
\cite{WMAP03,Planck13_16}.

\subsection{Baryon acoustic oscillations} \label{4s4a}

We noted in the previous section that, before decoupling, baryons and photons 
were tightly coupled and the baryon-photon fluid underwent a series of 
acoustic oscillations, which have left imprints 
in the CMB: the characteristic distance scale is the sound horizon, which is 
the comoving distance that sound waves could travel from the Big Bang until 
recombination at $z=z_*$: 
\begin{equation}
\label{rs}
r_s = \int_0^{t_*} {c_s(t) \over 1+z} dt = \int_{z_*}^{\infty} {c_s(z) \over H(z)} 
dz \ ,
\end{equation}
where $c_s$ is the sound velocity. This distance has been recently
measured with precision by the Planck collaboration to be $r_s = 144.96 \pm 
0.66$ Mpc~\cite{Planck13_16}. 

We have until now followed the fate of photons after 
decoupling. Similarly, once baryons decouple from the radiation, their 
oscillations freeze in, which leads to specific imprints in the galaxy power 
spectrum, such as the characteristic scale $r_s$. Indeed, remember that, 
until recombination, baryons and photons 
were tightly coupled (but not dark matter). Thus a given matter density 
perturbation may have travelled a distance $r_s$ in the case of baryons under 
the influence of radiation pressure (to which the photons are sensitive), 
whereas it did not move in the case of dark matter.
This will lead, once (dark and baryonic) matter has collapsed into galaxies, 
to a secondary peak a 
distance $r_s$ away in the distribution of separations of pairs of galaxies. 
In 2005, Eisenstein 
and collaborators~\cite{Ei05}, using data 
from the Sloan Digital Sky Survey, have indeed identified such a baryon 
acoustic 
peak in the matter power spectrum (Fourier transform of the two-point 
correlation function) on scales of order $105 h^{-1} \sim 150$ Mpc.
Figure~\ref{4-5} shows how 
the $\Lambda$CDM model just described fares with respect to observations by 
comparison with a model with $\Omega_\Lambda = 0$, a small value of the Hubble 
parameter and a small fraction of 
matter which does not cluster on small scale (relic neutrinos or 
quintessence).  

\begin{figure}[h]
\begin{center}
\includegraphics[scale=0.5]{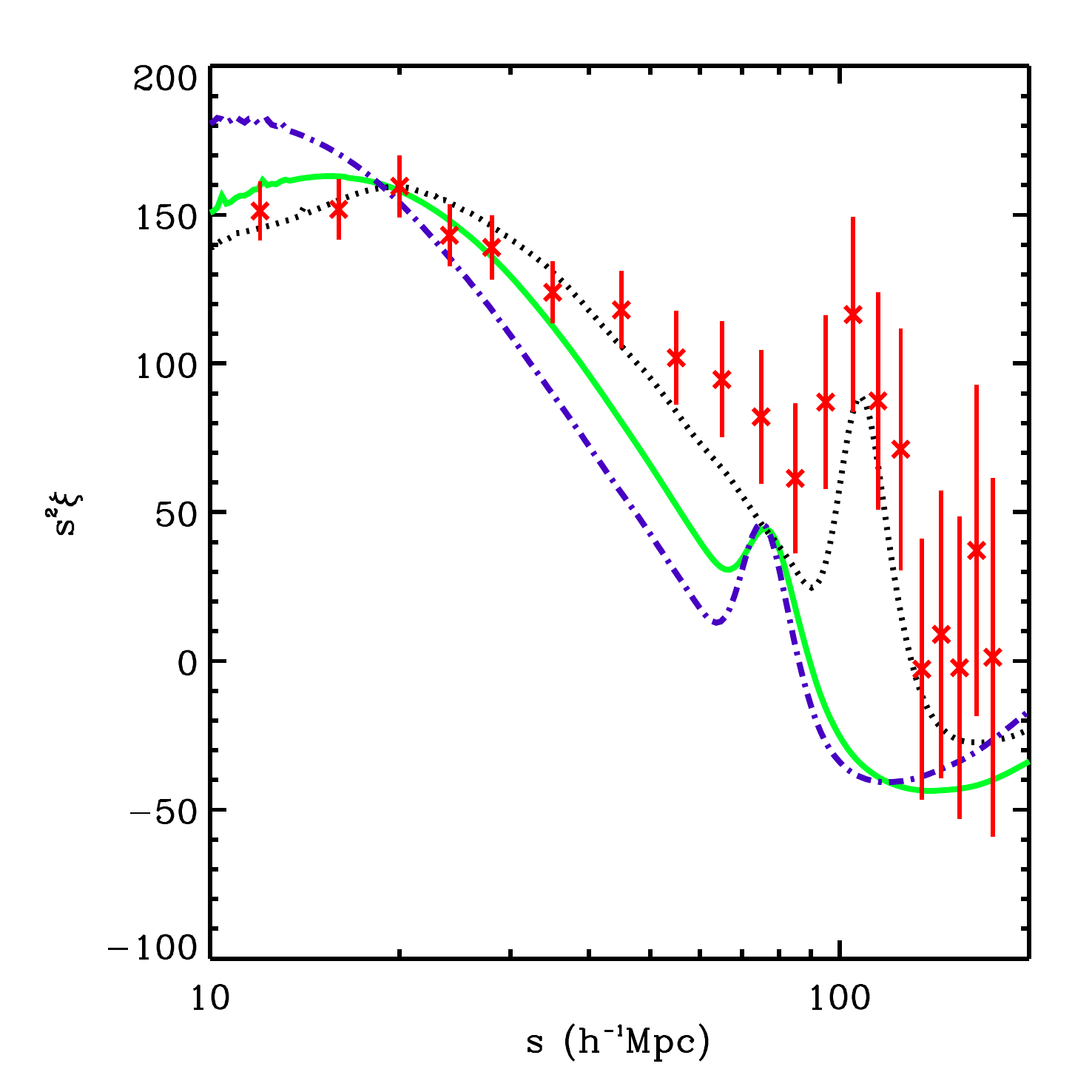}
\end{center}

\caption{Correlation function in redshift space for the best-fit power-law 
$\Lambda$CDM model (dotted line) and for the best-fit model with
$\Omega_\Lambda=0, H_0 = 46$ km/s/Mpc and a relic neutrino component 
$\Omega_\nu = 0.12$ (solid line) or a quintessence component 
$\Omega_Q = 0.12$ (dot-dashed line)~\cite{BDRS06}.} 
\label{fig4-5}
\end{figure}
The acoustic peak provides a standard ruler which can be used for measuring 
distances:
measurements along the line of sight depend on $H(z)r_s$ whereas measurements 
transverse to the line of sight depend on the angular diameter distance 
$d_A(z)/r_s$. In fact, because  the analysis rests on spherically-averaged 
two-point statistics, the distance scale determined is $d_V$ defined as:
\begin{equation}
\label{dV}
\left( {d_V(z) \over r_s} \right)^3 \equiv \left( {d_A(z) \over r_s}\right)^2
{cz \over H(z) r_s} \ .
\end{equation}
Measuring the acoustic scale at $z=0$ provides a standard distance ruler which 
allows to identify $H_0$, and then $\Omega_{_M}$ (from  $\Omega_{_M} h^2$). 
Going to higher $z$,  
this allows to put constraints on the recent history of the 
Universe, and thus on the evolution of the dark matter component.

\subsection{Inflation} \label{5s1}

The inflation scenario has been proposed to solve a certain number of 
problems faced by the cosmology of the early universe~\cite{Gu81}. 
Among these one may cite:
\begin{itemize}
\item the flatness problem

If the total energy density $\rho_T$ of the universe is presently close to 
the critical 
density, it should have been even more so in the primordial universe. 
Indeed, we can write (\ref{2-8}) of Section~\ref{chap:2} as
\begin{equation}
\label{5-1}  
{\rho_T(t) \over \rho_c(t)} -  1 = {k \over \dot a^2} \quad ,
\end{equation}
where $\rho_c(t) = 3H^2(t)/(8\pi G_{_N})$ and the total energy density
$\rho_T$ includes the vacuum 
energy. If we take for example the radiation-dominated era where $a(t) \sim 
t^{1/2}$, then (\ref{5-1}) can be written as ($\dot a \sim t^{-1/2} 
\sim a^{-1}$)
\begin{equation}
\label{5-2}
{\rho_T(t) \over \rho_c(t)} -  1 = \left[{\rho_T(t_{_U}) \over \rho_c(t_{_U})} 
-  1 \right] \left({a(t) \over a(t_{_U})}\right)^2 =  \left[{\rho_T(t_{_U}) 
\over \rho_c(t_{_U})} -  1 \right] \left({k T_{_U} \over kT} \right)^2 \quad ,
\end{equation}
where we have used the fact that $T(t) \propto a(t)^{-1}$ and 
we have taken as a reference point the epoch 
$t_{_U}$ of the grand unification phase transition. This means that, 
if the total energy density is close to the critical density at 
matter-radiation equality (as can be inferred
from the present value), it must be even more so at 
the time of the grand unification phase transition: by a factor
$\left( 1 \hbox{eV}/10^{16} \hbox{GeV}\right)^2 \sim  10^{-50}$! 
Obviously, the choice $k=0$ in the spatial metric ensures $\rho_{_T} = \rho_c$ 
but the previous estimate shows that this corresponds to intial conditions 
which are highly fine tuned.

\item the horizon problem

We have stressed in the previous Sections the isotropy and homogeneity of
the cosmic microwave background and identified its primordial origin. It
remains that the horizon at recombination is seen on the present sky under an 
angle of $2^\circ$. This means that two points opposite on the sky were 
separated by about 100 horizons at the time of recombination, and thus not 
causally connected. It is then 
extremely difficult to understand why the cosmic microwave background 
should be isotropic and homogeneous over the whole sky.

\item the monopole problem

Monopoles occur whenever a simple gauge 
group is broken to a group with a $U(1)$ factor. This is precisely what 
happens in grand unified theories. In this case their mass is of order
$M_{_U}/g^2$ where $g$ is the value of the coupling at grand unification. 
Because we are dealing with stable particles with a superheavy mass, there is
a danger to overclose the universe, i.e. to have an energy density much larger 
than the critical density.. 
We then need some mechanism to dilute the relic density of monopoles.
\end{itemize}

Inflation provides a remarkably simple solution to these problems: it consists 
in a period of the evolution of the universe where the expansion is
exponential. Indeed, if the energy density of the universe is dominated by 
the vacuum energy $\rho_{{\rm vac}}$ (or by some constant form of energy), then 
the Friedmann equation reads 
\begin{equation}
\label{5-8}
H^2 = {\dot a^2 \over a^2} = {\rho_{{\rm vac}} \over 3 m_{_P}^2} \  .
\end{equation}
where $m_{_P} \equiv \left( 8 \pi G_{_N} \right)^{-1/2}$ is the reduced Planck
mass. If $\rho_{{\rm vac}}>0$,this is readily solved as
\begin{equation}
\label{5-9}
a(t) = H_{{\rm vac}}^{-1} e^{H_{{\rm vac}}t}    \quad 
\hbox{with} \ H_{{\rm vac}} \equiv \sqrt{{\rho_{{\rm vac}} \over 3 m_{_P}^2}} 
\ .
\end{equation}
Such a behaviour is in fact observed whenever 
the magnitude of the Hubble parameter changes slowly with time i.e. 
is such that $\left| \dot H \right| \ll H^2$.

As we have seen in (\ref{deSitter}), such a space was first proposed by de 
Sitter~\cite{dS17a,dS17b} 
with very different motivations and is thus called de Sitter space.

Obviously  a period of inflation will ease the horizon problem. 
Indeed, the particle horizon size during inflation reads, 
following (\ref{4-38a})
\begin{equation}
\label{5-10}
\left. d_{ph}(t)\right|_{{\rm de \ Sitter}} = a(t) \int_{t_i}^t {cdt' \over a(t')}
= {c \over H_{{\rm vac}}} \ e^{H_{{\rm vac}}(t-t_i)} \ \hbox{for} \  H_{{\rm vac}}(t-t_i)
\gg 1 . 
\end{equation} 
It follows  that a period of inflation extending from $t_i$ 
to $t_f=t_i + \Delta_t$ contributes to the particle horizon size a value
$c e^{H_{{\rm vac}}\Delta t} / H_{{\rm vac}}$, which can be very large\footnote{We also 
note that, in a pure de Sitter space, the particle horizon diverges as we take  
$t_i \rightarrow -\infty$. This reflects the fact that, in a de Sitter space, 
all points were in causal contact.}

We note that de Sitter space also has a finite event horizon. This is the 
maximal distance that comoving particles can travel between the time $t$ where 
they are produced and $t=\infty$ (compare with (\ref{4-38a}):
\begin{equation}
\label{5-10a}
d_{eh}(t) = a(t) \int_t^{\infty} {c dt' \over a(t')} \ .
\end{equation}
In the case of de Sitter space, this is simply 
\begin{equation}
\label{5-10b}
\left. d_{eh}(t) \right|_{{\rm de \ Sitter}} = {c \over H_{{\rm vac}}} 
= R_H \ ,
\end{equation}
i.e. it corresponds to the Hubble radius (constant for de Sitter spacetime).
This allows to make an analogy between de Sitter spacetime and a black hole:
we will see in Section~\ref{2s6} that a 
Schwarzschild black hole of mass $M$ has an event horizon at the 
Schwarzschild radius 
$R_S = 2 G_{_N} M$ (see also Exercise 1-3 of Section~\ref{chap:1}
for a comparison between the 
Schwarzschild and the de Sitter metric in its static form). Thus, just as 
black holes evaporate by emitting
radiation at Hawking temperature $T_H = 1 /( 4 \pi R_S)$ (see Eq.~(\ref{3-TH})),  an 
observer in de Sitter spacetime feels a thermal bath at temperature 
$T_H = H / (2\pi)$.

We see that it is the event horizon that fixes here the cut-off scale of 
microphysics. Since it is equal here to Hubble radius, and since the
Hubble radius is of the order of the particle horizon for matter or 
radiation-dominated universe\footnote{In an open or flat universe, the 
event horizon~\ref{5-10a} is infinite.}, it has become customary to compare 
the comoving scale associated to physical processes with the Hubble radius 
(we already 
did so in our discussion of acoustic peaks in CMB spectrum; see  Fig.~\ref{fig4-3}, and  Fig.~\ref{fig5-1} below). 

A period of exponential expansion of the universe may also solve 
the monopole problem by diluting the concentration of monopoles by a
very large factor. It also dilutes any kind of matter. Indeed, a sufficiently 
long period of inflation ``empties'' the universe. However matter and 
radiation may be produced at the end of inflation by converting the energy 
stored in the vacuum. This conversion is known as reheating (because the 
temperature of the matter present in the initial stage of inflation 
behaves as $a^{-1}(t) \propto e^{-H_{{\rm vac}} t}$, it is very cold at 
the end of inflation; the new matter produced is hotter). 
If the reheating temperature is lower than the scale of grand unification, 
monopoles are not thermally produced and remain very scarce.

Finally, it is not surprising that the universe comes out very flat after a 
period of exponential inflation. Indeed, the spatial curvature term in the 
Friedmann equation  is then damped by a factor  $a^{-2} \propto
e^{-2  H_{{\rm vac}}\Delta t}$. For example, a value $H_{{\rm vac}}\Delta t 
\sim 60$ (one refers to it as $60$ $e$-foldings) 
would easily account for the huge 
factor $10^{50}$ of adjustment that we found earlier.

Most inflation models rely on the dynamics of a scalar field in its 
potential. Inflation occurs whenever the scalar field evolves slowly enough
in a region where the potential energy is large. The set up necessary to 
realize this situation has evolved with time: from the initial proposition 
of Guth~\cite{Gu81} where the field was trapped in a local minimum to  
``new inflation'' with a plateau in the scalar potential~\cite{Li82,AS82}, 
chaotic inflation~\cite{Li83} where the field is trapped at values 
much larger than the Planck scale and more recently  hybrid 
inflation~\cite{Li91} with at least two scalar 
fields, one allowing an easy exit from the inflation period. 

The equation of motion of a homogeneous scalar field $\phi(t)$  with potential
$V(\phi)$ evolving in a Friedmann-Robertson-Walker universe is:
\begin{equation}
\label{5-11}  
\ddot \phi + 3 H \dot \phi = -  V'(\phi).
\end{equation}
where $V'(\phi) \equiv  d V / d \phi$. The term $3H\dot\phi$ is a friction 
term due to the expansion. The corresponding energy density 
and pressure are:
\begin{eqnarray}
\rho &=& {1 \over 2} \dot \phi^2 + V(\phi) \ , \label{5-12} \\ 
p &=&  {1 \over 2} \dot \phi^2 - V(\phi) \ . \label{5-13}
\end{eqnarray}
We may note that the equation of conservation of energy $\dot \rho =
- 3H (p+\rho)$ takes here simply the form of the equation of motion 
(\ref{5-11}).
These equations should be complemented with the Friedmann equation (\ref{5-8}).
 
When the field is slowly moving in its potential, the friction term dominates 
over the acceleration term in the equation of motion  (\ref{5-11}) which reads:
\begin{equation}
\label{5-14}  
3 H \dot \phi \simeq -  V'(\phi) \quad .
\end{equation}
The curvature term may then be neglected in the Friedmann equation (\ref{5-8})
which gives
\begin{equation}
\label{5-15}
H^2 \simeq {\rho \over 3 m_{_P}^2} \simeq {V \over 3 m_{_P}^2} \ .
\end{equation}
Then the equation of conservation  $\dot \rho = - 3H (p+\rho)= - 3H \dot 
\phi^2$ simply gives
\begin{equation}
\label{5-16}
\dot H \simeq - {\dot \phi^2 \over 2 m_{_P}^2} \ .
\end{equation}
It is easy to see that the condition $|\dot H| \ll H^2$ amounts to
$\dot \phi^2/2  \ll \rho/3 \sim V(\phi)/3$, i.e. a kinetic energy 
for the scalar field much smaller than its potential energy. Using (\ref{5-14})
and (\ref{5-15}), the latter condition then reads
\begin{equation}
\label{5-17}
\epsilon \equiv {1 \over 2} \left(  {m_{_P} V' \over V} \right)^2 \ll 1 
\ .
\end{equation}
The so-called slow roll regime is characterized by the two equations 
(\ref{5-14}) and (\ref{5-15}), as well as the condition (\ref{5-17}).
It is customary to introduce another small parameter:
\begin{equation}
\label{5-18}
\eta \equiv {m_{_P}^2 V'' \over V}  \ll 1 \quad ,
\end{equation}
which is easily seen to be a consequence of the previous 
equations\footnote{Differentiating (\ref{5-14}), one obtains 
\begin{equation}
\label{5-18a}
\eta = \epsilon - \ddot \phi/(H\dot \phi) \ .
\end{equation}}\footnote{{\em Note that one finds also in the literature the slow roll 
coefficients defined from the Hubble parameter~\cite{LLKCBA97}
\begin{equation}
\label{slowrollH}
\epsilon_H \equiv 2m_{_P}^2 \left({H'(\phi) \over H(\phi)}\right)^2 = \epsilon \ , \ \ 
\eta_H \equiv 2m_{_P}^2 {H''(\phi) \over H(\phi)} = \eta - \epsilon \ .
\end{equation}}}.

An important quantity to be determined is the number of Hubble times elapsed 
during inflation. From some arbitrary time $t$ to the time $t_e$ marking the 
end of inflation (i.e. of the slow roll regime), this number is given by
\begin{equation}
 \label{5-19}  
N(t) = \int_t^{t_e} H(t) dt \ .
\end{equation}
It gives the number of e-foldings undergone by the scale factor $a(t)$ during 
this period (see (\ref{5-9}). Since $dN = -Hdt = -Hd\phi/\dot \phi$, one 
obtains from (\ref{5-14}) and (\ref{5-15})
\begin{equation}
\label{5-20}  
N(\phi) = \int_{\phi_e}^\phi {1 \over m_{_P}^2}{V \over V'} d\phi \ .
\end{equation}

During the inflationary phase, the scalar fluctuations of the metric
may be written in a conformal Newtonian coordinate system as:
\begin{equation}
\label{5-21}   
ds^2 = a^2 \left[ (1+2\Phi) d\eta^2 - (1-2 \Phi) \delta_{ij}dx^idx^j 
\right] \ ,
\end{equation}
where $\eta$ is conformal time ($a d\eta = dt = da/\dot a$). We may write the
correlation function in Fourier space ${\cal P}_S(k)$ by
\begin{equation}
\label{5-21a}
\langle \Phi_{{\bf k}} \Phi^*_{{\bf k'}} \rangle = 2 \pi^2 k^{-3}
{\cal P}_S(k) \delta^3\left( {\bf k}-{\bf k'} \right) \ .
\end{equation}
The origin of fluctuations is found in the quantum fluctuations of the scalar 
field during the
de Sitter phase. Indeed, if we follow a given comoving scale $a(t)/k$ with 
time (see Fig.~\ref{fig5-1}), we have seen in Section~\ref{4s3} 
that, some time during the matter-dominated phase, it enters the Hubble 
radius. Since $a(t)$ is growing (even exponentially during inflation) whereas
the Hubble radius is constant during inflation, this means that 
at a much earlier time, it has emerged from the Hubble radius of the de Sitter 
phase. In this scenario, the origin of the fluctuations is thus found in the 
heart of the de Sitter event horizon: using quantum field theory in curved 
space, one may compute the amplitude of the quantum fluctuations of the scalar 
field; their wavelengths evolve as $a(t)/k$ until they outgrow the event
horizon i.e. the Hubble radius; they freeze out and continue to evolve 
classically. The fluctuation spectrum produced is given by
\begin{equation}
\label{5-23}  
{\cal P}_S(k) = \left[ \left({H^2 \over \dot \phi^2}\right) 
\left({H \over 2\pi}\right)^2 \right]_{k=aH} = {1 \over 12 \pi^2 m_{_P}^6}
\left({V^3 \over V'^2}\right)_{k=aH} \ ,
\end{equation}
where the subscript $k=aH$ means that the quantities are evaluated at Hubble 
radius crossing, as expected. We also note that $H$ i.e. $R_H$ sets the scale 
of quantum fluctuations in the de Sitter phase ({\em see Appendix~\ref{app:E}
for details, in particular Eq.~(\ref{E-40})}): $R_H$ is indeed the dynamical scale 
associated with the physics of fluctuations (which happens to coincide with
either of the kinematical scales which are the particle horizon for the 
matter dominated phase, or the event horizon in the inflation phase).

\begin{figure}
\begin{center}
\includegraphics[scale=0.45]{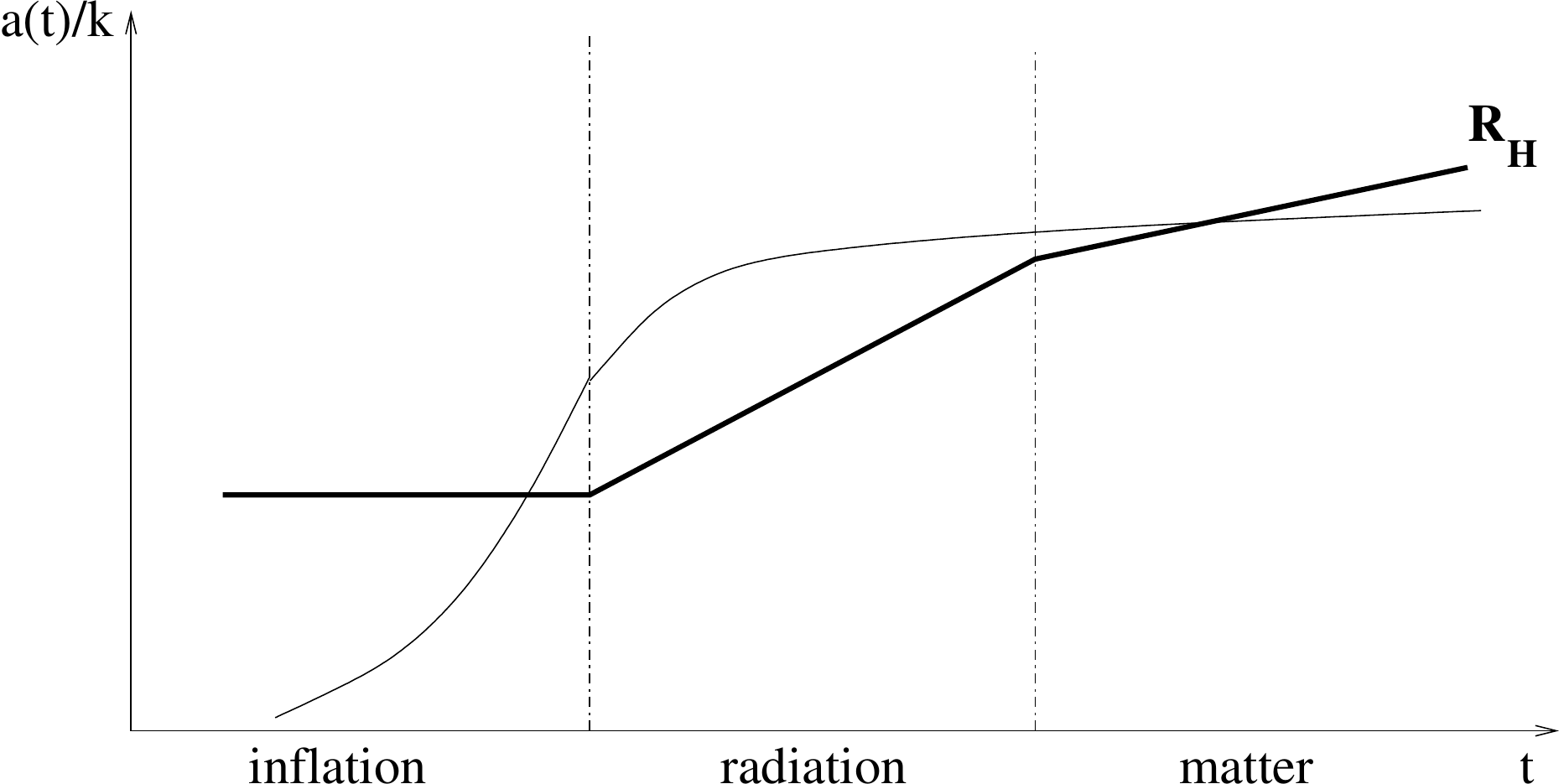}
\end{center}
\caption{Evolution of a physical comoving fluctuation scale with respect to 
the Hubble radius during the inflation phase ($R_H(t) =  H_{{\rm vac}}^{-1}$),
the radiation dominated phase ($R_H(t) = 2t$) and matter dominated phases 
($R_H(t) = 3t/2$).}
\label{fig5-1}
\end{figure}

The scalar spectral index $n_S(k)$ is computed to be ({\em see Exercise D-1 in
Appendix~\ref{app:E}}):
\begin{equation}
\label{5-24}
n_S(k) - 1 \equiv {d \ln {\cal P}_S (k) \over d \ln k} 
= -6 \epsilon + 2 \eta \ .
\end{equation}
Thus, because of the slow roll, the fluctuation spectrum is almost scale 
invariant, a result that we have alluded to when we discussed the origin of
CMB fluctuations. One of the highlights of the Planck cosmology results 
\cite{Planck13_16,Planck13_22} is the confirmation that the spectrum is not 
scale invariant i.e. $n_S$ is different from $1$:
\begin{equation}
\label{Planckns}
n_S = 0.9603 \pm 0.0073 \ .  
\end{equation}
In other words, we are really in a slow roll phase, i.e. an unstable phase 
which is
crucial since eventually one has to get out of inflation and reheat.
 
The observation by the COBE satellite of the largest scales  has
set an important constraint on inflationary models by putting an important 
constraint on the size of fluctuations (see the caption of Figure 
\ref{COBE_CMB}). Specifically, in terms of the value of the scalar potential
at horizon crossing, this constraint known as COBE normalization, reads
(see (\ref{5-23}):
\begin{equation}
\label{5-22}
{1 \over m_{_P}^3} {V^{3/2} \over V'} = 5.3 \times 10^{-4} \ .
\end{equation}
Using the slow roll parameter introduced above in (\ref{5-17}), the COBE 
normalization condition can be written as 
\begin{equation}
\label{5-22a} 
V^{1/4} \sim 0.03 \ \varepsilon^{1/4} \ m_{_P} \ .
\end{equation}

Besides scalar fluctuations, inflation produces fluctuations which have a 
tensor structure, i.e. primordial gravitational waves. They can be written as 
perturbations of the metric of the form
\begin{equation}
\label{5-25}
ds^2 = a^2 \left[ \eta_{\mu\nu} + h^{TT}_{\mu\nu}\right] \ ,
\end{equation}
where $h^{TT}_{\mu\nu}$ is a traceless transverse tensor (which has 
two physical degrees of freedom i.e. two polarizations). The corresponding 
tensor spectrum is given by
\begin{equation}
\label{5-26}  
{\cal P}_T(k) = {8  \over m_{_P}^2} \left( {H \over 2 \pi} \right)^2 \ ,
\end{equation}
with a corresponding  spectral index
\begin{equation}
\label{5-27}
n_T(k)  \equiv {d \ln {\cal P}_T (k) \over d \ln k} 
= -2 \epsilon  \ .
\end{equation} 
 We note that the ratio ${\cal P}_T/{\cal P}_S$ depends only on 
$\dot \phi^2/H^2$ and thus on $\epsilon$, which yields the consistency 
condition:
\begin{equation}
\label{5-27a}
r \equiv { {\cal P}_T \over  {\cal P}_S} = { 8 \dot \phi^2 \over m_{_P}^2 H^2}
= 16 \epsilon = -8 n_T \ .
\end{equation}

\subsection{Inflation scenarios} \label{5s2}

We conclude this discussion by reviewing briefly the main classes of 
inflation models. 
Let us note that, for an inflationary model, the whole observable universe 
should be within the Hubble radius at the beginning of inflation. This
corresponds to a scale
\begin{equation}
\label{infsc1}
k = a_0 H_0 = \left. a H\right|_{h.c.} \ .
\end{equation}
where $h.c.$ stands for ``horizon crossing''. This puts  a constraint 
on the number of e-foldings (\ref{5-20}) between horizon crossing and the end 
of inflation (i.e. end of the slow roll regime) necessary for the inflation to 
be efficient. More generally, one defines~\cite{LL93} $N(k)$ as the number 
of e-foldings 
between the time of horizon crossing of the scale $k$ ($t_{k,h.c.}$ at which
$k = aH(t_{k,h.c.})$) and the end of inflation ($t_e$):
\begin{equation}
\label{infsc2}
N(k) \equiv \ln \left(a(t_e) /a(t_{k,h.c.})\right) \ .
\end{equation}
Distinguishing the time when the universe reheats ($t_{rh}$) and the time of 
matter-radiation equality ($t_{eq}$), we have
\begin{eqnarray}
\nonumber
{a(t_{k,h.c.}) \over a_0} &=& e^{-N(k)} {a(t_e) \over a(t_{rh})} \cdot 
{a(t_{rh}) \over a(t_{eq})} \cdot {a(t_{eq}) \over a_0} \\
\label{infsc3}
&=& e^{-N(k)} \left({\rho(t_{rh}) \over \rho(t_e)}\right)^{1/3} 
\left({\rho(t_{eq}) \over \rho(t_{rh})}\right)^{1/4}{a(t_{eq}) \over a_0} \\
\nonumber
&=& e^{-N(k)}  \left({\rho(t_{rh}) \over \rho(t_e)}\right)^{1/3} 
\left({\rho_0 \over \rho(t_{rh})}\right)^{1/4}\left({a(t_{eq}) \over a_0}
\right)^{1/4} \,
\end{eqnarray}
where we used $\rho(t_{eq}) = \rho_0 (a_0/a(t_{eq}))^3$ and we have assumed 
matter domination between the end of inflation and reheating. Using $k = 
aH(t_{k,h.c.})$ and  $H(t_{k,h.c.})/H_0 = \left(\rho_{k,h.c.}/\rho_0\right)^{1/2}$, 
one obtains
from (\ref{infsc2}), with transparent notations:
\begin{equation}
N(k) = 62 - \ln {k\over a_0 H_0} - \ln {10^{16} \ \hbox{GeV} \over 
V_{k,h.c.}^{1/4}} +
\ln {V_{k,h.c.}^{1/4} \over V_e^{1/4}} - {1 \over 3} \ln {V_e^{1/4} \over 
\rho_{rh}^{1/4}} \ .
\end{equation}
In the rather standard case where $V_{k,h.c.} \sim V_e \sim  \rho_{rh} \sim
\left(10^{16} \ \hbox{GeV}\right)^4$, this requires $60$ e-folding for 
inflation to be efficient at the scale of the observable Universe. A length 
scale corresponding to $200$ Mpc (hence $k=2\pi/200$ Mpc) corresponds to $N(k)
\sim 50$.

The three main classes of 
inflation models (see Figs.~\ref{fig5-2} and~\ref{figPlanck_inflation})
are:

\begin{figure}
\begin{center}
\includegraphics[scale=0.3]{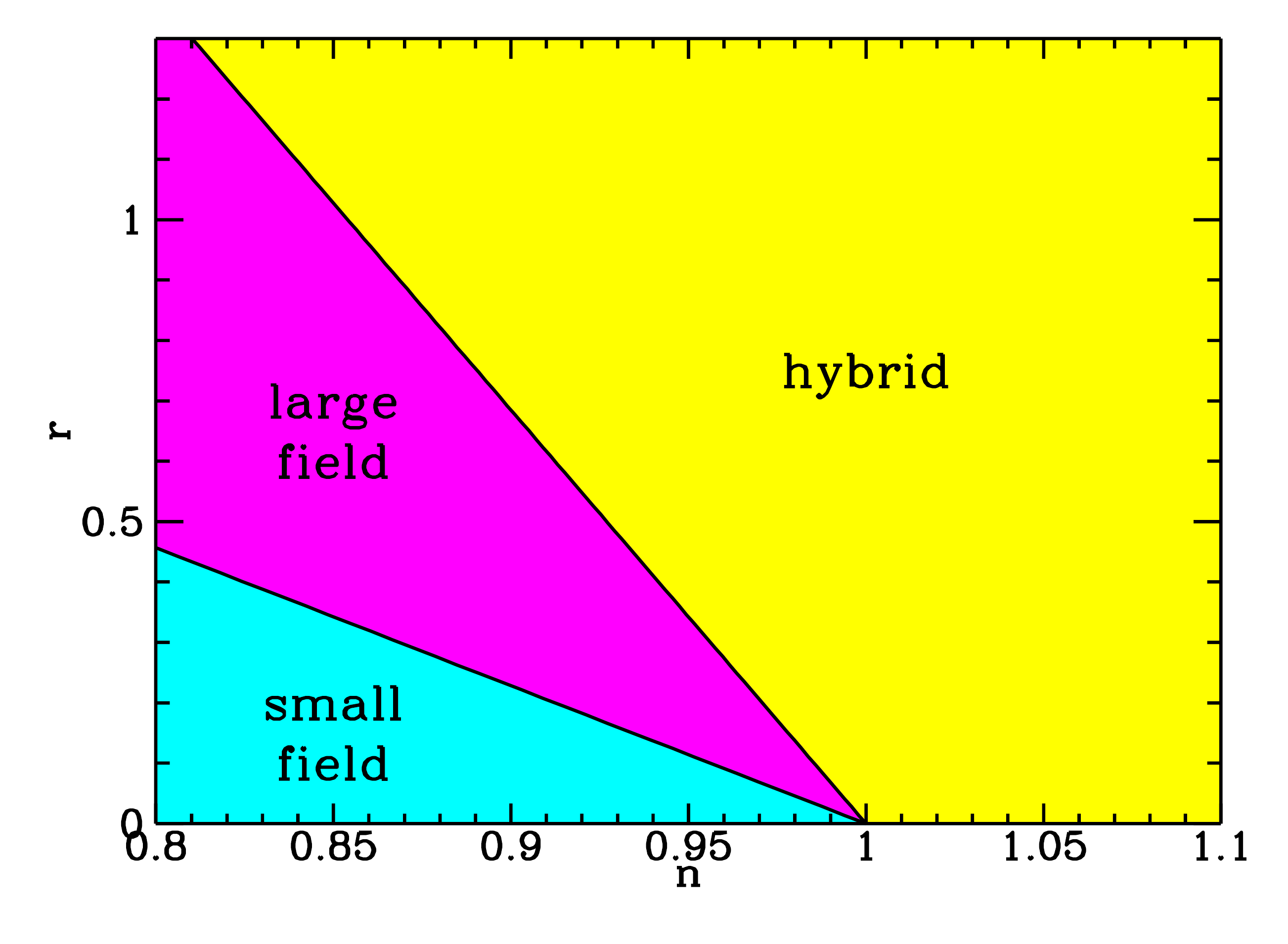}
\end{center}
\caption{Regions corresponding to the different inflation models in the
plot $r$ vs $n_s$~\cite{DKK97}}
\label{fig5-2}
\end{figure}

\begin{itemize}
\item convex potentials or large field models ($0<\eta < 2 \epsilon$)

The potential is typically a single monomial potential:
\begin{equation}
\label{chaotic}
V(\phi) = \lambda m_{_P}^4 \left({\phi \over m_{_P}}\right)^n \ ,
\end{equation}
with $n > 1$. Since the slow roll parameters are $\epsilon = 
n^2 (m_{_P}/\phi)^2/2$ and\footnote{Note that $\epsilon < \eta < 2\epsilon$.} 
$\eta = n(n-1) (m_{_P}/\phi)^2$, the slow roll regime 
corresponds to $\phi \gg m_{_P}\sqrt{n(n-1)}$ (for $n>2$). Because the field has a 
value larger than the Planck scale (hence the name ``large field model''), this might 
seem out of the reach of the effective low energy gravitation theory. But A. Linde
\cite{Li83} argued that the criterion is rather $V < m_{_P}^4$; in fact, he suggested 
that the scalar field emerges from the Planck era with a value $\phi_0$ such that
$V(\phi_0) \sim m_{_P}^4$ i.e. $\phi_0 \sim m_{_P}/\lambda^{1/n}$. This corresponds to the 
chaotic inflation scenario, the simplest example of which being a quadratic potential 
\cite{Li83}. A difficulty is that the COBE normalisation imposes an unnaturally small 
value for the $\lambda$ coupling: $\lambda \sim (5.3\times 10^{-4} n)^{2n/(n-2)}$.
 Another drawback is the large value
of the field which makes it necessary to include all non renormalisable
corrections of order $\left( \phi / M_{_P} \right)^{n+p}$, unless they are forbidden by 
some symmetry. 

The limit case in this class is the exponential potential 
\begin{equation}
\label{powerlaw}
V(\phi) = V_0 \exp (-\lambda \phi/m_{_P})
\end{equation}
which leads to a power law inflation~\cite{LM85}:
$a(t) \propto t^{2/\lambda^2}$.
This model yields $\eta = 2 \epsilon = \lambda^2$, hence $r = -8(n_S-1)$. It is 
incomplete since inflation does not end.

\begin{figure}
\begin{center}
\includegraphics[scale=0.6]{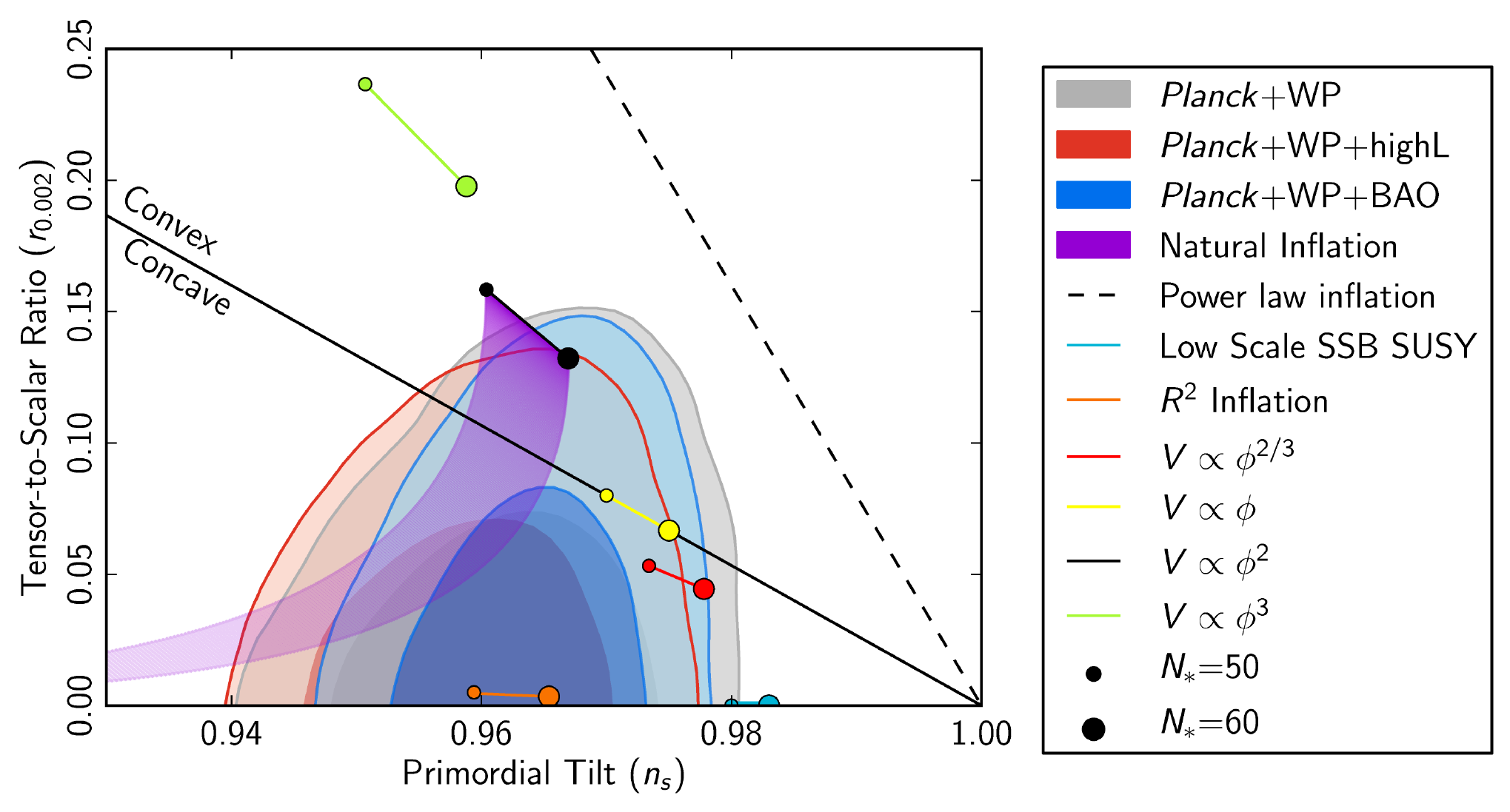}
\end{center}
\caption{Constraints set by Planck data on various inflation models in the
plot $r$ (evaluated at the pivot scale $k_* = 0.002$ Mpc${}^{-1}$) 
vs $n_s$~\cite{Planck13_22}. Small dots correspond to models with $50$ 
e-foldings, large dots to $60$ e-foldings.}
\label{figPlanck_inflation}
\end{figure}

\item concave potential or small field models ($\eta<0$)

In this class, illustrated first by the new inflation scenario~\cite{Li82,AS82},
the field $\phi$ starts at  a small value and rolls along an almost flat 
plateau (where $V''(\phi)<0$) before falling to its ground state. This type of 
potential, often encountered in symmetry breaking transitions may be 
parametrized, during the phase transition by:
\begin{equation}
\label{hilltop}
V(\phi) = V_0 \left[ 1 - \left({\phi \over \mu}\right)^p + \cdots \right] \ ,
\end{equation}
where the dots indicate higher order terms not relevant for inflation.
In the same class appears the so-called ``natural inflation'' potential 
\cite{BG86,FFO90}
\begin{equation}
\label{natural}
V(\phi) = V_0 \left[1 + \cos (\phi/f) \right] \ .
\end{equation}

A difficulty shared by the class of small field models is the unnaturalness of 
the initial conditions: why start at the height of the potential, in a plateau 
region or close to an unstable extremum? In the case of a symmetry breaking
potential, the rationale could be thermal: the restoration of the symmetry at 
high temperature naturally leads to start at the unstable 
``false vacuum''.

The first model proposed for inflation~\cite{St80}  was based on a modification 
of gravity described by the following action:
\begin{equation}
\label{Star}
S = {m_{_P}^2 \over 2}\int d^4 x \sqrt{-\tilde g} \left( \tilde R + 
\alpha \tilde R^2 \right)  \ ,
\end{equation}
where $\tilde R$ is the Ricci scalar associated with the metric $\tilde 
g_{\mu\nu}$. It may be proved that this is equivalent to standard Einstein 
gravity plus a scalar
field with a potential that falls in the same class as we just discussed 
({\em see Exercise 2-1}).

\item hybrid models ($0<2\epsilon<\eta$)

The field rolls down to a minimum of large vacuum energy (where $V''(\phi)<0$)
from a small initial value. Inflation ends because, close to this minimum, 
another direction in field space takes over and brings the system to a 
minimum of vanishing energy~\cite{Li91}. Such models, which thus require 
several fields, were constructed in order to allow inflation at scales much 
smaller than the Planck scale. In this case (see (\ref{5-22a}), $\epsilon$ can 
be very small and thus $r$ as well. This class of model is the only one that 
can accomodate values of $n_S$ larger than $1$ (i.e. a blue spectrum). 

We note that, for most models, the inflation scale is much larger than the TeV 
scale. This should in principle lead us to consider inflation models in the 
context of supersymmetry, in order to avoid an undesirable fine tuning of 
parameters. In fact,  supersymmetric (and superstring) theories are plagued 
with the presence of numerous
flat directions: this might be a blessing for the search of inflation 
potential\footnote{{\em One possible difficulty arises from the condition  
(\ref{5-18}) which may be 
written as a condition on the mass of the inflaton field
\begin{equation}
\label{5-22b} 
m^2 \ll H^2 \ .
\end{equation}
Any fundamental theory with a single dimensionful scale (such as string theory)
runs into the danger of having to fine tune parameters in order to satisfy
this constraint. This is known as the $\eta$ problem.}}. Two such types of 
potentials rely on the properties of basic supersymmetry multiplets: they are 
called $F$-term~\cite{CLLSW94,St95} and $D$-term~\cite{BD96,Ha96} inflation and 
fall in the category of hybrid inflation. 
They are not presently 
favoured by Planck data because they give too large values of $n_S$ (of the 
order of $0.98$).

\end{itemize}

\vskip .5cm
{\em \underline{Exercise 2-1}~: We show that a certain class of 4-dimensional 
models which extend Einstein theory are equivalent to Einstein gravity with a 
scalar field coupled to the metric (i.e. a scalar-tensor theory). This class 
of models is described by the following action:
\begin{equation}
\label{4-1-1}
S = {m_{_P}^2 \over 2}\int d^4 x \sqrt{-\tilde g} f(\tilde R) + S_m(\psi, 
\tilde g_{\mu\nu}) \ ,
\end{equation}
where $\tilde R$ is the Ricci scalar associated with the metric $\tilde 
g_{\mu\nu}$, and $f(\tilde R)$ is a general function of this Ricci scalar. 
We note that the Starobinsky action (\ref{Star}) corresponds to $f(\tilde R) 
= \tilde R + \alpha \tilde R^2$.

Let us consider the more general action:
\begin{equation}
\label{4-1-2}
S = {m_{_P}^2 \over 2} \int d^4x  \sqrt{-\tilde g} \left[ f(\chi) + 
{df \over d\chi} (\tilde R - \chi) \right] + S_m (\psi, \tilde g_{\mu\nu})
\ .
\end{equation}

a) Show that the variation of (\ref{4-1-2}) with respect to $\chi$ leads to  
(\ref{4-1-1}).

b) Redefining the metric and the scalar field through
\begin{equation}
\label{4-1-3}
g_{\mu\nu} \equiv {df \over d\chi} \tilde g_{\mu\nu} \ \  , \ \ 
\phi \equiv -\sqrt{{3 \over 2}} m_{_P} \log {df \over d \chi} \ ,
\end{equation}
show that one recovers the familiar form of the scalar-tensor gravity:
\begin{equation}
\label{4-1-4} 
S = \int d^4 x \sqrt{-g} \left[{m_{_P}^2 \over 2} R - {1\over 2} 
\partial^\mu \phi \partial_\mu \phi - V (\phi)\right] + S_m (\psi, A^2(\phi)
g_{\mu\nu})  \ ,
\end{equation}
where one will express the potential $V(\phi)$ in terms of $\chi$, $f(\chi$ and 
$df/d\chi$, and give the explicit form of $f(\phi)$.

c) Identify the potential $V(\phi)$ in the case of the Starobinsky function
$f(\chi) = \chi + \alpha \chi^2$.

\vskip .3cm 
Hints:
a) $\chi = \tilde R$, under the condition that $f'' \not = 0$.

b) $V \equiv m_{_P}^2 {\chi df/ d\chi-f \over 2(df/d\chi)^2}$
and $A(\phi) = \exp \left[\phi/(m_{_P}\sqrt{6})\right]$.
}
\section{Light does not say it all (1): the violent Universe} 
\label{chap:3}

The Universe is the siege of many violent phenomena; one may cite explosions 
like supernovae, gamma ray bursts (GRB) or the emission of energetic particles 
by active galaxy nuclei (AGN), quasars, blazars... The time constants $\tau$
associated 
with the phenomena are very short on the scale of the Universe. For example, a 
GRB may be visible on the sky only for a few seconds. This means that
the distance scales $c\tau$ involved are very small: the distance that light 
travels in $10$ seconds is only 3 million km, that is $0.002$ astronomical unit 
($1$ a.u. is the Sun-Earth distance). Indeed, very compact 
objects, such as neutron stars or black holes, are at the heart of such violent 
phenomena. We will start by reviewing the origin of such compact astrophysical 
objects, which appear at the end of the life of a star.

\subsection{The end of the life of a star: from white dwarfs to neutron stars and
black holes}    \label{2s4}

The evolution of a generic gravitational system such as a star is governed by 
two competing processes: gravitational forces which tend to contract the system 
and  thermal pressure which is due to the thermonuclear reactions within, which
tendto expand the sytem. In a stable star like our Sun at present, the two 
processes balance each other. But when the nuclear fuel is exhausted, the (core 
of the) star starts to collapse under the effect of gravity; the gravitational
energy thus released heats up the outer layers of the star, which produces the 
explosive phenomena that we observe. 

But what is the fate of the collapsing core? Gravitational pressure is 
eventually counterbalanced by quantum degeneracy pressure. Let us explain the 
nature of this 
pressure. Since matter is made of fermions of spin 
$1/2$, Pauli principle applies: two fermions cannot be in the same state.
Fermionic matter will thus resist at some point to excessive pressure. 

{\em Let us be more quantitative. 
Since there are $4\pi p^2 dp / (2\pi \hbar)^3$ levels per unit volume with 
momentum between $p$ and $p+dp$ and two spin states per level, the number of 
fermions per unit volume is given in terms of the maximal momentum by 
\begin{equation}
\label{2-23}   
n = {2 \over (2 \pi \hbar)^3} \int_0^{p_F} 4 \pi k^2 dk = {p_F^3 \over 3 \pi^2 
\hbar^3} \ .
\end{equation}
The energy of the highest level, or Fermi energy $\epsilon_F$, is 
therefore given in 
terms of the number density $n$. If the particles are non-relativistic, then
\begin{equation}
\label{2-24}
\epsilon_F = {p_F^2 \over 2 m} = {1 \over 2} \left( 3 \pi^2 \right)^{2/3} 
\hbar^2 {n^{2/3} \over m} \ .
\end{equation}

On the other hand,  the gravitational energy per nucleon of a system of size 
$R$ and mass $M$ (with $N= 4\pi R^3 n_N/3 = M/m_N$ nucleons) is
\begin{equation}
\label{2-25}
\epsilon_g = {G_{_N}M m_N \over R} = G_{_N} m_N^2 {N \over R} 
= \left( {4 \pi \over 3} \right)^{1/3} G_{_N} m_N^2 N^{2/3} n_N^{1/3} \ .
\end{equation}
The Fermi energy starts to dominate over the gravitational energy for
\begin{equation}
\label{2-26}
n_N^{1/3} > {2 \over (3\pi^2)^{2/3}} \left( {G_{_N} m_N^2 m \over \hbar^2}
\right) N^{2/3} \nu^{2/3}, 
\end{equation}
where $\nu = n_N/n$ ($\nu$ depends on the species of the fermions that are 
degenerate; see below), or
\begin{equation}
\label{2-27}
RM^{1/3} < {1 \over \alpha_G} {\hbar \over mc} m_N^{1/3} \nu^{-2/3} \ ,
\end{equation} 
where, as above, $\alpha_{_G} \equiv (G_{_N} m_N^2/ \hbar c) \sim 6 \times 
10^{-39}$.

We see from (\ref{2-26}) that} gravitational collapse is first stopped by
the quantum degeneracy of electrons: the corresponding astrophysical objects 
are known as white dwarfs. {\em Writing thus $m=m_e$ and $\nu = 2$ 
(two nucleons per electron), we find that $RM^{1/3} \sim 10^{-2} R_\odot
M_\odot^{1/3}$.} A white dwarf with $M=M_\odot$ has radius $R \sim 10^{-2}
R_\odot$ and density $\rho \sim 10^6 \rho_\odot$. It is more compact than a 
star.

If density continues to increase, the value of the Fermi energy is such that 
the fermions are relativistic: {\em it follows from (\ref{2-23}) that 
$p_F > mc$ 
reads $\left( 3 \pi^2 \right)^{1/3} \hbar n^{1/3} > mc$ or, 
using $n = 3N/\left( 4\pi \nu R^3 \right)$,
\begin{equation}
\label{2-28}
R < \left( {9 \pi \over 4} \right)^{1/3} {\hbar c \over mc^2}  N^{1/3} 
\nu^{-1/3} \ .
\end{equation}
But, since $\epsilon_F \sim p_F c = \left( 3 \pi^2 
\right)^{1/3} \hbar c n^{1/3}$, both $\epsilon_F$ and $\epsilon_g$ 
scale like $n^{1/3}$.} Quantum degeneracy pressure can overcome gravitational 
collapse only for $N < 3 \sqrt{\pi} \alpha_G^{-3/2}/(2\nu^2)$, or 
\begin{equation}
\label{2-29}
M < 3 \sqrt{\pi} \alpha_G^{-3/2}m_N/(2 \nu^2) \sim 1 \ M_\odot/\nu^2 \ . 
\end{equation}
This bound is the well-known Chandrasekhar limit for white dwarf masses 
(a more careful computation gives a numerical factor of $5.87$~\cite{Wei}).  
The radius of the object then satisfies ({\em see (\ref{2-28})}) 
\begin{equation}
\label{2-30}
R < {3 \sqrt{\pi} \over 2} \alpha_G^{-1/2} {\hbar c \over mc^2} {1 \over \nu} 
\ .
\end{equation}
Setting $m=m_e$ gives a limit value of some $10^4$ km.

For even higher densities, most electrons and protons are converted into 
neutrons through inverse beta decay ($p+e^- \rightarrow n + \nu$). A new 
object called neutron star
forms when the neutron Fermi energy balances the gravitational energy. 
Writing $m=m_n$ instead of $m_e$ in (\ref{2-27}), we now have $RM^{1/3} 
\sim 10^{-5} R_\odot
M_\odot^{1/3}$: a neutron star  with $M=M_\odot$ has radius $R \sim 10^{-5}
R_\odot$ and density $\rho \sim 10^{15} \rho_\odot$. 

The bound (\ref{2-29}) obtained above in the case of relativistic fermions
(neutrons in this case)  is called the Oppenheimer-Volkoff bound: more 
precisely, the maximal mass of a neutron star is $M=0.7 \ M_\odot$, with
a corresponding radius $R = 9.6$ km (cf. (\ref{2-30}) with $m=m_n$).
 If the mass is larger, the star undergoes 
gravitational collapse and forms a black hole. 

\subsection{Gravitational collapse: black holes} \label{2s6}

Let us first backtrack a little and return to Einstein's equations 
(\ref{2-Einstein}). Because they are non-linear, there are few solutions known.
The first exact non-trivial solution was found in late 1915 by Schwarzschild,
who was then fighting in the German army, 
within a month of the publication of Einstein's theory and presented on his 
behalf by Einstein at the Prussian Academy in the first days of 1916
\cite{Sc16}, just 
before Schwarzschild death from a illness contracted at the front.
It describes static isotropic regions of empty spacetime, such as the ones 
encountered in the exterior of a static star of mass $M$ and radius $R$.

The Schwarzschild solution reads, for $r>R$ ({\em see Exercise 3-1}),
\begin{equation}
\label{2-Schwarz'}
ds^2 = \left( 1 - {2G_{_N} M \over r} \right) dt^2 - 
\left( 1 - {2G_{_N} M \over r} \right)^{-1} dr^2 - r^2 d\theta^2 - r^2 \sin^2 
\theta d\phi^2 \ .
\end{equation}

The Schwarzschild solution is singular at $r=R_S \equiv 2G_{_N} M$, a distance 
known as the Schwarzschild radius. This is not a problem as long
as $R_S < R$ since this solution describes the exterior region of the star.
A different metric describes the interior. On the other hand, we will see 
in Section~\ref{2s6} that, in the case where $R<R_S$, i.e. $2 G_{_N} M/R >1$,
the system undergoes gravitational collapse and turns into a black hole.

\vskip .5cm
{\em \underline{Exercise 3-1}~: 
In this exercise, we derive the Schwarzschild solution (\ref{2-Schwarz'}). 
Because we look for static isotropic solutions, we may always write the 
spacetime metric as\footnote{We have absorbed a general function $e^{2\mu(r)}$ 
in front of the last term by redefining the variable $r$.}:
\begin{equation}
\label{2-17a}
ds^2 = e^{2\nu(r)} dt^2 - e^{2\lambda(r)} dr^2 -r^2 \left( d\theta^2 +
\sin^2 \theta d\phi^2 \right) \ .
\end{equation}
In other words, the only non-vanishing elements of the metric are:
\begin{equation}
\label{2-17b}
g_{tt} = e^{2\nu(r)}, \ g_{rr} = -e^{2\lambda(r)}, \ g_{\theta\theta} =
- r^2, \ g_{\phi\phi} = - r^2 \sin^2 \theta \ . 
\end{equation}

a) Work out the Christoffel symbols
from (\ref{2-1a}) and the Ricci tensor components from (\ref{2-1c}).

b) Show that the Einstein's equations in the vacuum simply amount to a 
condition of vanishing Ricci tensor: 
\begin{equation}
\label{2-17e}
R_{\mu\nu} = 0 \ .
\end{equation}

c) From the vanishing of $R_{tt}$ and $R_{rr}$ and the fact that,
at large distance from the star, space should be flat, both $\lambda$
and $\nu$ should vanish at spatial infinity. Hencededuce that
\begin{equation}
\label{2-17f}
\lambda = - \nu \ .
\end{equation}

d) From the vanishing of $R_{\theta\theta}$, deduce that
\begin{equation}
\label{2-17h}
g_{tt} = e^{2 \nu} = 1 - {2G_{_N} M \over r} \ . 
\end{equation}

\vskip .3cm 
Hints: 
a)\begin{eqnarray}
\label{2-17d}
R_{tt} &=& \left( \nu'' + \nu'^2 - \lambda'\nu' + {2 \nu' \over r} \right) 
e^{2(\nu-\lambda)} \ , \nonumber \\
R_{rr} &=& - \nu'' -  \nu'^2 + \lambda'\nu' + {2 \lambda' \over r} \ , 
\nonumber \\
R_{\theta\theta} &=& 1 -\left( 1 + r \nu' - r \lambda' \right) e^{-2\lambda}
\ , \nonumber \\
R_{\phi\phi} &=& R_{\theta\theta} \sin^2 \theta \ .
\end{eqnarray}

b) (\ref{2-Einstein}) reads $R_{\mu\nu} - {1 \over 2} g_{\mu\nu} R = 0$. 
Contracting with $g^{\mu\nu}$ yields $R =0$.

d) The constant of integration is identified with the mass $M$ because, in the
Newtonian limit, $g_{tt}= 1 + 2 \Phi$ where $\Phi$ is the Newtonian potential.}

\vskip .5cm
\underline{Exercise 3-2}~: What is the Schwarzschild radius of the sun? of an 
astrophysical object of mass $3\times 10^6 \ M_\odot$?

\vskip .3cm 
Hints: Do not forget that we have set $c=1$. Otherwise, $R_S = 2 G_{_N}M/c^2$,
that is $2.95$ km for the sun, $8.85 \times 10^9$ m $= 0.06$ au for an object 
of mass $3\times 10^6 \ M_\odot$.  

\vskip .3cm
We now understand that, when (the core of) a star of mass $M$ in gravitational 
collapse  overcomes the degeneracy pressure of neutrons to reach a size 
$R<R_S = 2 G_{_N} M$, nothing seems to drastically change for observers 
located at distances $r>R_S$. However, the behaviour of the Schwarzschild 
metric~\ref{2-Schwarz'} appears to be singular: $g_{tt}$ vanishes and $g_{rr}$ 
diverges. It took some time (Lema\^{i}tre
again!) to realize that this was not the sign of a real singualrity but was just
an artifact of the choice of coordinates: other choices lead to a regular 
behaviour (see Exercise 3-3). The true singularity lies at $r=0$ where the 
collapsing matter ends up.

In order to understand the nature of the surface at $r=R_S$, let us keep for 
a moment longer the Schwarzschild coordinates and consider sending a light 
signal radially from some point $r_1$ to 
$r_2>r_1$ where it is received a time $\Delta t$ later. Since $ds^2=0$ 
(as well as $d\theta=d\phi=0$), we have simply 
\begin{equation}
\label{2-49a}
\Delta t = \int_{r_1}^{r_2} {dr \over \left( 1-R_S/r \right)} \ .
\end{equation}
If $r_1<R_S$, this is finite only for  $r_2<R_S$, in which case it is 
simply $r_2 - r_1 + R_S \ln \left[(R_S -r_2)/(R_S -r_1)\right]$. In other
words, signals emitted from within the Schwarschild radius never reach the 
outside. There is really a breach of communication. Indeed, the surface 
$r=R_S$ is  an event horizon (see Section~\ref{4s3}).  

Let us take this opportunity to present a classical interpretation of the
Schwarzschild radius. Remember that the existence of black holes was conceived 
by Michell~\cite{Mi1784} and Laplace~\cite{Laplace} centuries earlier than 
general relativity. Indeed, the classical condition for escape a body of mass 
$m$ and velocity $v$ from a spherical star of mass $M$ and radius $R$ is
\begin{equation}
\label{2-40}
{1 \over 2} m v^2 > {G_{_N} mM \over R} \ .
\end{equation}
Thus, not even light ($v=c$) can escape the attraction of the star if 
$R < 2 G_{_N}M/c^2$, the Schwarzschild radius. 

We note that the Schwarzschild horizon is a fictitious surface, in the sense 
that an observer crossing this surface would not experience anything particular 
(we said that there exist coordinates where the behaviour at $R_S$ is regular),
except deformations due to tidal forces because it comes closer to a very 
massive object. But once it has crossed this fictitious surface, there is no 
way to backtrack: the further information that might be gained is lost for ever 
to the outside world. A useful picture is the one of a person swimming in a 
river with a waterfall downstream: swimming in the river involves no danger as 
long as one is safely far from the waterfall, but, at some point the swimmer 
crosses a fictitious line (the ``horizon'' of the waterfall) which is the point 
of no return: even the best swimmer is attracted towards the ``singularity'' of
the waterfall.

So far, our description has been purely classical. Quantum mechanical processes 
change this picture. Indeed, S. Hawking~\cite{Ha75} pointed out that black 
holes emit radiation through what is known as the process of evaporation. 
Indeed, it can be shown 
that an accelerated observer sees a thermal bath of particles at a temperature 
$T =\hbar a/(2\pi)$ ($a$ being the acceleration): this is the so-called Unruh 
effect~\cite{Un76}. Now, an observer who is at
a fixed distance $r >R_S$ from the horizon of a black hole in Schwarzschild 
coordinates has an acceleration $a=R_S(1-R_S/R)^{1/2}/(2r^2)$ ({\em see Exercise
3-4}). For $r \sim R_S$, it thus observes a thermal bath of particles at a 
temperature measured by an observer at infinity to be 
\begin{equation}
\label{3-TH}
T_H = {\hbar \over 4 \pi R_S}
\end{equation}
The Hawking evaporation process is important to understand the non-observation 
of primoridal black holes, which would be due to fluctuations of density during 
the Planck era: such primordial black holes have evaporated.

A final comment using the Schwarzschild coordinates (\ref{2-Schwarz'}): we see
that, when $r$ crosses $R_S$, the respective signs of $g_{tt}$ and $g_{rr}$ 
changes. In other words, $t$ becomes a spatial coordinates and $r$ becomes time:
the movement towards the central singularity is the clock that ticks. 

\vskip .5cm
\underline{Exercise 3-3}~: Define the Kruskal coordinates $(v,u,\theta,\phi$)
related to the Schwarzschild coordinates $(t,r,\theta,\phi)$ through 
\cite{Kr60}: 
\begin{eqnarray}
\label{2-K}
\hbox{for} \ r>R_S \ , && u=\left( r/R_S-1 \right)^{1/2} e^{r/2R_S} 
\cosh(t/2R_S) \ , \nonumber \\
&&  v=\left( r/R_S-1\right)^{1/2} e^{r/2R_S} \sinh(t/2R_S) \ , \nonumber \\
\hbox{for} \ r<R_S \ , && u=\left( 1-r/R_S \right)^{1/2} e^{r/2R_S} 
\sinh(t/2R_S) \ ,  \nonumber \\
&& v=\left( 1-r/R_S \right)^{1/2} e^{r/2R_S} \cosh(t/2R_S) \  .
\end{eqnarray}
Deduce from (\ref{2-Schwarz'}) the form of the metric in Kruskal coordinates:
\begin{equation}
\label{2-Kruskal}
ds^2 = {4R_S^3 \over r} e^{-r/R_S} \left( dv^2 - du^2 \right) - r^2 
\left( d\theta^2 + \sin^2 \theta d\phi^2 \right) , 
\end{equation}
where $r$ is given as an implicit function of $u$ and $v$:
\begin{equation}
\label{2-K'}
\left( {r \over R_S} -1 \right) e^{r/R_S} = u^2 - v^2 \ .
\end{equation}   

\vskip .5cm
{\em In order to be more quantitative, let us follow the analysis of 
Oppenheimer and Snyder~\cite{OS39} who were the first to discuss the collapse 
into a black hole.
We consider a fluid of negligible pressure, thus described by 
the energy-momentum tensor (see (\ref{2-4})) $T_{\mu\nu} = \rho U_\mu U_\nu$,
and study its spherically symmetric collapse. 

It turns out that we have already studied this system when we discussed the 
evolution of a homogeneous and isotropic universe in Section~\ref{sect:1-3}.
The metric is given by 
\begin{equation}
\label{2-41}
ds^2 = d\hat t^2 -  a^2(\hat t) \left( {d\hat r^2 \over 1-k \hat r^2} + 
\hat r^2 d \hat \theta^2 + \hat r^2 \sin^2 \hat \theta d\hat \phi^2 \right) 
\ , 
\end{equation} 
as in (\ref{2-2a})\footnote{except that we do not normalize $k$ to $\pm 1$ 
or $0$ because we are looking at  a different system. We will see just below 
that it is fixed by initial conditions. We add a hat to this system of 
coordinates to distinguish it from the Robertson-Walker coordinates, as well 
as from the Schwarzschild coordinates that we will use later.} and the 
Einstein tensor components are the 
same as in (\ref{2-3a},\ref{2-3b}). We normalize the 
coordinate $\hat r$ so that $a(0)=1$. Thus 
\begin{equation}
\label{2-42}
\rho(\hat t) = \rho(0)/a^3(\hat t)
\end{equation} 
and Einstein's equations simply read:
\begin{eqnarray}
\dot a^2 + k &=& {8 \pi G_N \over 3} {\rho(0) \over a} \ , \label{2-43a} \\
\dot a^2 + 2 a \ddot a + k &=& 0 \ . \label{2-43b}
\end{eqnarray}
Assuming that the fluid is initially at rest ($\dot a=0$), we obtain from
(\ref{2-43a})
\begin{equation}
\label{2-44}
k = {8 \pi G_N \over 3} \rho(0) \ .
\end{equation}
Thus, (\ref{2-43a}) simply reads
\begin{equation}
\label{2-45}
\dot a^2(\hat t) = k \left[ a^{-1}(\hat t) -1 \right] \ .
\end{equation}
The solution is given by the parametric equation of a cycloid:
\begin{eqnarray}
\label{2-46}
\hat t &=& {\psi + \sin \psi \over 2 \sqrt{k}} \ , \nonumber \\
a &=& {1 + \cos \psi \over 2} \ .
\end{eqnarray}
We see that $a$ vanishes for $\psi = \pi$, that is after a time
\begin{equation}
\label{2-47}
\tau = {\pi \over 2 \sqrt{k}} = {\pi \over 2} \left( {3 \over 8 \pi G_{_N} 
\rho(0)} \right)^{1/2} \ .
\end{equation}
Thus a sphere initially at rest with energy density $\rho(0)$ and negligible 
pressure collapses to a state of infinite energy density in a finite 
time $\tau$.

In the case of a star of radius $R$ and mass $M$, this solution for the 
interior of the star should be matched with the Schwarzschild solution
(\ref{2-Schwarz'}) describing the exterior.
The correspondence between the interior and exterior coordinates is simply
$r = R a(\hat t)$, $\theta =\hat  \theta$ and $\phi = \hat \phi$, with 
a more complicate relation between $t$ and $\hat t$ (see Ref.~\cite{Wei} section 
11.9). The first relation ensures that
\begin{equation}
\label{2-49} 
k = {2 M G_{_N} \over R^3} = {R_S \over R^3} \ ,
\end{equation}
in agreement with (\ref{2-44}) and $M=(4\pi/3)\rho(0) R^3$.}

{\em
\vskip .5cm
\underline{Exercise 3-4}~: We consider an observer at rest outside the horizon 
of a black hole described by the Schwarzschild metric (\ref{2-Schwarz'}) which 
we write (see Exercise 3-1):
\begin{equation}
\label{2-17aa}
ds^2 = e^{2\nu(r)} dt^2 - e^{-2\nu(r)} dr^2 -r^2 \left( d\theta^2 +
\sin^2 \theta d\phi^2 \right) \ .
\end{equation}
The observer velocity is $U^\mu= \xi^\mu/\sqrt{\xi^2}$ with $\xi^\mu \equiv 
\delta_0^\mu$.

a) Show that $\xi_{\mu ;\nu} = \xi_\mu w_\nu - \xi_\nu w_\mu$, where 
$w_\mu \equiv \partial_\mu \nu$ (the covariant derivative $\xi_{\mu ;\nu}$ is
defined in (\ref{covder}) of Appendix~\ref{app:B}). 

b) Deduce that the acceleration $A^\mu \equiv U^\rho \nabla_\rho U^\mu$ of the 
observer is simply $A^\mu = - w^\mu$.

c) Show that the acceleration $a^2 \equiv - w^\mu w_\mu$ is given for the 
observer at fixed $r$, $\theta$ and $\phi$ by
\begin{equation}
\label{Ex3-4-1}
a = {G_{_N} M \over r^2 (1- 2 G_{_N} M/r)^{1/2}} \ .
\end{equation}
}

We finally note that, at large distance, the black hole is only caracterized by
its mass $M$. Black holes are indeed very simple objects, somewhat similar to 
particles: Schwarzschild black holes are only characterized by their mass. 
Other more complex solutions were found later but it was realized that 
one can only add spin (rotating or Kerr black holes)
and charge (charged black holes) but no other independent characteristics:
in the picturesque language used by Wheeler, it is said that 
black holes can have no hair. In a sense, the black holes of general relativity
are very similar to fundamental particles,
which are caracterized by a finite set of numbers (including mass, 
spin, and electric charge). 

Astrophysical black holes are somewhat more complex because of their material 
environment, as we will now see.

\subsection{Astrophysical black holes} \label{sect:3-3}

\begin{figure}
\begin{center}
\includegraphics[scale=0.8]{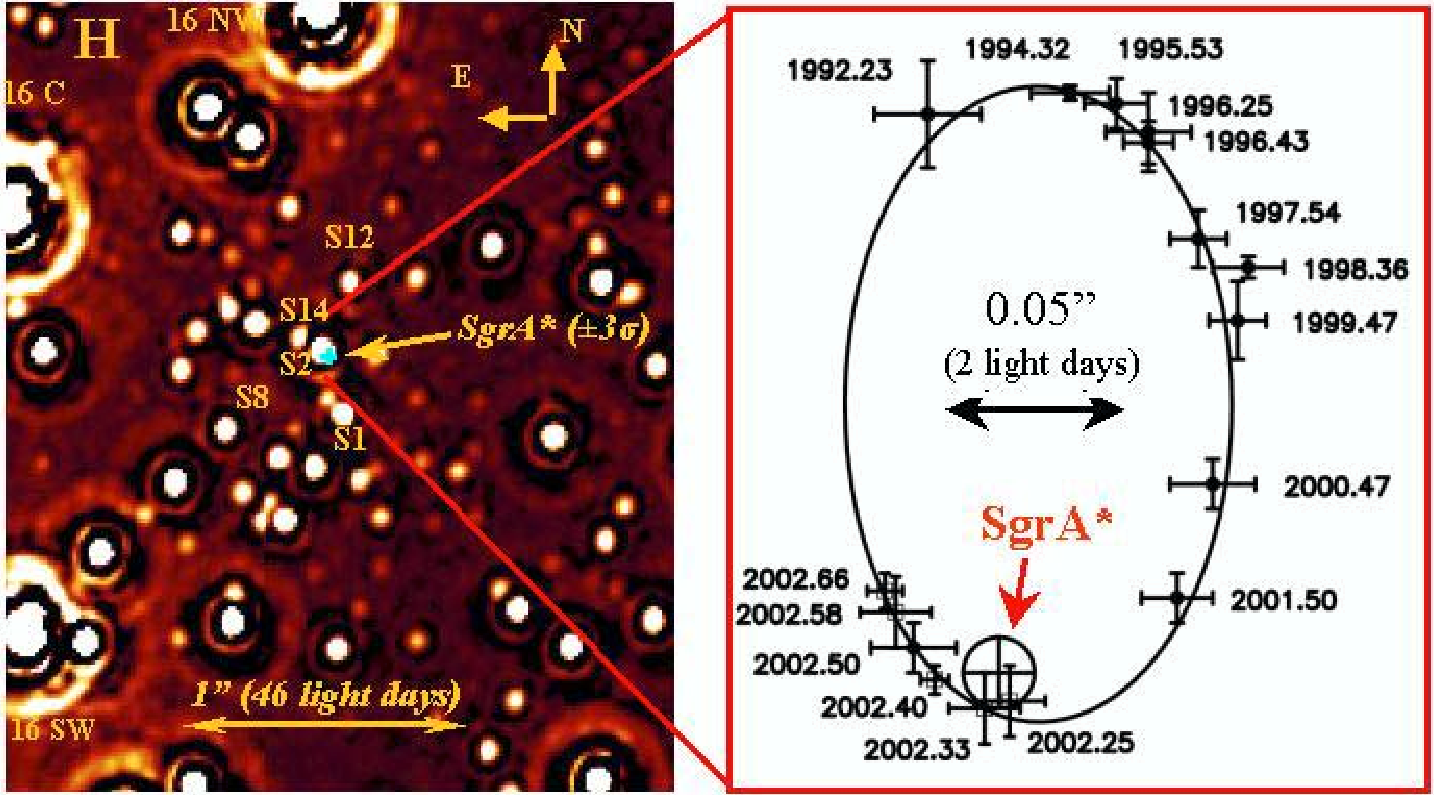}
\end{center}
\caption{Left: NAOS/CONICA image of the central 2'' (Sgr A$^*$ in light blue).
Right: orbit of the star S2 around Sgr A$^*$~\cite{Ot03}.}
\label{fig3-0a}
\end{figure}

For a long time, black holes were considered as a curiosity of general 
relativity and did not have the status of other stellar objects. This has 
changed in the last decade which has seen mounting evidence that,
at the center of our own galaxy (Milky Way) cluster, there is  a massive black 
hole associated with the compact radio source Sagittarius A$^*$.  
Observations of the motions of nearby stars by the  imager/spectrometer 
NAOS/CONICA working in the infrared~\cite{Ot03} 
have indeed confirmed the presence of a very massive object 
($(2.6 \pm 0.2). 10^6$ solar mass) localized in a very small region (a fraction 
of an astronomical unit, see 
Fig.~\ref{fig3-0a}), which seems only compatible with a black hole.

\begin{figure}
\vskip -6cm
\begin{center}
\includegraphics[scale=0.8]{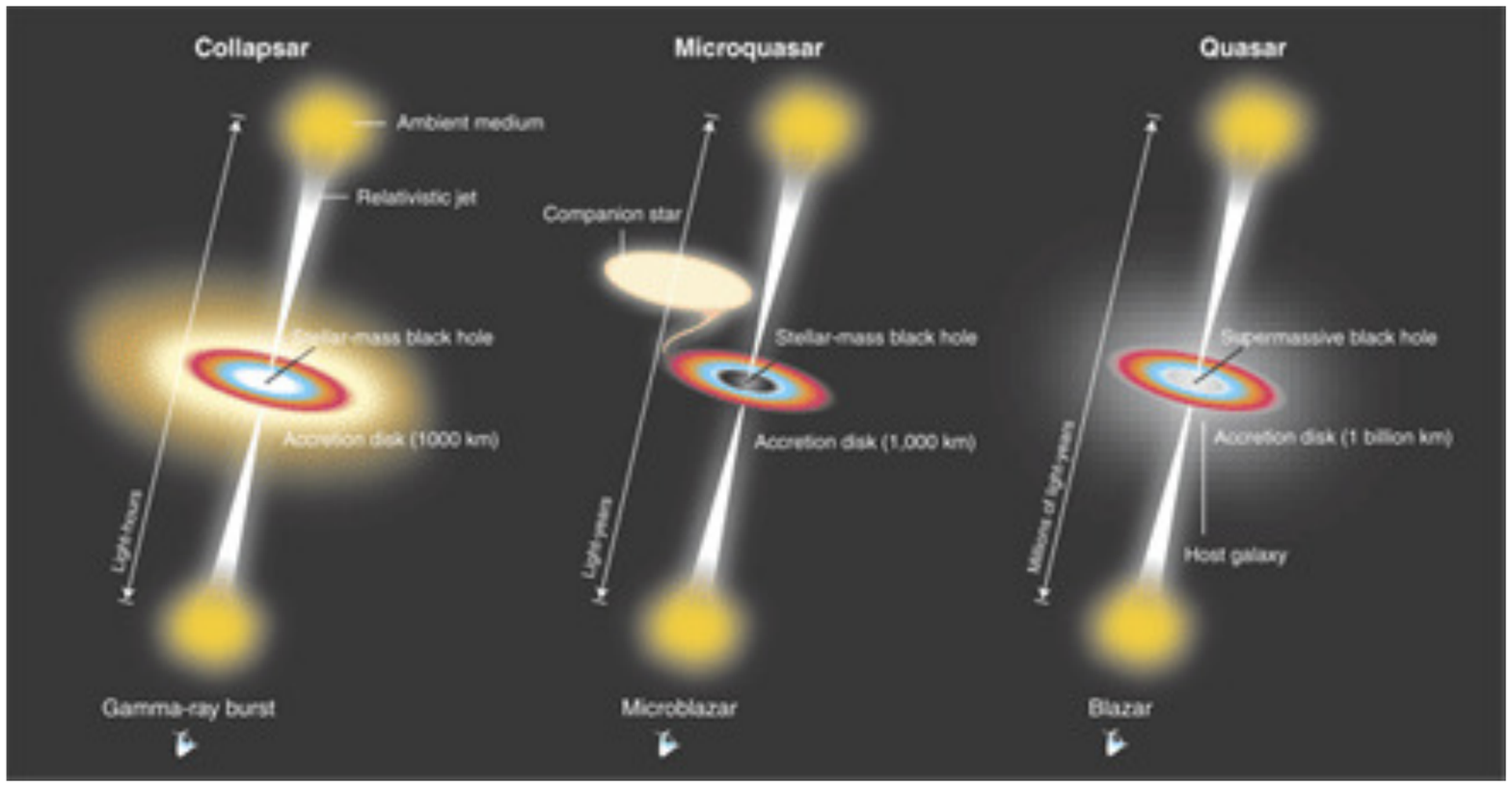}
\end{center}
\vskip -7cm
\caption{Unified picture of the system surrounding an astrophysical black hole 
in the case of a gamma ray burst (left), a microblazar (centre) and a blazar
(right). In the first two cases, the central object is a stellar mass black 
hole and the size of the accretion disk is about $1000$ km.
In the latter (AGN), it is a supermassive black hole of potentially several 
million solar masses and the accretion disk is of the order of $1$ billion kms.}
\label{astroBH}
\end{figure}

Since then, black holes have been identified in many instances and their role 
might be central in many phenomena. We have seen that they are very simple 
gravitational objects. But these simple objects accrete matter, and are thus 
associated with very diverse phenomena. A picture has emerged, which seems to 
be valid at very diverse scales (see Fig.~\ref{astroBH}) of a black hole 
surrounded by an accretion disk and a torus of dust, with two opposite 
relativistic jets, which 
are supposedly formed during the gravitational collapse through a recombination 
of the magnetic fields. Of course, at the centre of this complex structure, 
lies the black hole surrounded by its horizon. But the complex phenomena that 
take place in this surrounding region allow to detect indirectly the black 
hole.      

Let us review some of the astrophysical occurences of black holes.

First, our galaxy is not the only onewhich has a cventral black hole. This is 
believed to be very general, and in many cases the blck hole and its 
environment is much more active than our own.
Active galaxies are galaxies where the dominant energy output is not due to 
stars. In the case of Active Galactic Nuclei (AGN), the non-thermal radiation 
comes from a central region of a few parsecs around the centre of the galaxy.
The most famous example of such AGNs is provided by quasi-stellar objects 
(QSOs) or quasars: these starlike objects 
turn out to be associated with the point-like optical emission from the 
nucleus of an active galaxy. 

The typology of active extragalactic objects is very complex: radio loud and 
radio quiet quasars, Seyfert galaxies, BL Lacs or blazars... There has been
an effort to build a unified picture~\cite{BBR84}: the apparent diversity in 
the observations would then result from the diversity of perspectives from 
which we observers see these highly non-isotropic objects. Typically, the model
for radio-loud AGNs includes (see Fig.~\ref{astroBH}, right panel): a central
engine, a pair of oppositely directed 
relativistic jets (cones of semi-angle around $1^\circ$) an accretion disk 
(of size of the 
order of $1$ parsec), and a torus of material (of size of the 
order of $100$ parsec) which obscures the central engine when one observes
it sideways. Depending on the relative angle between the line of sight and the 
jet axis, observation may vary in important ways.

Gamma ray bursts (GRB) are the most luminous events observed in the universe. 
They were 
discovered accidentally by the American military satellites VEGA
which were designed to monitor the nuclear test ban treaty of 1963.
The first burst was found in 1969, buried in gamma-ray data from 1967: 
two Vela satellites had detected more or less identical signals, showing the 
source to be roughly the same distance from each satellite~\cite{KSO73}.

A GRB explosion can be as luminous as objects which are in our vicinity, such 
as the Crab nebula, although they are very distant. 
The initial flash is short (from a few seconds to
a few hundred for a long GRB, a fraction of a second for a short one). 
From 1991 to 2000, BATSE (Burst and Transient Source Experiment) has allowed
to detect some 2700 bursts and showed that their distribution is isotropic, a 
good argument in favor of their cosmological origin.
In 1997 (February 28), the precise determination of the position of a GRB 
(hence named 970228) 
by the Beppo-SAX satellite allowed ground telescopes to discover a rapidly
decreasing  optical counterpart, called afterglow. Typically in the afterglow, 
the photon energy decreases with time as a power law (from X ray to optical, 
IR and radio) as well as the flux: it stops after a few days or weeks. 
The study of afterglows gives precious information on the dynamics of GRBs.
The launch of the SWIFT satellite on 20 November 2004 has started a new era for
the understanding of GRBs. 

Given the time scales involved (a few milliseconds for the rise time of the 
gamma signal), the size of the source must be very small: it cannot exceed the 
distance that radiation can travel in the same time interval, 
i.e. at most a few 
hundred kilometers.  Energy must have been ejected in an ultra-relativistic 
flow which converted its kinetic energy into radiation away from the source:
the Lorentz factors involved are typically of the order of $100$! 

This flow is
collimated and forms a jet of half opening angle $\theta$. The observational 
evidence for 
this collimation is an achromatic break in the afterglow light curve: for 
$t > t_{{\rm jet}}$, it decreases faster than it would in the spherical case
\cite{Rh99,SPH99}. If we assume that the relativistic jet, after emitting a 
fraction $\eta_\gamma$ of its kinetic energy into prompt $\gamma$ rays, hits a 
homogeneous medium with a constant number density $n$, the break appears in 
the afterglow light curve when the Lorentz factor $\gamma$ becomes of the 
order of $1/\theta$. This gives a relation between the half opening angle  
$\theta$ and the break time $t_{{\rm jet}}$~\cite{SPH99}:
\begin{equation}
\label{3-35}
\theta = {0.161 \over (1+z)^{3/8}} \left( {t_{{\rm jet}} \over 1 \ {\rm day}} 
\right)^{3/8}  \left( {10^{52} \ {\rm ergs} \over E_{\gamma,{\rm iso}}} 
\right)^{1/8} \left( {n \over 1 \ {\rm cm}^3} \right)^{1/8}
\eta_\gamma^{1/8} \ ,
\end{equation}
where $E_{\gamma,{\rm iso}}$ is the isotropic equivalent gamma ray energy.

In the collapsar model of 
Woosley~\cite{Wo93}, long GRBs are associated with  the explosion of a 
rapidly rotating massive star which collapses into a spinning black hole
(see Fig.~\ref{astroBH} left panel). 
The burst and its afterglow have been successfully explained by the 
interaction  of a highly relativistic jet with itself (internal shocks 
\cite{PX94,RM94}) and with the circumstellar medium  (external shocks
\cite{RM92}).  
Typically, 
one expects per day $10^6$ collapses of massive stars in the Universe;
$10^3$ give rise to a GRB and approximately $1$ of these is pointing towards us
its jet. Hence, one may observe from earth about  one GRB per day.

\vskip .5cm
Supernovae explosions also provide very bright events in the sky, some of 
them being visible to the naked eye. Supernovae explosions  were thus recorded
in 1006, 1054, 1181, 1572 and 1604 . The Sn 1987 A explosion allowed the 
detection of neutrinos and gamma emission.

The modern theory of supernovae was initiated in the 30s by Baade and Zwicky 
\cite{BZ34}.

Supernovae follow a classification according to spectroscopy. In type I 
supernovae, hydrogen lines are absent whereas they are present in type II.
Moreover, type I has subclasses: for example, type Ia involves intermediate 
mass elements (Si). Each type corresponds to a different mechanism for the 
explosion. In particular, type II and type Ia have a completely different 
interpretation.

We will focus first on type II supernovae.

Presupernova stars ($M> 8 M_\odot$) 
have an onion-like structure. From the outer to the inner layers, one finds 
increasingly heavy elements: $H$, $He$, $C$, $O$, $Ne$, $Si$ and $Fe$. 

As $Si$ is consumed by nuclear reactions, the mass of the $Fe$ core increases . 
The resulting
density increase then turns the electrons relativistic and makes electronic
capture ($p+e \rightarrow n+\nu$) energetically favorable. This diminishes the 
degenerate electron pressure and leads to the collapse of the core.  Since
$\rho_{core} \sim 10^{12}$ kg.m$^{-3}$, the collapse time is typically 
$\left( G_{_N} \rho \right)^{-1/2} \sim 0.1$ s. 

This time, neutrinos produced as electrons are turned into neutrons,
are trapped in the imploding core. The critical density 
for which neutrinos are trapped, is
typically $\rho \sim 2\times 10^{14}$ kg/m$^3$.
As the core is crushed to higher densities, the density approaches that of a 
neutron star ($\rho \sim 2 \times 10^{17}$ kg/m$^3$) and matter becomes almost 
incompressible. If the process was elastic, the kinetic energy would be enough 
to bring it back to the initial state. Typically
\begin{equation}
\label{3-34}
E \sim G_{_N} M_{core} \left( {1 \over R_{NS}} - {1 \over R_{WD}} \right) 
\sim {G_{_N} M_{core}  \over R_{NS}} \sim 3\times 10^{46} \ \hbox{J} \ .
\end{equation}

This is not completely so but there is a rebound of the core which 
sends a shock wave outward. Meanwhile, the stellar matter has started to free 
fall since it is no longer sustained by its core. The falling matter meets the 
outgoing shock wave and turns it into an accretion wave. 

Neutrinos emitted from the core heat up and expand the bubble thus formed. 
Convection and neutrino heating thus convey a fraction of the order of one 
percent of the neutron star gravitational mass (\ref{3-34})
to the accretion front. This is enough to make it explode.

One word of caution however: numerical models that try to reproduce supernovae 
explosions have been until now unable to explode the supernovae! One needs
to start the explosion artificially. It therefore remains possible that one is 
still missing a key ingredient in the recipe.

The bulk of the star blown off by the explosion makes what is known as a
supernova remnant. It sweeps the interstellar medium at great velocity ($10 000
$ km/s) and may remain visible for $10^5$ to $10^6$ years. A large fraction of
the interstellar medium is thus swept by supernovae remnants (see Exercise 
3-1). This is 
important since this is believed to be the way the heaviest nuclear elements 
are scattered in the  universe (primordial nucleosynthesis produces no 
element heavier than ${}^7 \hbox{Li}$).

\vskip .5cm
\underline{Exercise 3-4}~: a) Assuming approximately one supernova explosion
every 30 years 
in our galaxy (assimilated to a disk of radius $15$ kpc and thickness 
$200$ pc), compute the corresponding rate ${\cal R}$ of supernovae explosions
per pc$^3$ and per year.

b) If every supernova leads to a remnant of radius $R=100$ pc that lasts for 
$t\sim 10^6$ yrs, what fraction of the galaxy volume is filled by the 
supernova remnant?

\vskip .3cm
Hints: a) ${\cal R}\sim 2.3\times 10^{-13}$ pc$^{-3}$.yr$^{-1}$.

b)$1-\exp\left[-(4\pi/3)R^3{\cal R}t\right] \sim 0.5$.

\vskip .5cm

SNIa events on the other hand are thermonuclear explosions of white dwarfs. 
More precisely, a 
carbon-oxygen white dwarf accretes matter (from a companion star or by 
coalescence with another white dwarf) which causes its mass to exceed the 
Chandrasekhar limit. 
The central core collapses, making the carbon burn and causing a wave of 
combustion to propagate through the star, disrupting it completely. 
The total production  of energy is thus almost constant. For a white dwarf of 
radius $1500$ to $2000$ km,  about $2\times 10^{51}$ ergs  is released 
in a few seconds during which takes place the acceleration of the material. 
This is followed by a period of free expansion. Virtually all the energy of the
explosion goes into the expansion. The luminosity of the supernova, on the 
other hand, finds its origin in the nuclear decay of the ${}^{56} \hbox{Ni}$ 
freshly
synthesized.  The energy release in the nuclear
decays ${}^{56} \hbox{Ni} \rightarrow {}^{56} \hbox{Co} \rightarrow {}^{56} 
\hbox{Fe}$, with 
respective lifetimes of $8.8$ and $111$ days,
represents a few percent of the initial energy release.

This model allows to understand the homogeneity of the observed type Ia 
supernovae explosions and why they have been used successfully as standard 
candles in cosmology (see Section~\ref{sect:1obs-1} of Chapter~\ref{chap:5}).
The structure of a white dwarf is determined by degenerate electrons and thus 
independent of detailed chemical composition (see Section~\ref{2s4}). The rate of expansion is set by the total energy 
available since the complete white dwarf is disrupted. {\em Finally, the 
absolute brightness is determined by the radioactive decay of ${}^{56} 
\hbox{Ni}$
produced during the explosion. Less Ni means a lower luminosity but also 
lower temperature in the gas and thus lower opacity and more rapid energy 
escape. Thus dimmer supernovae are quicker i.e. have narrower light curves.}

\subsection{High energy cosmic particles} \label{sect:3-4}

As explained above, compact objects and the violent phenomena associated with 
their formation are important to understand the origin of high energy cosmic 
particles. It is important to identify the potential sites of acceleration.
Obviously, the jets described in the preceding Section are sources of energetic 
particles. Shock fronts, such as supernovae remnants, are also the siege of 
acceleration for particles whose multiple scattering off magnetized clouds
lead to multiple encounters with the shock front.   

M. Hillas~\cite{Hi84} has proposed a general discussion of potential 
acceleration sites, in terms of the magnetic fields $B$ availalble and the size 
$R$ of the site. The Larmor radius of the particle $r_L = E/(qBc)$ 
(in relativistic regime) may, 
with increasing energy $E$, become larger than the dimension $R$ of the 
accelerating site. We thus have the condition ($q=Ze$)
\begin{equation}
\label{3-64}
E < E_{{\rm max}} = qBcR = Z \left({B \over 1 \ \mu\hbox{G}}\right)
\left({R \over 1 \ \hbox{Mpc}}\right) 9.3 \ 10^{20} \hbox{eV} \ .
\end{equation}
Note that this is the work of the electric field ${\cal E}=Bc$ over the maximal
distance $R$. In the case of acceleration on magnetic clouds or a shock wave
(where ${\cal E}= BV$), the maximal energy reads:
\begin{equation}
\label{3-65}
 E_{{\rm max}} = ZeB V  R \ .
\end{equation}
In the case  where acceleration involves large Lorentz factors $\gamma$, an 
extra factor $\gamma$ should be included to account for the energy being 
measured in the lab frame (also ${\cal E}= \gamma BV$). 

\begin{figure}
\begin{center}
\includegraphics[scale=0.6]{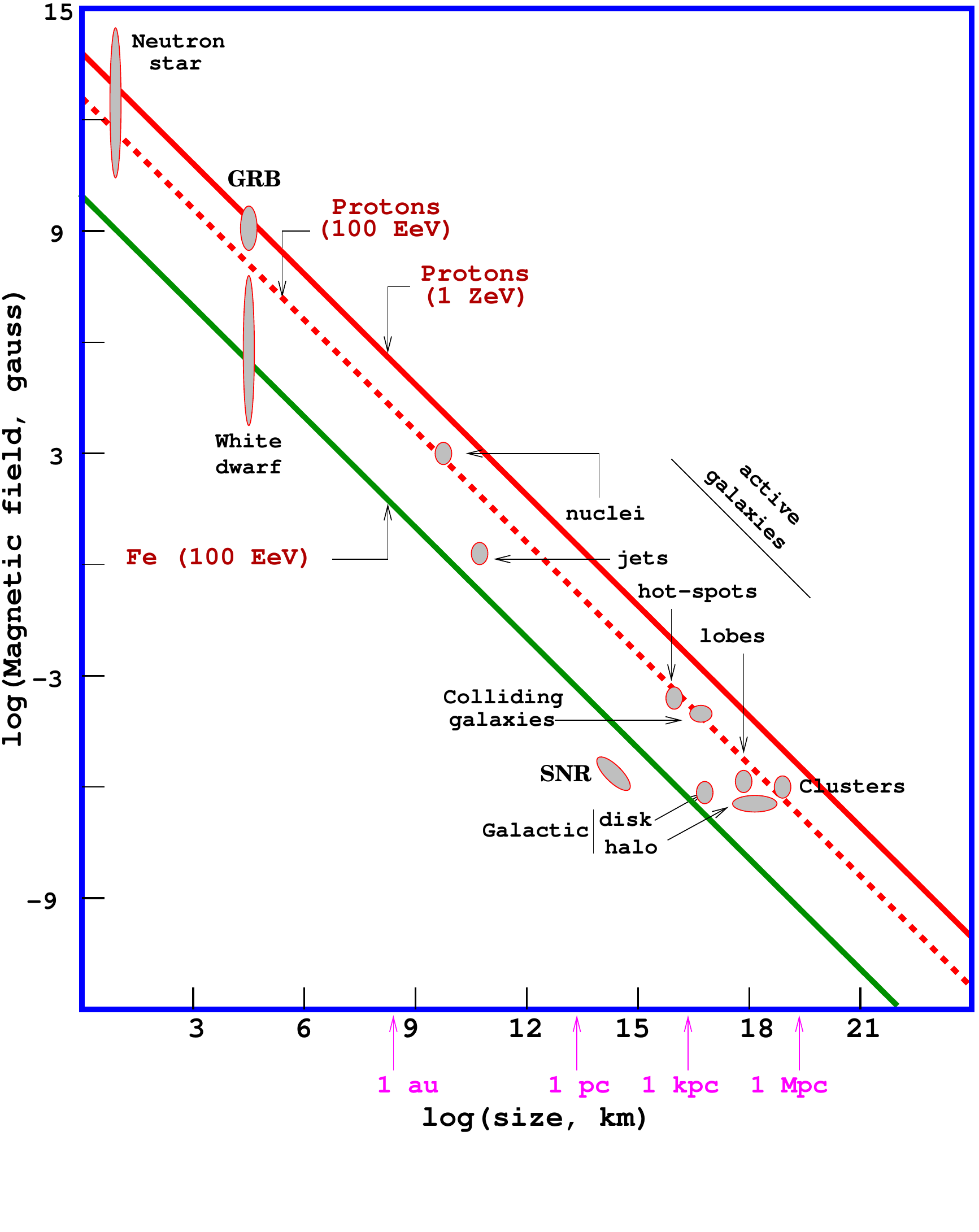}
\end{center}
\vskip -1cm
\caption{Hillas diagram showing size and magnetic fields of potential 
acceleration sites. Sites below the diagonal lines cannot accelerate 
protons above $10^{21}$ eV, protons above $10^{21}$ eV and {\em Fe} nuclei
above $10^{20}$ eV, respectively from top to bottom.}
\label{fig3-1b}
\end{figure}

This general criterion allows to draw the 
now classical Hillas diagram which identifies the possible 
acceleration sites in a plot $\log \left( B/1\ 
\hbox{G}\right)$ vs. $\log \left( R/1\ \hbox{km}\right)$. As can be seen on 
Fig.~\ref{fig3-1b}, given species of 
cosmic particles accelerated at given energies are represented by diagonal 
lines (from top to bottom on the figure: protons of $10^{21}$ eV, protons
of $10^{20}$ eV and iron nuclei of $10^{20}$ eV). 

   
\section{Light does not say it all (2): dark matter} 
\label{chap:4}

We have known for a long time that the Universe has a dark component, and is not
only luminous matter and radiation: already in 1933, by studying the velocity 
distribution of galaxies, Zwicky~\cite{Zw33} discovered that the Coma cluster 
had $400$ times more mass 
than expected from its luminosity. Through the XX$^{{\rm th}}$ century, it was 
realized that non-luminous matter, dark matter, is needed at all scales from 
galactic to cosmological. This pleads for a form of matter which is 
not included in the Standard Model, hence for physics beyond the Standard Model.
Indeed, the detection of dark matter might be the first sign of physics beyond 
the 
Standard Model. Hence the programs of direct or indirect dark matter detection 
are of key importance not only for astrophysics but also for high energy 
physics.

\subsection{The observational case}\label{4s2a}

As we alluded to above, dark matter was first identified by Fritz Zwicky 
\cite{Zw33,Zw37}
in 1933 when studying the velocity distribution of galaxies in the Coma cluster.
Using the virial theorem
\begin{equation}
\label{virial}
2 \langle E_{{\rm kin}} \rangle = - \langle E_{{\rm pot}} \rangle 
\end{equation}
(where $\langle \cdots \rangle$ indicates time averaging), he concluded that 
there is $400$ times more mass than expected from the luminosity.

\begin{figure}[h]
\begin{center}
\includegraphics[scale=0.45]{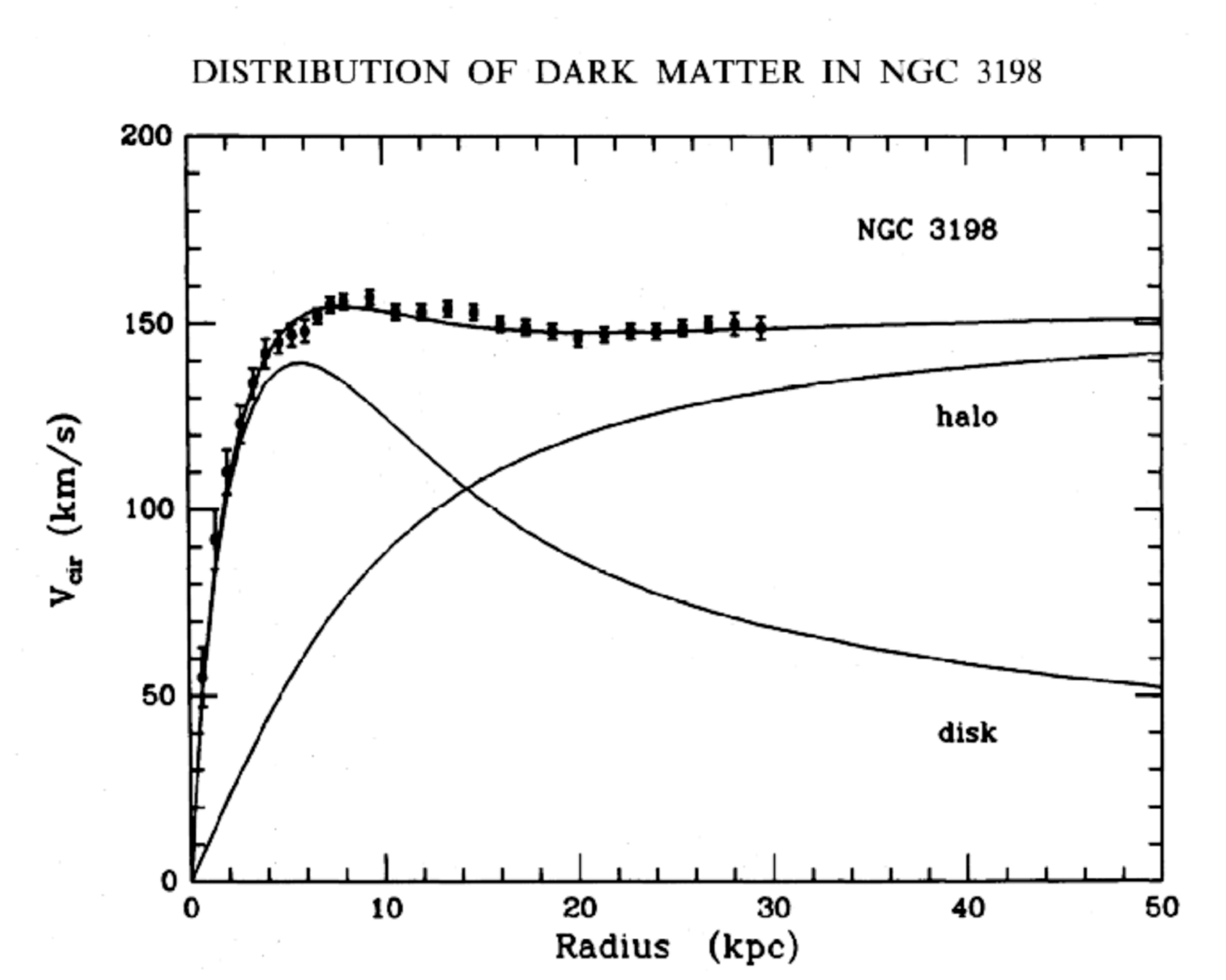}
\end{center}
\caption{Rotation curve $v(r)$ for the galaxy NGC 3198. The curve labeled 
``disk'' 
indicates the curve due to the stars in the galaxy, which extend only to $10$ 
kpc; the curve labeled ``halo'' 
is the one that would be due solely to a spherical halo of dark matter.}
\label{DM1}
\end{figure}

This was consistently confirmed by the study of the rotation curves of galaxies
i.e. the velocity $v(r)$ of stars as function of their distance $r$ to the 
centre of the galaxy. Using again the virial theorem (\ref{virial}), we have for stars in the outer regions of the galaxy, 
since $E_{{\rm kin}} \sim mv^2$ and $E_{{\rm pot}} \sim G_{_N}mM/r$ 
($m$ is the mass of 
the star, $M$ of the galaxy),
\begin{equation}
\label{rotation}
v \propto \sqrt{{G_{_N} M\over r}}  \ .
\end{equation}
Thus, one should see the velocity decrease as $r^{-1/2}$ for stars at the 
border of the galaxy\footnote{Note that for stars in the interior of the galaxy
one should replace $M$ by the mass of the sphere of radius $r$ ($M(r)
\propto \rho_{{\rm gal}} r^3$) which makes $v(r)$ increase with $r$.}. 
This is not what is observed. Indeed, the rotation curves of galaxies, 
which were studied thoroughly through the 60s and 70s showed (see Figure
\ref{DM1}) 
that the velocities do not start decreasing at the border of the luminous 
galaxy, as if there was more matter beyond. In fact, one needs a factor of 
order $10$ more matter in spiral galaxies. The extra matter forms a halo
that extends beyond the luminous galaxy.

For example, the modern picture of our own Galaxy, the Milky Way, is one of  a 
luminous bulge of a few kpc at the centre of a disk of radius $12.5$ kpc and 
thickness $0.3$ kpc, containing some $10^{11}$ stars, surrounded by a nearly 
spherical halo of dark matter of typical radius $30$ kpc (see Appendix 
\ref{app:A} for the definition of a parsec).

But dark matter is not only present in galaxies. We have seen that it was first 
identified by Zwicky in clusters of galaxy. This is now confirmed by many 
observation of clusters. X-ray studies have revealed the presence of large 
amounts of intergalactic gas which is very hot, and hence emits X-rays. 
The total mass of the gas is greater than that of the galaxies by roughly a 
factor of two. However this is not enough mass to keep the galaxies within 
the cluster. Since this gas is in approximate hydrostatic equilibrium with the 
cluster gravitational field, the mass distribution can be determined,
which leads to a total mass estimate  approximately six times larger than the 
mass of the individual galaxies or of the hot gas. 

\begin{figure}[h]
\begin{center}
\includegraphics[scale=0.99]{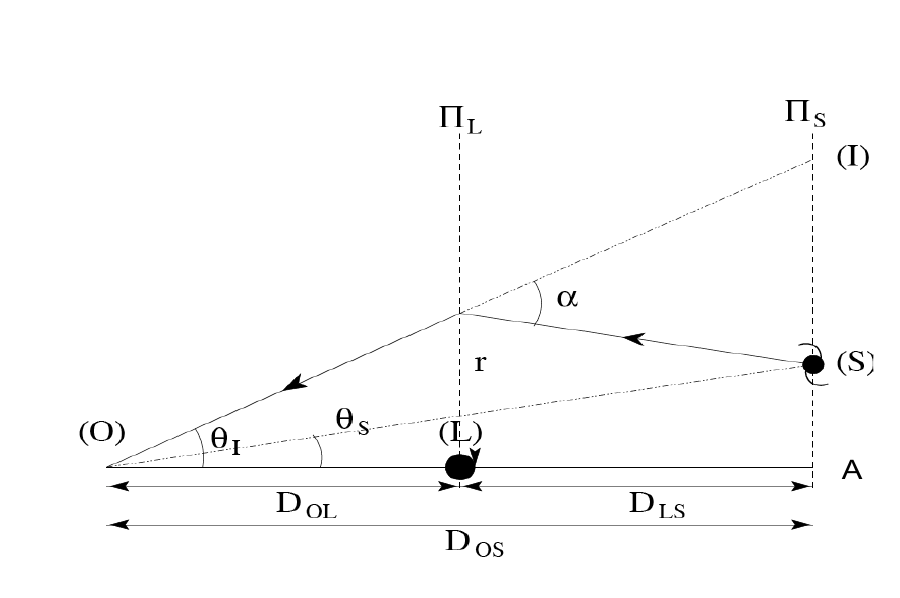}
\end{center}
\vskip -.75cm
\caption{Configuration of a gravitational lens: $S$ is the astrophysical source,
$O$ is the observer, $L$ is the (point) lens and $I$ is the image. The source 
lies at an angle $\theta_S$ from the axis $OL$ but is seen at an angle 
$\theta_I$: the deviation angle is thus $\alpha$.}
\label{fig1obs-2}
\end{figure}

A powerful tool for mapping dark matter is gravitational lensing, which is 
based on the deflection 
of light by matter: the light of a distant galaxy is deflected by an 
accumulation of matter present on the line of sight, just as it is by a lens
(see Fig.~\ref{fig1obs-2}). The deviation of light rays depends on the ratio 
of distances between observer, lens and source, as well as on the mass of the 
deflector. 

More precisely, the ``lens equation'' may be with written as (see Fig. 
\ref{DM1} for notations):
\begin{equation}
\label{1obs-15}
{\bf \theta}_I = {\bf \theta}_S + {D_{LS} \over D_{OS}} {\bf \alpha} \ .
\end{equation}
It is well-known that the deflection angle of a light ray passing an object of 
mass $M$ with an impact parameter $b$ is $\alpha = 4 G_{_N}M/(bc^2)$ ({\em 
see for example~\cite{Schutz} p.286}). Hence the lensing effect allows to detect
the mass distribution. 

Massive clusters may induce multiple images of background galaxies. One then 
talks of strong lensing~\cite{WCOT95}. The shape of a single galaxy can also be 
deformed into an arclet through lensing. In the case of weak lensing, 
the effect is measured through the deformation of the shape of galaxies 
but, because galaxies do not have a 
circular shape, it can only be measured statistically: galaxies tend 
through lensing to have aligned shapes~\cite{Me99}. 

\begin{figure}[h]
\begin{center}
\includegraphics[scale=0.45]{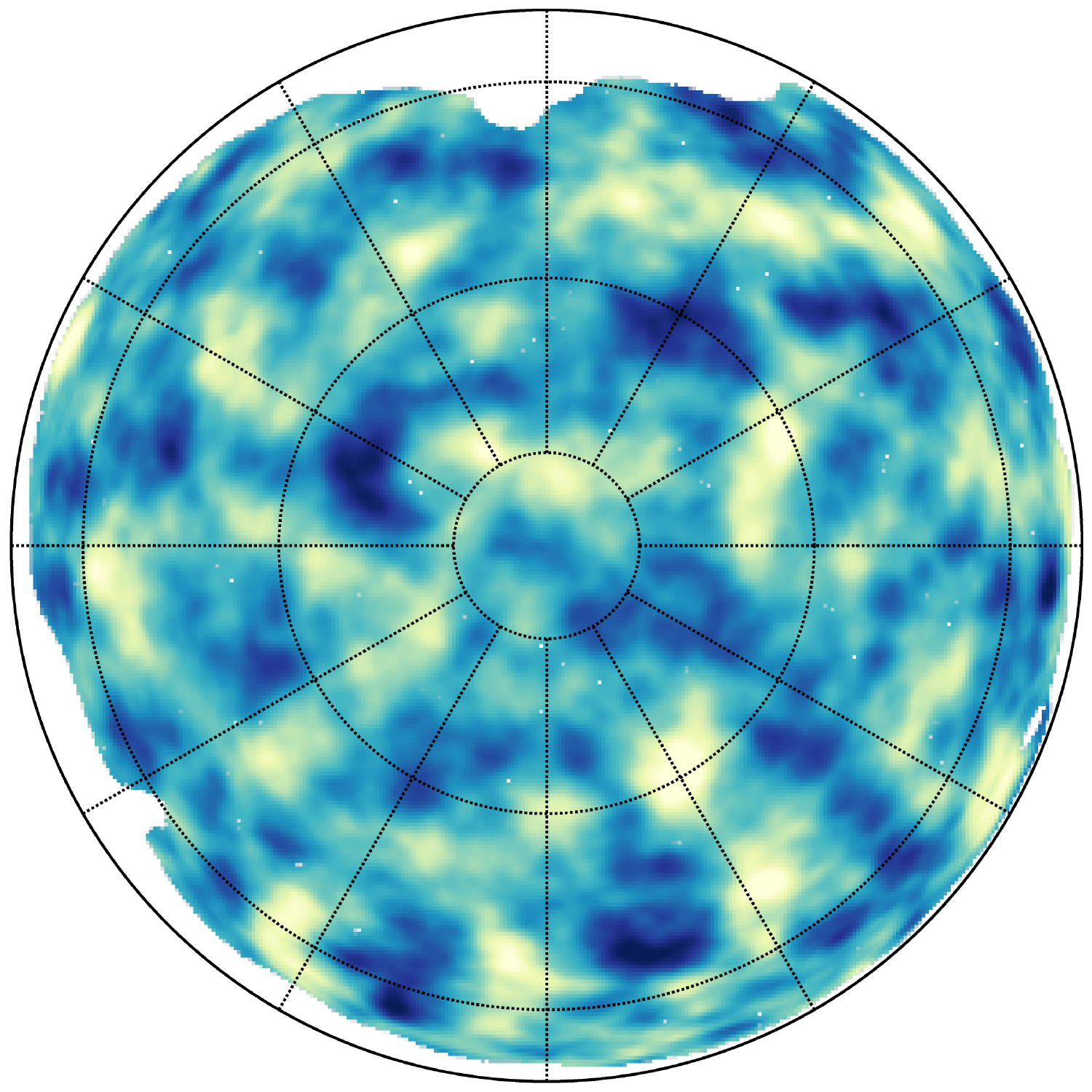}
\end{center}
\caption{Large scale structures detected by Planck through their lensing 
potential (galactic North)~\cite{Planck13_17}}
\label{DM2}
\end{figure}

Finally, we have seen in Section~\ref{4s1} that cosmological data shows that 
dark matter is also needed at the largest scales. An illustration of this is the
map of dark matter that the Planck collaboration could draw using the 
gravitational lensing of the CMB light (see Fig.~\ref{DM2}).
 
Since the only proof of existence of dark matter is gravitational (rotation 
curves, lensing,...), one may wonder whether the observed phenomena are due
to a modification of gravity, which would then be different from what general 
relativity predicts at the corresponding scales. Besides the difficulty 
of finding a theory that encompasses all the successes of general relativity, 
the problem is to modify gravity at all the scales where we see signs of dark
matter, that is galaxies, clusters of galaxies and cosmological scales.For 
example, the MOND theory~\cite{Mi83a,Mi83b} has been proposed to explain the 
rotation curves of galaxies but is only Newtonian  and requires to be 
generalized~\cite{Be04a} in order to be valid at the scale of the Universe.

\begin{figure}[h]
\begin{center}
\includegraphics[scale=0.45]{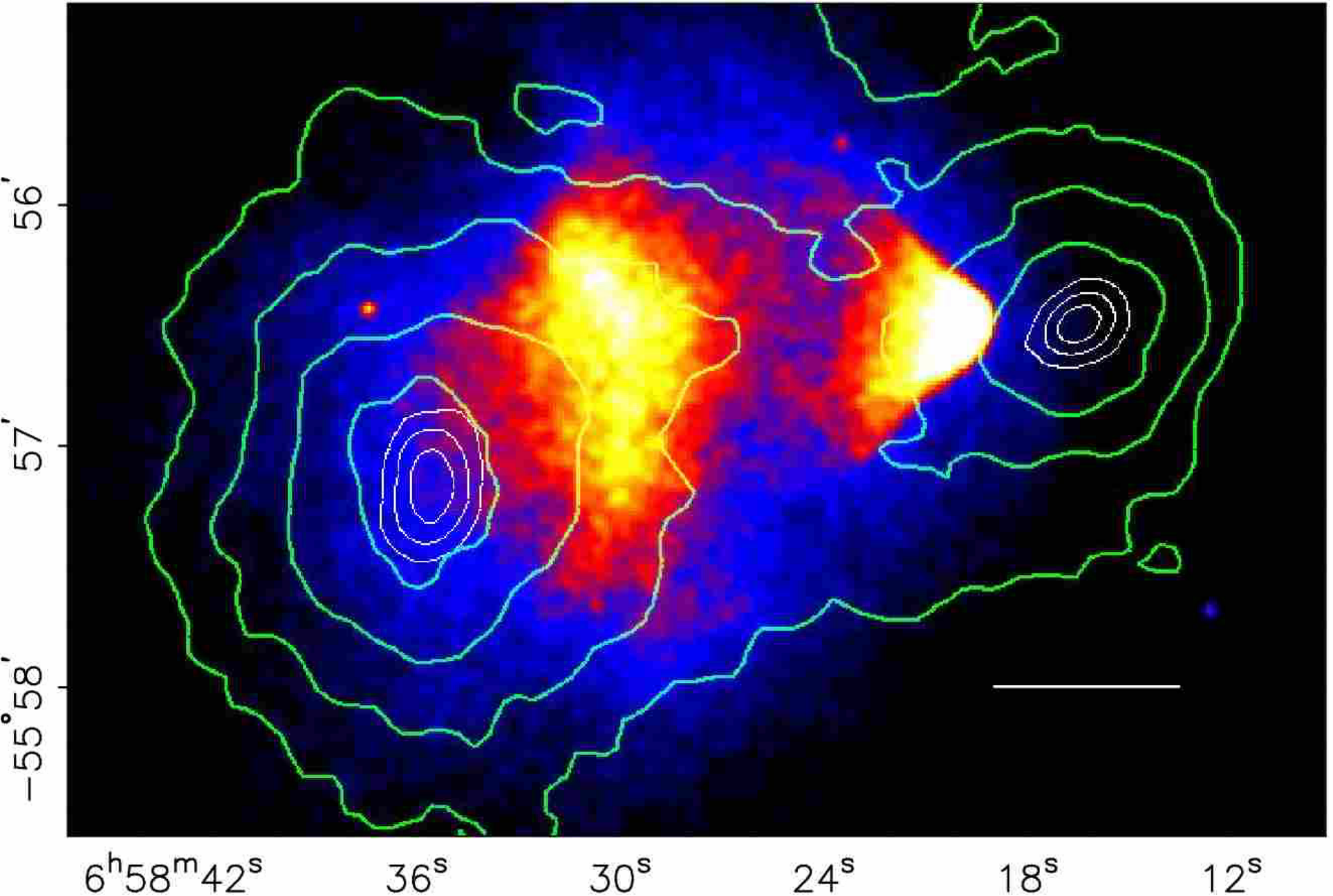}
\end{center}
\caption{X-ray image of the merging cluster 1E0657-558 obtained by the 
satellite Chandra, over which is superimposed 
in green contours the weak lensing 
reconstruction of the dark matter halos~\cite{Cl06}.}
\label{DM3}
\end{figure}

The existence of the bullet cluster (see Fig.~\ref{DM3}) where two galaxies 
collide has been presented as a support of dark matter~\cite{Cl06} because the 
luminous parts of the galaxies
are displaced with respect to their halos:  because dark matter is weakly 
coupled, the halos (detected through gravitational lensing) continue their way 
during the collision whereas their luminous matter counterparts (detected 
through their X-ray emission) are deformed. 

\subsection{Dark matter particles} \label{4s2b}

What is dark matter? Since it is nonluminous it is not electrically charged, 
and the only possible candidate within the Standard Model is the neutrino. 
But the random motion of neutrinos (the technical term is ``free streaming'') 
would wash out any density fluctuation and prevent the formation of galaxies;
one expresses this by saying that neutrinos are hot dark matter. Instead, we 
need cold dark matter i.e. particles with smaller free streaming length.

Moreover, we need dark matter particles in sufficient quantity. This means that 
they cannot be in thermal equilibrium today: they must have decoupled from the
thermal history of the Universe at some early time. Typically, there are two 
competing effects to modify the abundance of a species $X$: 
$X \bar X$ annihilation and
expansion of the Universe. Indeed, the faster is the dilution associated
with the expansion, the least effective is the annihilation because the
particles recede from one another. When the temperature drops below the mass 
$m_{_X}$, the annihilation rate 
becomes smaller than the expansion rate and there is a 
freezing of the number of particles in a covolume. In more quantitative term, 
this reads for the freezing temperature $T_f$:
\begin{equation}
\label{freeze}
n_{_X} (T_f) <\sigma_{{\rm ann}} v> \sim H(T_f) \ ,
\end{equation}
where $<\sigma_{{\rm ann}}v>$ is the thermal average of the $X \bar X$ 
annihilation cross-section times the relative velocity of the two particles
annihilating. One finds for the 
present density (in units of $\rho_c$ as usual) ({\em see Ref.~\cite{Binetruy}
Section 5.5})
\begin{equation}
\label{5.DM5}
\Omega_{_X} h_0^2 \sim {1.07 \times 10^{9} \ {\rm GeV}^{-1} \over g_*^{1/2}  
M_{_P}} \ {x_f \over <\sigma_{{\rm ann}} v>}
\quad . 
\end{equation}
where  $x_f \equiv m_{_X}/(kT_f \sim 20$ and $g_*$ is the total number of 
relativistic degrees of freedom present in the universe at the time of 
decoupling.

We note that the smaller the annihilation cross section 
is, the larger is the relic density.
We find $\Omega_X \sim  (100 \ \hbox{TeV})^{-2} 
(~<\sigma_{{\rm ann}} v~>~)^{-1} \sim 0.1 \ \hbox{pb}
/<\sigma_{{\rm ann}} v >$ (in units where ${/ \hskip - 2 truemm h} = c = 1$,
$1$ pb $= 2.5 \times 10^{-9}$ GeV$^{-2}$). This should be compared with the
latest result $\Omega_{DM} = 0.1187 \pm 0.0017$ coming from Planck 
\cite{Planck13_16}.

Thus $\Omega_{_X}$ will be of the right order of magnitude
if $<\sigma_{{\rm ann}} v>$ is of the order of a picobarn, which is a 
typical order of magnitude for an electroweak process. 
Also writing dimensionally
\begin{equation}
\label{5.DM6}
<\sigma_{{\rm ann}} v > \ \sim {\alpha^2 \over m_X^2} \ ,
\end{equation}
where $\alpha$ is a generic coupling strength, we find that 
$\Omega_{_X}$
is of order 1 for a mass $m_X \sim \alpha \times 1000 \ \hbox{TeV}$, i.e. in 
the TeV
range. This is why one is searching for a weakly interacting massive particle
(or wimp). 

There is also the possibility that dark matter particles have produced 
non-thermally, e.g. from the decay of heavy particles.

A puzzle which is not addressed by the wimp scenario is why dark matter and 
baryonic matter densities are basically of the same order: 
\begin{equation}
\label{darktobaryon}
{\rho_{DM} \over \rho_B} \sim 5 \ .
\end{equation} 
Indeed, baryon density arises from baryogenesis (see section
\ref{sect:1-5}) and thus results from a small mismatch between baryons and 
antibaryons, as seen from (\ref{asymmetry}). On the other hand, the wimp 
density results from the freezing regime described by (\ref{freeze}). There is 
no reason that the two scenarios lead to similar energy densities, as in 
(\ref{darktobaryon}). This puzzle is addressed by the Asymmetric Dark Matter 
scenarios (see the review by C. Zurek~\cite{Zu13} and references therein). 
The idea is that dark matter has an asymmetry in the number density of matter 
over anti-matter\footnote{This obviously precludes models where the dark matter 
particle is its own antiparticle, as often the case for wimps (e.g. the 
neutralino).} similar to the one for baryons:
\begin{equation} 
\label{ADM}
n_X - n_{\bar X} \sim n_b - n_{\bar b} \ .
\end{equation}
The abundance is therefore approximately one part in $10^{10}$ in comparison 
with the thermal abundance (see (\ref{asymmetry})). Eq. (\ref{ADM}) suggests 
that $m_X$ is typically $5$ times the proton mass, as a typical baryon mass. 
Thus generic Asymmetric Dark Matter models tend to favor light dark matter 
particles.

\begin{figure}
\begin{center}
\includegraphics[width=\linewidth]{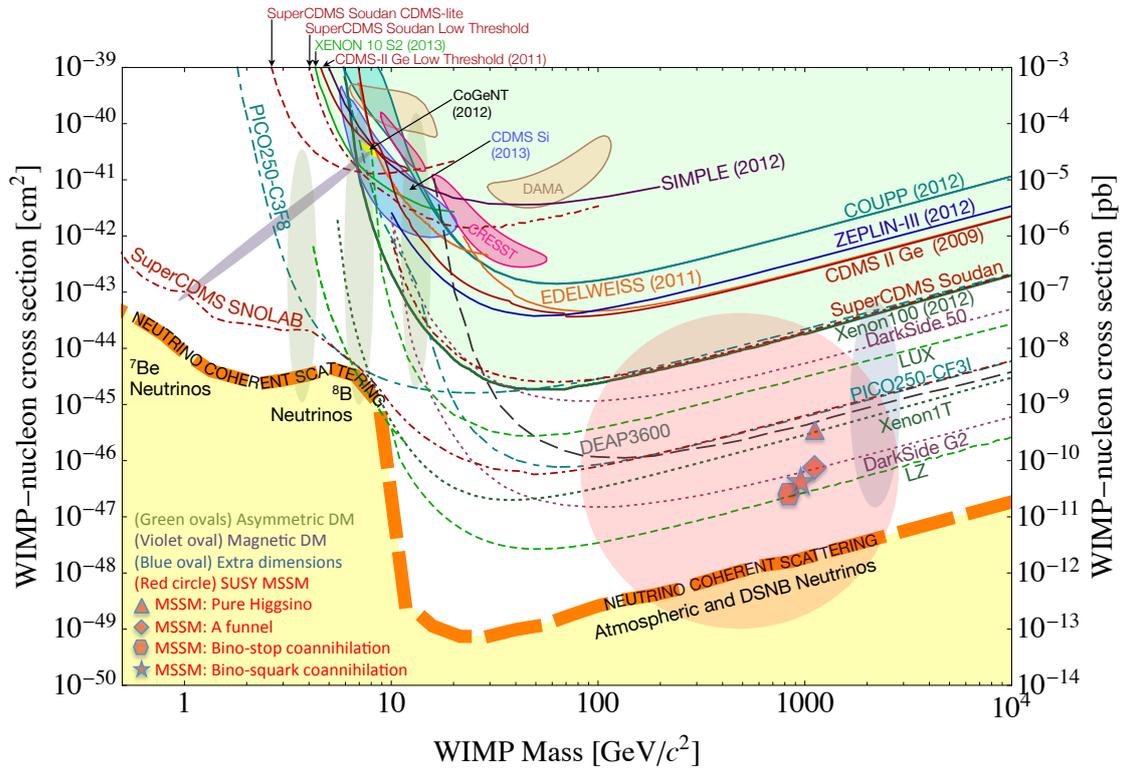}
\vskip -1.5cm
\caption{Sensitivities of some running and planned direct detection dark matter 
experiments to the spin-independent elastic scattering cross-section. Full 
curves correspond to limits from existing experiments, dashed curves to 
predicted sensitivities of future experiments. The full brown, pink, blue 
and yellow regions correspond respectively to the regions allowed by 
potential signals observed by the DAMA, CREST, CDMS and CoGeNT experiments. 
The light red disk corresponds to the region favoured by supersymmetric 
models. The thick yellow line corresponds to the irreducible neutrino 
background.}
\label{DM5}
\end{center}
\end{figure}

One may search for dark matter particles through direct detection using their 
elastic collisions with nuclei $XN \rightarrow XN$ in ultra-low background 
detectors. The energy of the recoiling nucleus is typically from a few keV to 
tens of keV. The recoil rate after integration over 
the dark matter velocity $v$ distribution is
\begin{equation}
\label{intrate}
R \sim 3.5\times 10^{-2} {\hbox{events} \over \hbox{kg.day}} {100 \over A}
\left[{100 \hbox{GeV} 
\over m_{_X}}
\times{\sigma_{XN} \over 1 \hbox{pb}}\times{\langle v \rangle \over 
220 \hbox{km.s}^{-1}}\times{\rho\over 0.3 \hbox{GeV.cm}^{-3}}\right] \ ,
\end{equation}
where $A$ is the atomic mass of the recoil nucleus, $\sigma_{XN}$ the 
cross-section 
for dark matter particle-nucleus elastic scattering and $\rho$ the local 
density of dark matter in our Galaxy. In the case of a neutralino wimp, 
$\sigma_{XN}$ can be as low as $10^{-12}$ pb, 
which yields a rate of $10^{-8}$ events/ton.year! Present, and future, 
experimental limits are shown on Fig.~\ref{DM5}. One should note that there
is below $10^{-12}$ pb ($10^{-8}$ pb for low-mass particles) 
an irreducible neutrino background corresponding to the reaction $\nu N
\rightarrow \nu N$. Experiments of the next decade should be able to reach 
this limit.

Dark matter particles may also be searched for through their annihilation 
products in massive celestal objects. This is known as indirect detection.
Indeed, because they are massive, they tend to accumulate in gravitational 
potentials, such as the centre of the Sun, or the centre of our Galaxy. They
annihilate there into pairs of energetic particles, the energy of which is 
directly connected with the mass $m_X$ of the dark matter particles. One
may therefore search for excess of energetic particles (positrons, photons, 
neutrinos) in the direction of galactic centres such as in our own Milky Way.
An excess of energetic positrons (energy of a few tens to few hundred GeV)
has actually been observed by the PAMELA, Fermi and AMS-02 experiments.
It remains to be seen if this arises from the annihilation of dark matter or 
from 
astrophysical sources, such as pulsars: we have seen at the end of Section 
\ref{chap:3} that a certain number of astrophysical sources produce energetic 
particles. This is indeed a limitation, at least at present, of the indirect 
detection of dark matter: it needs to be complemented by either direct 
detection or detection at colliders. 

\subsection{WIMPs and physics beyond the Standard Model}
\label{4s2c}

{\em We would like to 
stress in this section that the presence of a WIMP in a theory is deeply 
connected with the naturalness of the electroweak scale.  

Let us start by recalling what is the naturalness problem (see for example 
\cite{Binetruy}). As is well-known, the Higgs squared mass $m_h^2$ receives 
quadratically divergent corrections. In the context of an effective theory 
valid up to a cut-off scale
$\Lambda$ where a more fundamental theory takes over, $\Lambda$ is the mass of
the heavy degrees of freedom of the fundamental theory. Their contribution in 
loops, quadratic in their mass, destabilizes the Higgs mass 
and thus the electroweak scale ($m_h^2 \sim \lambda v^2$ where $\lambda$ is the
scalar self-coupling and $v \sim 1/(G_{_F}\sqrt{2})^{1/2} \sim 250 \
\hbox{GeV}$ is the Higgs vacuum expectation value. 
More precisely, we have at one loop
\begin{equation}
\label{4-101}
\delta m_h^2 = {3 m_t^2 \over 2 \pi^2 v^2} \Lambda_t^2 - {6 M_{_W}^2 + 
3 M_{_Z}^2 \over 8 \pi^2 v^2} \Lambda_g^2 - {3 m_h^2 \over 8 \pi^2 v^2} 
\Lambda_h^2 \ ,
\end{equation}
where for completeness we have assumed different cut-offs for the top loops
($\Lambda_t$), the gauge loops ($\Lambda_g$)  and the scalar loops 
($\Lambda_h$)~\cite{BHR06}. The naturalness condition states that the order of 
magnitude of the Higgs mass is not destabilized by the radiative corrections
i.e. $\left| \delta m_h^2 \right| < m_h^2$. This translates into the 
conditions: 
\begin{eqnarray}
\Lambda_t &\sim& \sqrt{{2 \over 3}} \  {\pi v \over m_t} m_h \sim 3.5 m_h \ , 
\label{4-102} \\
\Lambda_g &\sim& {2 \sqrt{2} \pi v \over \sqrt{6 M_{_W}^2 + 3 M_{_Z}^2}} m_h
\sim 9 m_h  \ , \label{4-103} \\
\Lambda_h &\sim& {2 \sqrt{2} \pi v \over \sqrt{3}}  \sim 1.3 \ \hbox{TeV}
\ .  \label{4-104}
\end{eqnarray}
Thus one should introduce new physics at a scale $\Lambda_t \sim 3.5 m_h$. 
We will illustrate our argument with two 
examples: supersymmetry and  extra dimensions. In the two 
cases, one introduces new physics at the scale $\Lambda_t$ (supersymmetric 
particles or Kaluza-Klein modes). 

Typically, these models require the presence of a symmetry that prevents 
direct coupling between the Standard Model (SM) fermions and the new fields 
that one has introduced: otherwise, such couplings introduce new mixing 
patterns incompatible with what is observed in flavor mixings 
(compatible with the Standard Model). This symmetry is 
usually a parity (i.e. a discrete symmetry) which is the low energy remnant 
of a continuous symmetry which operates at the level of the underlying 
fundamental theory: SM fermions are even under this parity whereas the new 
fields are odd.  Among these new fields, the lightest odd-parity particle
(we will refer to it as the LOP) is stable: it cannot decay into SM fermions 
because of the parity; it cannot decay into the new fields because it is the 
lightest. It is massive and weakly interacting. It thus provides an
adequate candidate for a WIMP.

Let us take our examples in turn. In the case of supersymmetry, the parity
operation is R-parity (which usually proceeds from a continuous 
R-symmetry broken by gaugino masses i.e. supersymmetry breaking). And the 
LOP is the Lightest Supersymmetric Particle, the famous LSP, the lightest 
neutralino in the simplest models. 

In the case of extra dimensions, say a $5$-dimensional model, the local 
symmetry is $5$-dimensional Lorentz invariance. It ensures conservation of 
the Kaluza-Klein levels: if $A^{(n)}$ is the $n$th Kaluza-Klein mode of 
the massless $5$-dimensional field $A$ (in other words, the $4$-dimensional 
field with mass $m = n/R$, where $R$ is the radius of the $5$th dimension), 
then in the reaction $A^{(n)}+B^{(p)} \rightarrow C^{(q)}+D^{(r)}$, we have
$n+p=q+r$. At energies smaller than $R^{-1}$, this turns into a 
Kaluza-Klein parity $(-1)^n$. The LOP is then the lightest Kaluza-Klein
mode, usually $B^{(1)}$, the first mode of the $U(1)_Y$ gauge boson
\cite{ST02}.  

In realistic models, there is often the possibility that other odd-parity 
fields are almost degenerate in mass with the LOP. This leads to the
possibility of co-annihilations, that is annihilations of the LOP against
these almost degenerate fields, and to a modification of the relic 
density in the corresponding region of parameter space. }

\begin{figure}
\begin{center}
\includegraphics[scale=0.5]{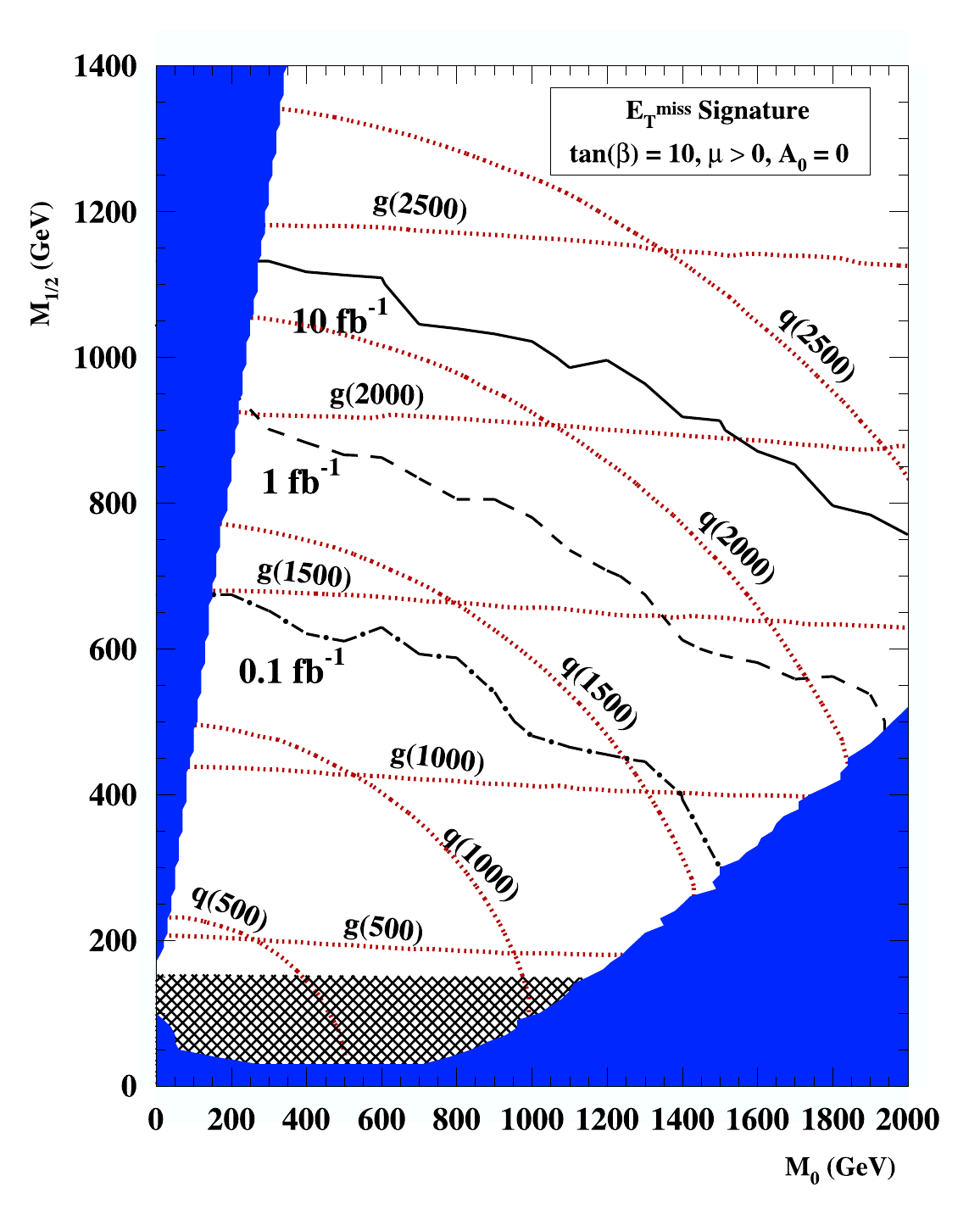}
\includegraphics[scale=0.5]{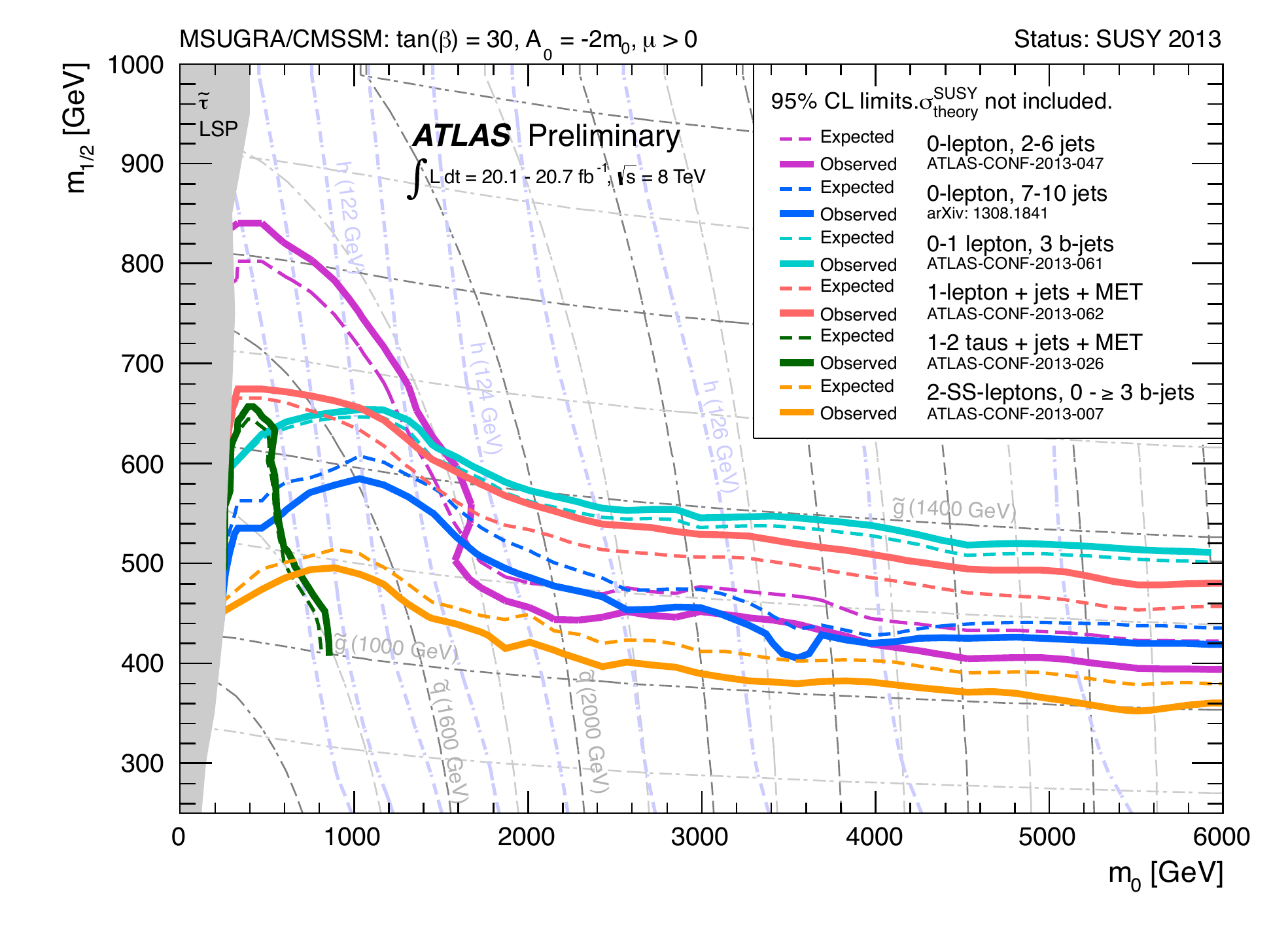}
\caption{Upper panel: Contours in a parameter space of supersymmetry models 
in the plot ($m_0$, $m_{1/2}$) for the 
discovery of the missing energy plus jets signature of new physics    
by the ATLAS experiment at the LHC.    
The three sets of contours correspond to levels of integrated      
luminosity at the LHC (in fb$^{-1}$), contours of constant squark      
mass, and contours of constant gluino mass~\cite{To02}.
\newline
Lower panel: Exclusion limits at $95\%$ CL for $8$ TeV, $20$ fb$^{-1}$ 
integrated luminosity analyses in the 
($m_0$, $m_{1/2}$) plane for the MSUGRA/CMSSM model with the remaining 
parameters set to $\tan\beta = 30, A_0 = -2m_0, \mu > 0$. Part of the model 
plane accommodates a lightest neutral scalar Higgs boson mass of 125 GeV
\cite{ATLAS_SUSY13}.}
\label{fig4-7a}
\end{center}
\end{figure}

Searches at LHC are based on the missing energy signal corresponding to the 
LSP (see Fig.~\ref{fig4-7a}). Since LSP are produced in pairs, 
they are difficult to reconstruct in all generality. But, in the case of a 
specific model, one may be able to reconstruct the mass of the LSP as well as 
the relic density. 

\subsection{Other candidates for dark matter}
\label{4s2d}
Among the many other candidates proposed for dark matter, one may single out the axion 
field since it is introduced to solve one puzzle of the Standard Model known as the 
strong CP problem, which remains to be solved. Moreover, from the point of view of 
cosmology,
this is an interesting illustration of a low-mass and wery weakly interacting particle, 
as we will encounter in the next Chapter.

As we have already seen, CP symmetry is violated in weak interactions.  But the 
symmetries of the Standard Model allow as well a CP-violating term in the QCD Lagrangian
$(\theta g^2/32 \pi^2) G^a_{\mu\nu} \tilde G^{a\mu\nu}$, where $G^a_{\mu\nu}$ is the gluon 
field 
strength and $g$ the QCD gauge coupling. For non-zero quark masses, this term leads to 
unobserved CP-violating effects in the strong sector\footnote{{\em To be more precise, 
it is $\bar \theta \equiv \theta - \hbox{arg det} \ m_q$, where $m_q$ is the quark mass 
matrix, which is observable. If $\bar \theta \not = 0$, then strong interactions 
violate P and CP. This is 
not compatible with the experimental upper bound on the neutron electric dipole moment 
unless $\left| \bar \theta \right| < 10^{-10}$.}}.

The most common way to solve the puzzle is to introduce a scalar field $a(x)$ called 
axion, with Lagrangian
\begin{equation}
\label{ax1}
{\cal L}_a = {1 \over 2} \partial^\mu a \partial_\mu a + {g^2 \over 32 \pi^2}
{a(x) \over f_a} G^a_{\mu\nu} \tilde G^{a\mu\nu} \ ,
\end{equation}
where $f_a$ is an energy scale, called the axion decay constant (by analogy with the 
pion decay constant). Strong interactions generate an effective potential for $a(x)$
whose minimum corresponds to no violation of CP\footnote{{\em i.e. a vanishing value of
$\bar \theta = a(x)/f_a -\hbox{arg det} \ m_q$}}. The non-renormalisable interaction 
$aG\cdot \tilde G$ is obtained as the low energy effect of the breaking of a $U(1)$ 
global symmetry, known as the Peccei-Quinn symmetry~\cite{PQ77a,PQ77b}, spontaneously 
broken at the scale $f_a$: the axion is the pseudo-Goldstone boson associated with 
this breaking~\cite{We78,Wi78}\footnote{{\em The axion is not a true Goldstone boson 
but a ``pseudo-Goldstone'' boson because the QCD vacuum which involves non-vanishing 
values for 
the quark condensates such as $\langle \bar u u \rangle$ and $\langle \bar d d \rangle$
indices a small explicit breaking of the $U(1)$ symmetry, hence a mass for the otherwise
massless Goldstone boson associated with the spontaneous breaking.}}.  

The axion mass is 
\begin{equation}
\label{ax2}
m_a \sim 6.10^{-6} \ \hbox{eV} \left({10^{12} \hbox{GeV}\over f_a}\right) \ . 
\end{equation}
Its coupling to ordinary matter is proportional to $1/f_a$ and can be calculated in 
specific models. It couples to leptons and to photons, the latter being of the form
\begin{equation}
\label{ax3}
{\cal L}_{a\gamma\gamma} =  -g_{a\gamma}{\alpha \over \pi} {a(x) \over f_a} \
{\bf E}.{\bf B} \ , 
\end{equation} 
where ${\bf E}$ and ${\bf B}$ are the electric and magnetic fields, $\alpha$ is the 
fine structure constant and $g_{a\gamma}$ is a model-dependent coefficient of order $1$.
Moreover, since $m_a \ll \Lambda_{QCD}$, the axion coupling to quarks should be described
through its coupling to hadrons, which occurs through small mixing with the $\pi_0$ and 
$\eta$ mesons. All of these interactions can play a role in searches for the axion, and 
allows the axion to be produced or detected in the laboratory and emitted by the sun or 
other stars. Its non-discovery leaves us with an axion window $10^{-6}$ eV 
$<m_a< 3.10^{-3}$ eV, or correspondingly, $2.10^9$ GeV $< f_a < 6.10^{12}$ GeV.

Let us describe briefly the cosmology of the axion field (see the review by P. Sikivie
\cite{Si08} for a more thorough treatment). The breaking of the $U(1)$ symmetry 
corresponds to a phase transition, known as the Peccei-Quinn phase transition,
at a temperature of order $f_a$. This phase transition is characterized by the formation 
of comic strings.

If the reheat temperature after inflation is smaller than $f_a$, then one starts the 
evolution in the reheated universe with an homogeneous axion field. When the temperature 
reaches the QCD scale, the effective potential turns on and the axion acquires a mass.
At a time $t_* \sim m_a^{-1}$ ($T_* \sim 1$ GeV), the axion starts to oscillate around 
its minimum. These oscillations do not dissipate into other forms of energy, and thus 
contribute to the cosmological energy density an amount
\begin{equation}
\label{ax4}
\Omega_a \left({h_0 \over 0.7}\right)^2 \sim 0.15 \left({f_a \over 10^{12} 
\hbox{GeV}}\right)^{7/6} \left({a(t_*) \over f_a} \right)^2 \ ,
\end{equation}
where $a(t_*)$ is the axion value at $t_*$, which measures the misalignment of the axion
with respect to its final minimum. Such a contribution is thus called vacuum realignment.

If the reheat temperature is larger than $f_a$, then the cosmic strings produced at 
the Peccei-Quinn phase transition become at time $t_*$ the boundaries of domain walls.
This leads to a potential domain wall problem (too much energy stored in the domain 
walls). There is a certain number of cases where this can be avoided. In this case
\cite{Si08}, 
\begin{equation}
\label{ax5}
\Omega_a \left({h_0 \over 0.7}\right)^2 \sim 0.7 \left({f_a \over 10^{12} 
\hbox{GeV}}\right)^{7/6}  \ .
\end{equation}

\section{Light does not say it all (3): dark energy} 
\label{chap:5}

We have seen in the introduction that the observation of the acceleration of 
the expansion of the Universe in 1998-1999 provided a way out of an 
increasingly uncomfortable tension between models and observation. On one hand,
the observed (luminous and dark) matter could account for only a fraction of 
the critical  density $\rho_c = 10^{-26}$ kg/m${}^3$ (other forms of energy being 
negligible at present time). On the other hand, the standard theory of 
inflation erased any curvature and naturally led to a spatially flat universe 
for which $\rho = \rho_c$. The discovery of the acceleration of the expansion
led to introduce a new component, named dark energy, of a type unknown so far 
since all known 
forms of energy (non-relativistic matter, radiation) decelerate the expansion.

In this last chapter, after briefly reviewing the observational case, we will 
illustrate dark energy models with the example of quintessence, identify
some of the problems posed by dark energy models, and the fundamental questions 
associated with the acceleration of the expansion of the Universe.

\subsection{Acceleration of the expansion of the Universe: supernovae of type 
Ia as standard candle}
\label{sect:1obs-1}

The approach that has made the first case for the acceleration of the 
expansion of the Universe  uses supernovae of type Ia as standard
candles (see Section~\ref{sect:3-3}). Two groups, the 
Supernova Cosmology
Project~\cite{SCP99} and the High-$z$ Supernova Search~\cite{HZS98a} 
have found that distant supernovae appear to be fainter than expected in a
flat matter-dominated Universe. If this is to have
a cosmological origin, this means that, at fixed redshift, they are at
larger distances than expected in such a context 
and thus that the Universe expansion is accelerating. 

\begin{figure}[h]
\begin{center}
\includegraphics[scale=0.65]{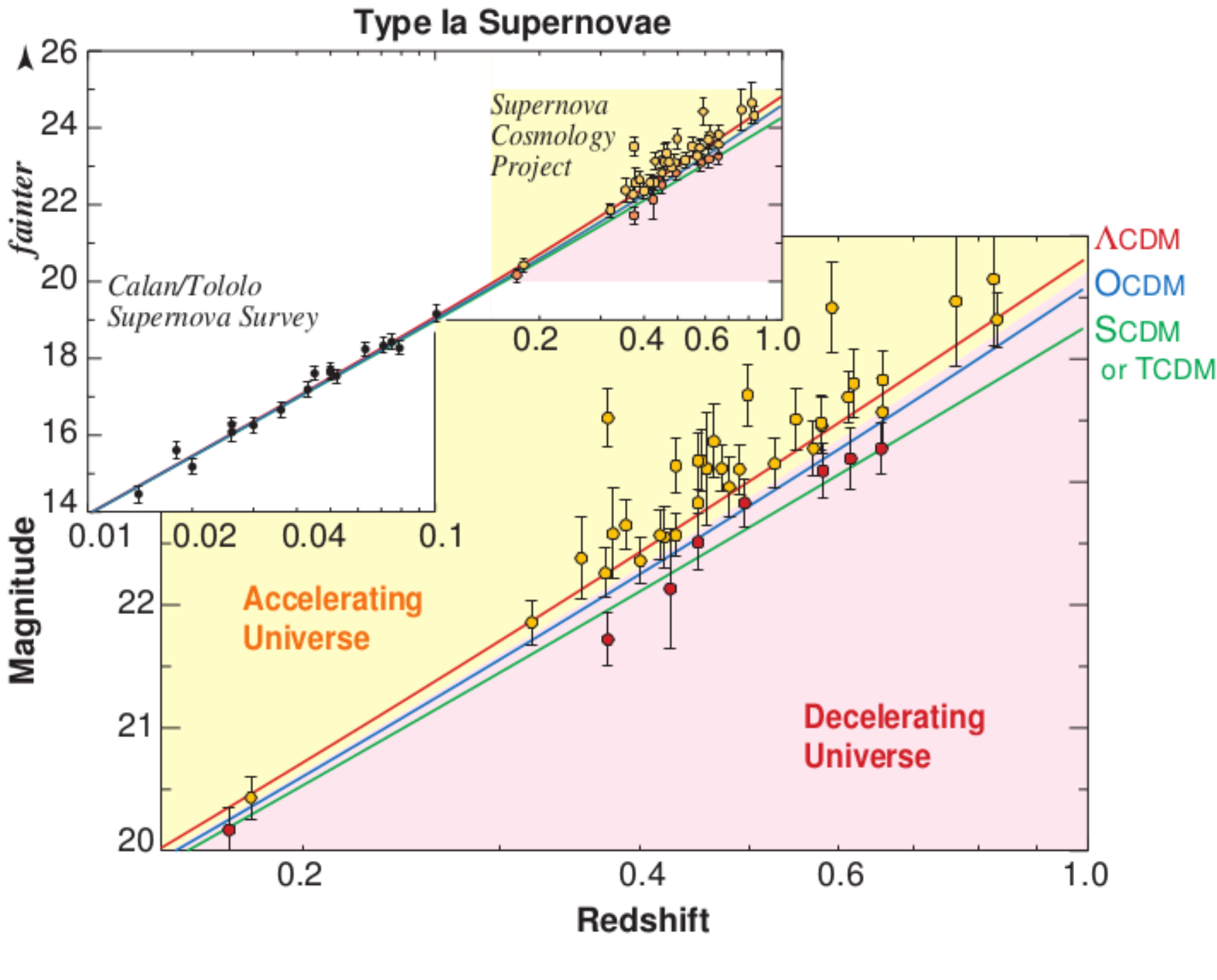}
\end{center}
\caption{Hubble plot (magnitude versus redshift) for Type Ia supernovae 
observed at low redshift by the Calan-Tololo Supernova Survey and at moderate 
redshift by the Supernova Cosmology Project.}
\label{fig4-1}
\end{figure}

More precisely, one uses the relation (\ref{4-34}) between the flux $\phi$ 
received on Earth and the luminosity $L$ of the supernova.
Traditionally, flux and luminosity are expressed  on a log scale as 
apparent magnitude $m_B$  and absolute magnitude $M$ (magnitude is 
$-2.5 \ \log_{10}$ luminosity + constant). The relation then reads 
\begin{equation} 
\label{4-39}
m_B = 5\log (H_0d_L) + M - 5 \log H_0 + 25. 
\end{equation} 
The last terms are $z$-independent,  if one assumes that supernovae
of type Ia are standard candles; they are then measured by using low $z$
supernovae. The first term, which involves the luminosity distance $d_L$,  
varies logarithmically with $z$ up to corrections which depend on the geometry,
more precisely on $q_0$ for small $z$ as can be seen from (\ref{4-37}).
This allows to compare with data  cosmological models with different components
participating to the energy budget, as can be seen from   Fig.~\ref{fig4-1}. 

\begin{figure}[htb] 
\begin{center}
\includegraphics[scale=0.45]{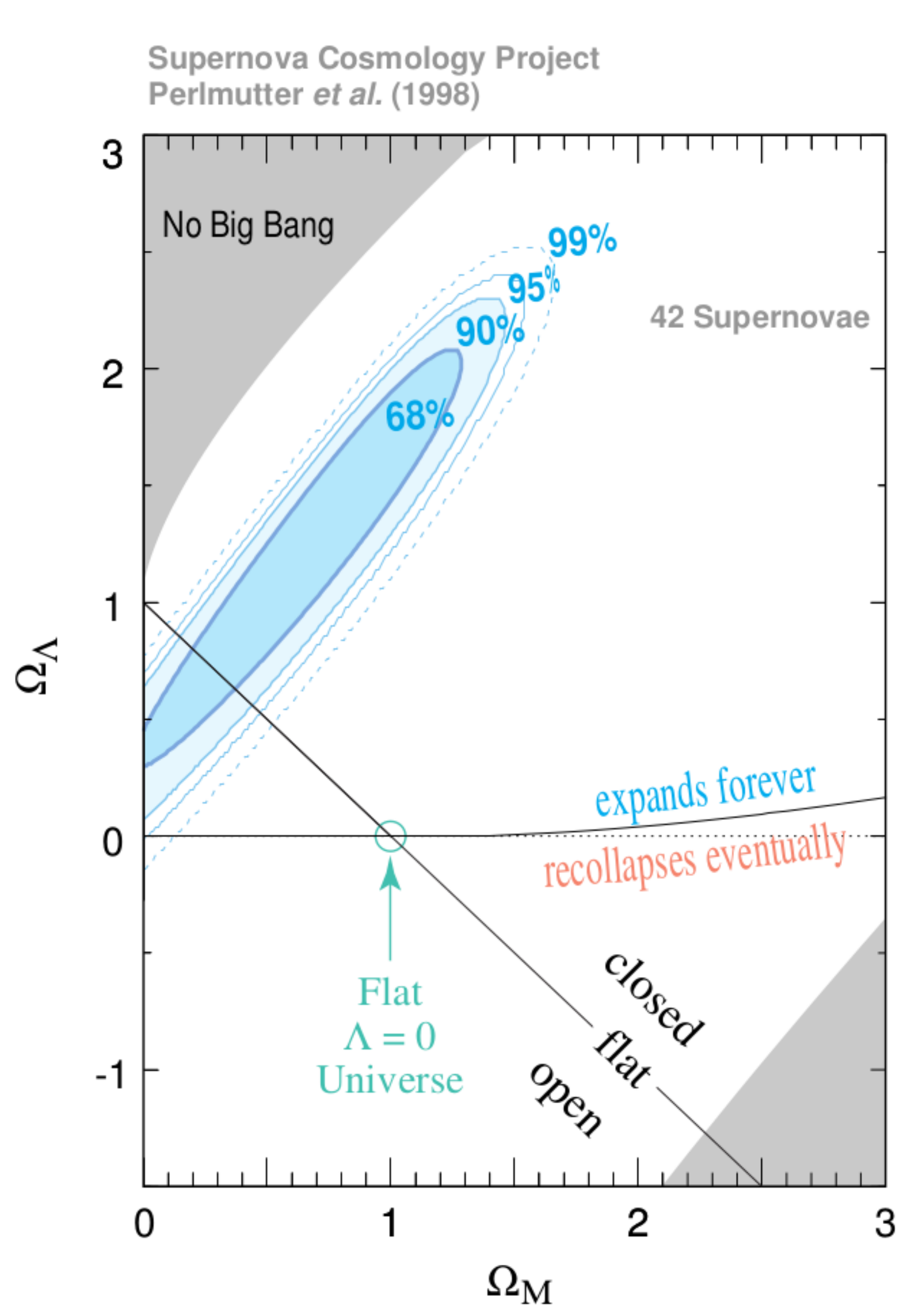}
\end{center}
\caption{ Best-fit coincidence regions in the $\Omega_M-\Omega_\Lambda$ 
plane, based on the analysis of 42 type Ia supernovae discovered by the 
Supernova Cosmology Project~\cite{SCP99}. } 
\label{fig4-2} 
\end{figure}

In the case of a model with matter and cosmological constant as dominant 
components, $q_0 = \Omega_{_M}/2 - \Omega_\Lambda$ and the measurement can be 
turned into a limit in the $\Omega_M-\Omega_\Lambda$ plane see Fig. 
\ref{fig4-2}).

Let us note that this combination $\Omega_{_M}/2 - \Omega_\Lambda$
is `orthogonal' to the combination $1-\Omega_k = \Omega_M + \Omega_\Lambda$ 
measured in CMB experiments. 
The two measurements are therefore complementary: this is
sometimes referred to as `cosmic complementarity'.

An important question raised by the analysis above is whether supernovae are 
truly standard candles. Otherwise, the observation could be interpreted 
as an history effect: for some reasons, older supernovae would be dimmer.
Indeed, strictly speaking, supernovae of type Ia are not standard candles:
dimmer supernovae 
are quicker  ({\em see Section~\ref{sect:3-3}}). In practice, one thus has to correct 
the light curves using a 
phenomenological stretch factor. It is thus more precise to state that 
supernovae of type Ia are standardizable candles.
Moreover, the type of measurement discussed above is sensitive to many possible
systematic effects (evolution besides the light-curve timescale
correction, presence of dust, etc.), and this  has fuelled a healthy debate 
on the significance of  supernova data as well as a thorough study of possible
systematic effects by the observational groups concerned.

\subsection{Cosmological constant and vacuum energy} \label{sect:5-2}

An obvious solution to the acceleration of the expansion is the introduction of 
a cosmological constant (see (\ref{4-8}) remembering that $q_0$ is a {\em 
deceleration} parameter).  But there is a severe conceptual problem associated 
with the cosmological constant, which we now describe.

Considering (\ref{2-8}) in flat space at present time implies the 
general following constraint on $\lambda$: 
\begin{equation}
\label{7-3}
|\lambda| \le H_0^2 \ .
\end{equation}
In other words, the length scale $\ell_{\Lambda} \equiv
|\lambda|^{-1/2}$ associated with the cosmological constant must be larger than 
the  Hubble length $\ell_{H_0} \equiv c H_0^{-1}=h_0^{-1}. 10^{26}$ 
m, and thus be a cosmological distance.

This is not a problem as long as one remains classical:
$\ell_{H_0}$ provides a natural cosmological scale for our present
Universe. The problem arises when one tries to combine gravity with
the quantum theory. Indeed, from Newton's constant {\em and} the
Planck constant $\hbar$, we can construct  the (reduced) Planck mass scale 
\begin{equation}
\label{7-4m}
m_{_P} = \sqrt{\hbar c / (8 \pi G_{_N})}=2.4 \times 10^{18}\; {\rm GeV/c}^2 \ .
\end{equation} 
The corresponding length scale is the Planck length  
\begin{equation}
\label{7-4}
\ell_{_P} = {\hbar \over m_{_P} c} = 8.1 \times 10^{-35} \; {\rm m} \ .
\end{equation}
The above constraint now reads:
\begin{equation}
\label{7-5}
\ell_{\Lambda} \equiv |\lambda|^{-1/2} \ge \ell_{H_0} = {c \over H_0} \sim
10^{60} \ \ell_{_P} \ . 
\end{equation}
In other words, there are more than sixty orders of magnitude
between the scale associated with the cosmological constant and
the scale of quantum gravity.

A rather obvious solution is to take $\lambda = 0$. This is as valid
a choice as any other in a pure gravity theory. Unfortunately, it is
an unnatural one when one introduces any kind of matter. Indeed,
set $\lambda$ to zero but assume that there is a nonvanishing vacuum
({\em i.e.} ground state) energy: $<T_{\mu\nu}> = \rho_{{\rm vac}}
g_{\mu\nu}$; then the Einstein equations (\ref{2-Einstein}) read
\begin{equation}
\label{7-6}
R_{\mu \nu} - {1 \over 2} g_{\mu \nu} R =  8 \pi G_{_N} T_{\mu \nu}
+ 8 \pi G_{_N} \rho_{{\rm vac}} g_{\mu \nu}\ .
\end{equation}
As first noted by Zel'dovich~\cite{Ze68}, the last term is interpreted as an 
effective cosmological constant (from now on, we set $\hbar = c = 1$):
\begin{equation}
\label{7-7}
\lambda_{{\rm eff}} = 8 \pi G_{_N} \rho_{{\rm vac}} \equiv {\Lambda^4 
\over m_{_P}^2} \ .
\end{equation}
Generically, $\rho_{{\rm vac}}$ receives a non-zero contribution from symmetry
breaking: for instance, the scale $\Lambda$ would be typically of the
order of $100$ GeV in the case of the electroweak gauge symmetry breaking 
or  $1$ TeV in the case of supersymmetry breaking. Moreover, it is divergent
in the context of the (non-renormalizable) theory of gravity, which would 
thus favour a value as large as the Planck scale.
But the constraint (\ref{7-5}) now reads:
\begin{equation}
\label{7-8}
\Lambda \le 10^{-30} \; m_{_P} \sim 10^{-3} \ {\rm eV}.
\end{equation}
It is this very unnatural fine-tuning of parameters (in explicit
cases $\rho_{{\rm vac}}$ and thus $\Lambda$ are functions of the parameters of
the theory) that is referred to as the  cosmological constant problem, or
more accurately the {\em vacuum energy problem}.

If the acceleration observed is indeed due to the 
cosmological constant, its value is as large as the upper bounds obtained in 
the previous subsection allow:
\begin{equation}
\label{7-9a}
\lambda \sim H_0^2 \ , \quad \ell_\Lambda \sim \ell_{H_0} \ , \quad 
\Lambda \sim 10^{-3} \ \hbox{eV} \ .
\end{equation}
Regarding the latter scale $\Lambda$, which characterizes the vacuum energy
($ \rho_{{\rm vac}} \equiv \Lambda^4$), one may note the interesting numerical 
coincidence:
\begin{equation}
\label{7-9b}
{1 \over \Lambda} \sim \sqrt{\ell_{H_0} \ell_{_P}} \sim 10^{-4} \ {\rm m}
 \ .
\end{equation}
This relation underlines the fact that the vacuum energy problem involves some 
deep connection between the infrared regime (the infrared cut-off being 
$\ell_{H_0}$) and the ultraviolet regime (the ultraviolet cut-off being
$\ell_{_P}$), between the infinitely large and the infinitely small.

{\em As an illustration of what such a relation could tell us about some very 
fundamental aspects of physics (if it is not a mere numerical coicidence), 
we will follow the approach developed by T. Padmanabhan~\cite{Pa08}.
As we will discuss later, the value of the cosmological constant 
might be related to the way that the short distance theory reacts to long 
distance fluctuations. Let us consider a 3-dimensional domain of size $L$ 
(e.g. the horizon $\ell_{H_0}$). From the point of view of the quantum theory, 
it consists of $N=(L/\ell_{_P})^3$ elementary cells. For each individual cell,
a ``natural value'' for the energy stored is provided by the scale $m_{_P}$
characteristic of quantum gravity. This yields
\begin{equation}
\label{120}
\rho \sim { m_{_P} \over \ell_{_P}^3} \sim {1 \over \ell_{_P}^4}
\ ,
\end{equation}
which is some 120 orders of magnitude larger than observed.

Alternatively, if a mechanism, yet to be determined, cancels this bulk energy, 
vacuum energy may be produced by the energy fluctuations.   
The Poissonian fluctuation in energy is $\Delta \epsilon \sim 1/\ell_{_P}$, 
which corresponds to an energy for overall fluctuations $\Delta E^2 \sim 
N/\ell_{_P}^2$, or an energy density 
\begin{equation}
\label{volume}
\rho \sim {\sqrt{N} \over \ell_{_P} L^3} \sim {1 \over \ell_{_P}^{5/2} L^{3/2}}
\ ,
\end{equation}
which again does not reproduce (\ref{7-9b}) ($\rho_{{\rm vac}} = \Lambda^4$).

If we make the further assumption that the relevant degrees of freedom 
lie on the 
surface (as the degrees of freedom of a black hole lie on the horizon), then
$N=(L/\ell_{_P})^2$ and
\begin{equation}
\label{surface}
\rho \sim {\sqrt{N} \over \ell_{_P} L^3} \sim {1 \over \ell_{_P}^2 L^2}
\ ,
\end{equation}
which is fully consistent with (\ref{7-9b}). 
Is this telling us something on the quantization of spacetime, hence of 
gravity? We will return to this question below.}

\subsection{Supersymmetry} \label{sect:5-3}

The most natural reason why  vacuum energy would be vanishing is a symmetry 
argument. It turns out that, among the various spacetime symmetries available, 
global supersymmetry is the symmetry intimately connected with the vanishing 
of the vacuum energy. 

We recall that supersymmetry is a symmetry between bosons and fermions which 
plays an important r\^ole in high energy physics, mainly because, through a 
cancellation between the boson and the fermion fields, it controls severely 
the quantum fluctuations. Many believe that the Standard Model is the 
effective theory of a more fundamental theory valid at higher energies, with 
more boson and fermion fields. These fields should, to some level, make 
themselves known through the quantum energy fluctuations to which they 
participe. The fact that we find no trace of them seems to suggest that 
these fluctuations are tightly constrained, i.e. that the underlying theory is 
supersymmetric.

Among the various spacetime symmetries, supersymmetry is rather unique. Indeed,
in the same way as the generator of time and space translations is the
4-momentum operator $P^\mu \equiv (P^0=H, P^i)$, and the generator of 
spacetime rotations is the tensor $M^{\mu\nu}$, the only other generators 
carrying a Lorentz index are the generators of supersymmetry  $Q_r$, where $r$ 
is a spinor index. In fact, the combination of two supersymmetry 
transformations is merely a translation in spacetime. This is expressed by the 
following algebra:
\begin{equation}
\label{SUSY}
\{ Q_r , \bar{Q}_s \} = 2 \gamma^{\mu}_{rs} \ P_{\mu} \ ,
\end{equation}
where $\bar{Q} \equiv Q \gamma^0$ and $\gamma^{\mu}$ are the gamma matrices 
introduced by Dirac to write a relativistic fermion (electron) 
equation\footnote{{\em The anticommutator in (\ref{SUSY}) arises 
from the fact that the supersymmetry transformation parameter is an 
anticommuting spinor.}}. Since the generator of time translations $P_0$ is the 
Hamiltonian $H$, 
we may easily infer  from (\ref{SUSY}) an expression for the Hamiltonian 
of the system\footnote{{\em Indeed, (\ref{SUSY}) reads explicitly
\begin{equation}
\label{SUSY-1}
\{Q_r, Q_t \} \gamma^0_{ts} = 2 \gamma ^{\mu}_{rs} \ P_{\mu} \ .
\end{equation}
Contracting with $\gamma^0_{sr}$, one obtains
\begin{equation}
\label{SUSY-2}
\sum_{r,t} \ \{Q_r, Q_t \} \left[ ( \gamma^0)^2 \right ]_{tr} 
= 2 \ {\rm Tr} \left ( \gamma^0 \gamma^{\mu}\right ) P_{\mu} \quad . 
\end{equation}
 Using $(\gamma^0)2 = {\bf 1}$ and ${\rm Tr} (\gamma^{0} \gamma^{\mu}) 
= 4 g^{0\mu}$, one obtains
\begin{equation}
\label{SUSY-3}
\sum_r Q_r^2 = 4P^0 = 4H \quad .
\end{equation}}}:
\begin{equation}
\label{SUSY-4}
H = {1 \over 4} \sum_r Q_r^2 \quad .
\end{equation}
It follows that the energy of the vacuum $| 0 \rangle$ can be expressed as:
\begin{equation}
\label{SUSY-5}
\langle 0 | H | 0 \rangle = {1 \over 4} \sum_r \parallel Q_r |0 \rangle 
\parallel^2 \ .
\end{equation}
Thus, the vacuum energy vanishes if and only if supersymmetry is a symmetry of 
the vacuum: $ Q_r |0 \rangle = 0$ for all $r$.\footnote{{\em Remember that a 
supersymmetry transformation $U$ is obtained by exponentiating the generators: 
$U |0 \rangle = |0 \rangle$.}}

The problem however is that, at the same time, supersymmetry predicts equal
boson and fermion masses and therefore needs to be broken since this is not 
observed in Nature. The amount of 
breaking necessary to push the supersymmetric partners high enough not to have 
been observed yet, typically $\Lambda \sim$ TeV, is incompatible with the 
limit (\ref{7-8}). 

Moreover, in the context of cosmology, we should consider supersymmetry in 
a gravity context and thus work with its local version, supergravity (following
(\ref{SUSY}), local supersymmetry transformations are associated with local 
spacetime translations which are nothing but the reparameterizations which play 
a central role in general relativity). In this context, 
the criterion of vanishing vacuum energy is traded for one of vanishing mass
for the gravitino, the supersymmetric partner of the graviton
(which allows to cancel the constant vacuum energy at the expense of generating
a mass $m_{_{3/2}}$ for the gravitino field; see e.g. Ref.~\cite{Binetruy}, section 
6.3 for a more complete treatment). Local supersymmetry  is then absolutely 
compatible with a 
nonvanishing vacuum energy, preferably a negative one (although possibly 
also a positive one). This is both a blessing and a problem: supersymmetry 
may be broken while the cosmological constant remains small, but we have lost 
our rationale for a vanishing, or very small, cosmological constant and 
fine-tuning raises again its ugly head.

In some supergravity theories, however, one may recover the vanishing 
vacuum energy criterion. 


\subsection{Why now?} \label{sect:5-4}

\begin{figure}[h]
\begin{center}
\includegraphics[scale=0.65]{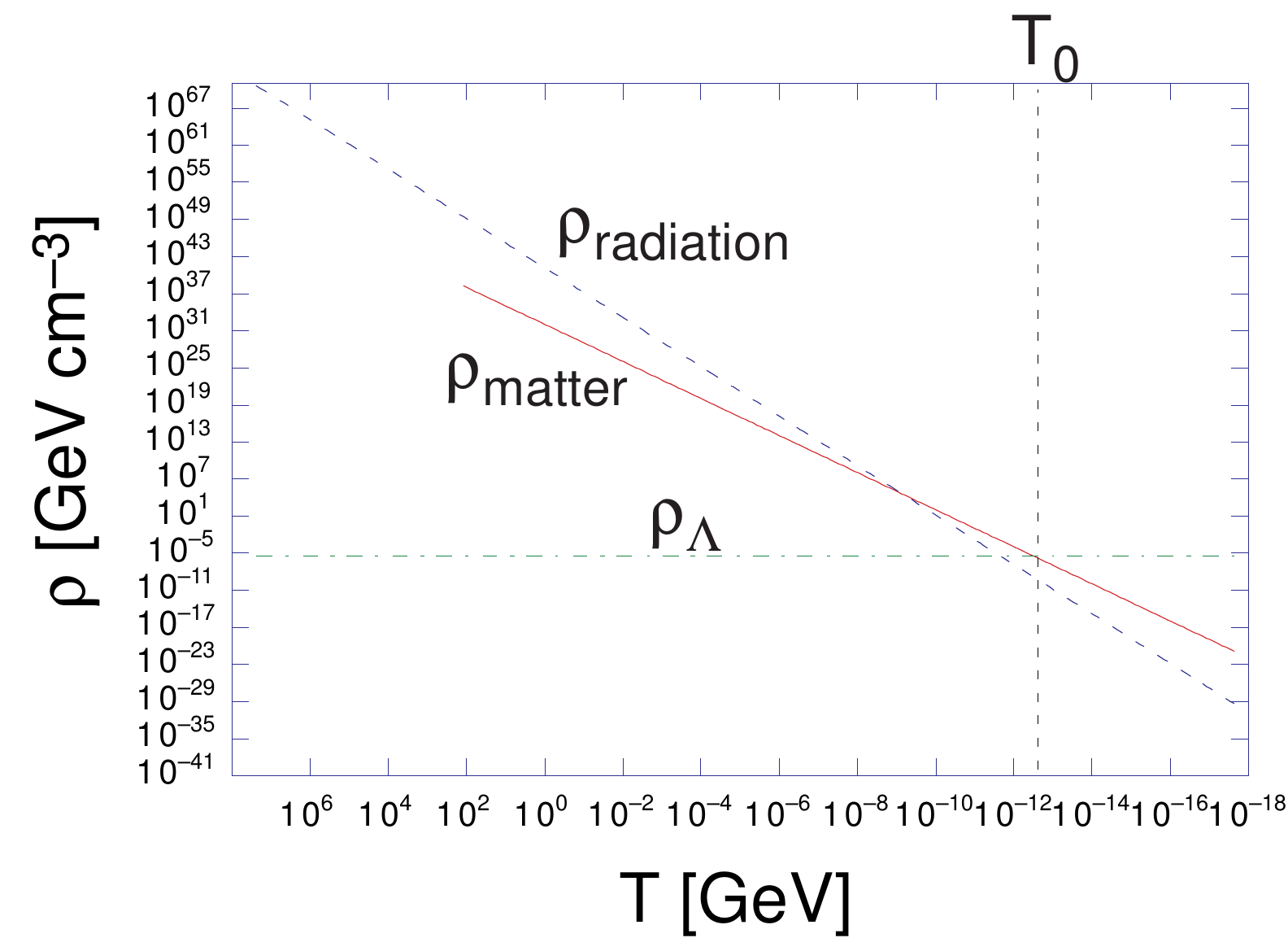}
\end{center}
\caption{The cosmic coincidence problem illustrated in the case of a 
cosmological constant}
\label{fig5-3}
\end{figure}

In the case where the acceleration of the expansion is explained by a 
cosmological constant, one has to explain why this constant contribution 
appears to start to dominate precisely now. This is the ``Why now?'' or cosmic 
coincidence problem summarized in Fig.~\ref{fig5-3}, the coincidence being 
between the onset of acceleration and the present time (on the scale of the age 
of the Universe).
In order to avoid any reference to us
(and hence any anthropic interpretation, see below), 
we may rephrase the 
problem as follows. Why does the dark energy starts to dominate at a time 
$t_\Lambda$ (redshift $z_\Lambda \sim 1$) which almost coincides with the 
epoch $t_G$ (redshift $z_G \sim 3$ to $5$) of galaxy formation?

{\em \vskip .5cm
\underline{Exercise 5-1}~: In this exercise, we will study the evolution of 
a flat ($k=0$) universe with a non-relativistic matter component (energy 
density $\rho_{_M}$) and a cosmological term (or equivalently vacuum energy 
density $\rho_\Lambda = \lambda/(8\pi G_{_N})$). The relevant equations 
of evolution are obtained from (\ref{2-6}), (\ref{2-11}) and (\ref{2-12}):
\begin{eqnarray}
\label{ex1-4_1}
H = {\dot a^2 \over a^2} &=& {8\pi G_{_N} \over 3} (\rho_{_M} +\rho_\Lambda ) 
\ , \\
\label{ex1-4_2}
{\ddot a \over a} &=& -{4\pi G_{_N} \over 3} (\rho_{_M} - 2 \rho_\Lambda) \ , \\
\label{ex1-4_3}
\dot \rho_{_M} &=& - 3H \rho_M \ .
\end{eqnarray}

a) Defining $H_\Lambda \equiv \left( 8\pi G_{_N} \rho_\Lambda/3\right)^{1/2}$, 
show that the solution to the system of equations is given by
\begin{equation}
\label{ex1-4_4}
a(t) = C \left( \sinh {3 \over 2} H_\Lambda t\right)^{2/3} \ ,
\end{equation}
where $C$ is a constant. Compute the Hubble parameter $H$ as well as the ratio
$\rho_{_M}/\rho_\Lambda$ as a function of time $t$ and $H_\Lambda$.

b) Compute the present Hubble constant $H_0$ in terms of $H_\Lambda$ and 
$\Omega_{_M}$. What is the age of the Universe $t_0$ in terms of $H_0$ and 
$\Omega_{_M}$?

c) Plot $a(t)/C$ and $\rho_{_M}/\rho_\Lambda$ in terms of time. When is the
equality between matter and vacuum energy reached?
\vskip .3cm 
Hints: a) $H = H_\Lambda / \tanh (3H_\Lambda t/2)$ and $\rho_{_M}/\rho_\Lambda
= \left[\sinh (3H_\Lambda t/2)\right]^{-2}$.

b) $H_0 = H_{\Lambda}/\sqrt{1-\Omega_{_M}}$,
\begin{equation}
\label{ex1-4_5}
t_0 = {2 \over 3 H_0}{1 \over \sqrt{1-\Omega_{_M}}} \ln {1 + \sqrt{1-\Omega_{_M}}
\over \sqrt{\Omega_{_M}}} \ .
\end{equation}}

\subsection{More dynamics: dark energy vs modification of gravity}
\label{sect:5-5}

An alternate possibility is that the cosmological 
constant is much smaller or even vanishing and that the acceleration is due to 
some new form of energy  --known as dark energy-- or some modifications of 
gravity. These two possibilities correspond to modifications of either sides 
of Einstein's equations (\ref{2-Einstein}). Let us envisage briefly these two 
cases. 

First, we may try to identify a new component $\rho_{_X}$ of the energy 
density with negative pressure: 
\begin{equation}
\label{5-43}
p_{_X} = w_{_X} \rho_{_X}, \; \; w_{_X} < 0 \ . 
\end{equation}
Note that the equation of state parameter $w_{_X}$ may not be constant and 
may thus evolve with time.

Observational data constrains such a dynamical component, referred to in the
literature as dark energy,  just as it
did with the cosmological constant. For example, in a spatially flat
Universe with only matter and this unknown component $X$, one obtains from 
(\ref{2-12}) with $\rho=\rho_M+\rho_{_X}$, $p=w_{_X} \rho_{_X}$ the following 
form for 
the Hubble parameter and the deceleration parameter (compare with (\ref{4-31})
and (\ref{4-33}))
\begin{eqnarray}
H^2(z) &=& H_0^2 \left[ \Omega_{_M} (1+z)^3 +  \Omega_{_X} (1+z)^{3(1+w_{_X})}
 \right] \ ,\label{5-44c} \\
q(z) &=& {H_0^2 \over 2H(z)^2} \left[  \Omega_{_M} (1+z)^3 
+  \Omega_{_X} (1+3w_{_X}) (1+z)^{3(1+w_{_X})}  \right] \ , \label{5-44a}
\end{eqnarray}
where $\Omega_X=\rho_{_X}/\rho_c$. At present time (compare with (\ref{4-8})), 
\begin{equation}
\label{5-44}
q_0 = {\Omega_M \over 2} + (1 + 3 w_{_X}) {\Omega_{_X} \over 2} \ .
\end{equation}
The acceleration of the expansion observed
requires that $\Omega_{_X}$ dominates\footnote{{\em One may easily obtain 
from (\ref{5-44a}) the time of the onset of the acceleration phase:
\begin{equation}
\label{5-44b}
1 + z_{{\rm acc}} =  \left[- (1 + 3 w_{_X}) {\Omega_{_X} \over \Omega_{_M}}  
\right]^{-1/(3w_{_X})} \ .
\end{equation}}} with $w_{_X}<-1/3$. 

\begin{figure}[h]
\begin{center}
\vskip 0.5cm
\includegraphics[scale=0.5]{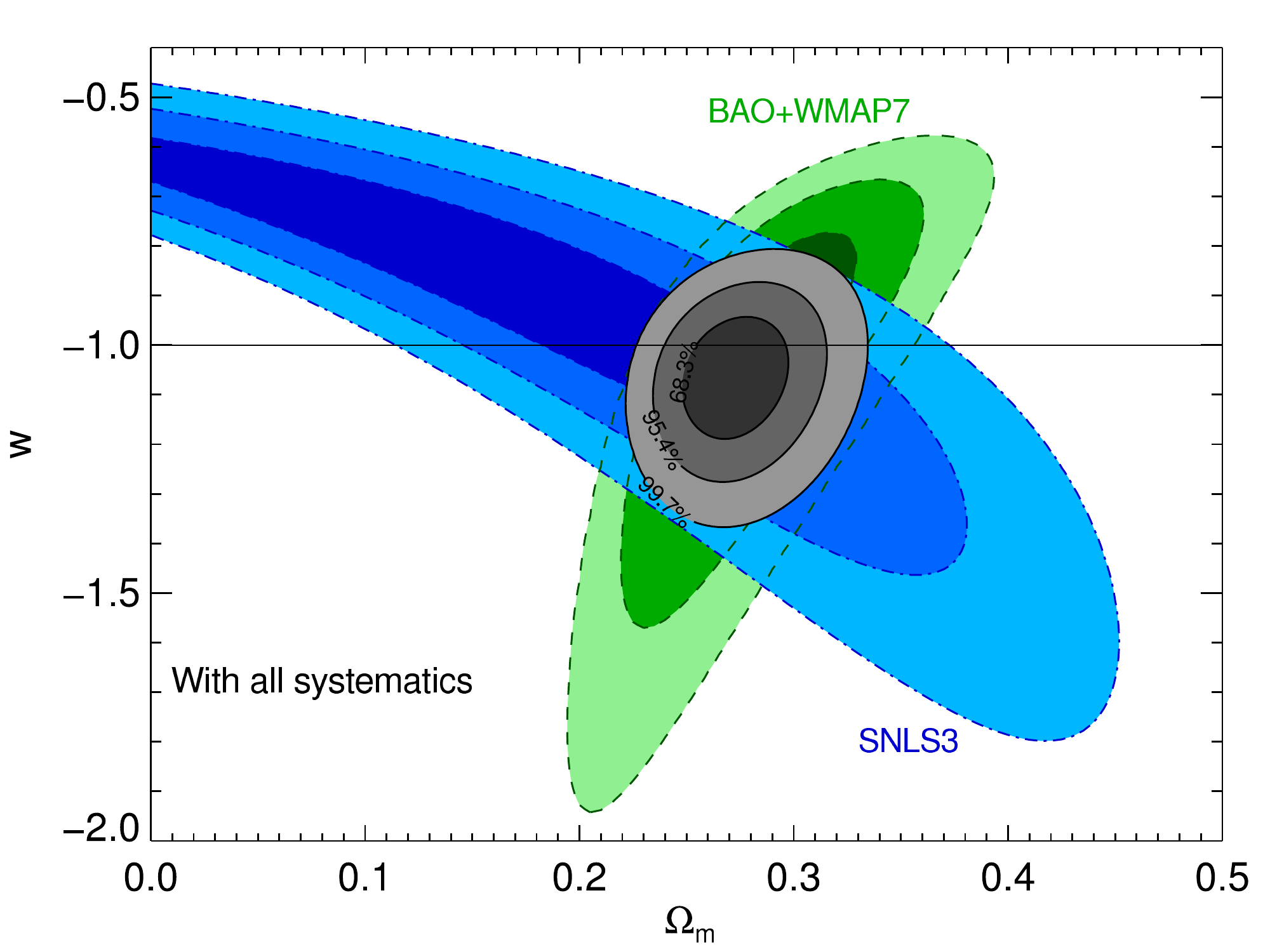}
\end{center}
\caption{Confidence contours in the $(\Omega_{_M}, w_{_X})$ plane arising from 
supernova results from SuperNova Legacy Survey (SNLS) 3 year results (in blue) 
and combined BAO/WMAP7 constraints (in green) [see Ref.~\cite{SNLS11}].}
\label{fig1obs-1}
\end{figure}

In Fig.~\ref{fig1obs-1}, we present constraints in the 
$(\Omega_{_M}, w_{_X})$ plane, obtained recently~\cite{SNLS11} from 
combining observations using supernovae and BAO as well as CMB results from
WMAP.


An important property of dark energy is that it does not appear to be 
clustered (just as a cosmological constant). Otherwise, its effects would 
have been detected locally, as is the case for dark matter. This points 
towards scalar fields, which generically have this property. 
Indeed, an attractive property of scalar fields is that they easily provide a 
diffuse background by resisting gravitational attraction. The key quantity 
when discussing gravitational clustering is the speed of sound defined
as
\begin{equation}
\label{5-46a}
c_s^2 \equiv {\delta p \over \delta \rho} \ .
\end{equation} 
It is a measure of how the pressure of the field resists gravitation
clustering. In most models of dark energy, we have $c_s^2 \sim 1$, 
which explains why such
scalar dark energy does not cluster: its own pressure resists gravitational 
collapse.

It should be stressed that, until recently, no fundamental scalar field had 
been observed in Nature. The discovery of a Higgs particle at the LHC
high energy collider has obviously promoted the status of fundamental scalar 
particles.

In the next section, we will illustrate the dynamics of dark energy on the 
example of a scalar field evolving with time along its potential. This is often 
referred to as quintessence. Such 
scalar fields turn out to be extremely light: the only dimensionful parameter 
in the problem being the Hubble constant $H_0$, their mass is $\hbar H_0
\sim 10^{-33} \ {\rm eV}$ (see e.g. (\ref{5-68}) below). The exchange of such 
fields leads to a long range force: the range is the inverse of the mass
(times $\hbar c$), typically $\ell_{H_0}$ i.e. the size of the observable 
universe. This force is therefore 
similar to gravity and can hardly be disentangled from it. Gravitational tests
such as the test of the equivalence principle thus apply not to the 
gravitational force alone but to the combination of gravity and this new 
force. This compels this force associated with dark energy to share many 
properties with gravity. One may thus talk of a gravitational type force. 

The second possibility is to modify the left-hand side of Einstein's equations,
i.e. to look for a modification of gravity (we already mentionned that 
possibility in Section~\ref{4s2a} for explaining observational data 
traditionally accounted fo by dark matter). This is a notoriously difficult 
task because the current theory of gravity, general relativity, has passed 
many stringent experimental and observational tests: equivalence 
principle, Lorentz invariance,.... Any alternative theory
should first pass these tests equally successfully before being further 
considered.
 
Again, the distinction with the previous case where one introduces a new 
dynamical component is not as clear-cut as it would first seem. Indeed, 
through field redefinitions, alternate theories of gravity may be rewritten 
as Einstein gravity plus a dynamical scalar field. Or if one goes to more 
spatial dimensions, the graviton of the higher-dimensional theory may be 
regarded as a standard 4-dimensional graviton plus a collection of scalar
(spin zero) or vector (spin one) fields.

Hence, the distinction between what is often presented as the two ways to 
account for the observed acceleration of the expansion is not so clear. 

\subsection{The example of quintessence}

A scalar field $\phi$ which has reached the minimum $\phi_0$ of its 
potential energy $V(\phi$) amounts to a cosmological constant in the form of 
vacuum energy: its kinetic energy is vanishing and its potential energy is the
constant $V(\phi_0)$. A more dynamical candidate for dark energy 
is a scalar field which is still slowly evolving in its potential
\cite{We88,PR88,RP88,CDS98}. One often 
refers to this field as a quintessence field.

To be more explicit, let us consider the general action which describes a real 
scalar field $\phi$ minimally coupled with Einstein gravity.
\begin{equation}
\label{5-49}
{\cal S}= \int d^4x \sqrt{-g} \left[ -{m_{_P}^2 \over 2} R + { 1 \over 2}
\partial^\mu \phi \partial_\mu \phi - V(\phi) \right] \ . 
\end{equation}
Computing the corresponding energy-momentum tensor, we obtain the pressure and
energy density (note the parallel with Section~\ref{5s1} where we discussed 
inflation models) 
\begin{eqnarray}
p_\phi &=& {1 \over 2} \dot \phi^2 - V( \phi) \quad , \label{5-45b} \\
\rho_\phi &=& {1 \over 2} \dot \phi^2 + V( \phi) \quad , \label{5-45a}
\end{eqnarray}
where, in the latter, we identify the field kinetic energy $\dot \phi^2/2$
and the potential energy $V(\phi)$.
The corresponding equation of motion is, if one neglects the spatial curvature 
($k \sim 0$),
\begin{equation}
\label{5-52}
\ddot \phi + 3 H \dot \phi = - {dV \over d \phi} \ ,
\end{equation}
where, besides the standard terms, one recognizes the friction term $3 H \dot 
\phi$ due to expansion. We deduce, as expected,
\begin{equation}
\label{5-53}
\dot \rho_\phi = -3H (p_\phi + \rho_\phi) \ .
\end{equation}

We have for the equation of state parameter
\begin{equation}
\label{5-46}
w_\phi \equiv {p_\phi \over \rho_\phi} 
= { {1 \over 2} \dot \phi^2 - V( \phi) \over  {1 \over 2} 
\dot \phi^2 + V( \phi)} \ge -1 \quad .
\end{equation}
If the kinetic energy is subdominant ($\dot \phi^2/2 \ll V(\phi)$), we
clearly obtain $-1 \le w_\phi \le 0$. In any case $-1 \le w_\phi \le +1$.


Let us look in more details at the dynamics of such a 
quintessence scalar field. For this purpose, it is useful to identify 
scaling solutions 
which we define as solutions where the $\phi$ energy density scales as a 
power of the cosmic scale factor: 
\begin{equation}
\label{5-46g}
\rho_\phi \propto a^{-n_\phi} \ , \quad n_\phi \ \hbox{constant} \ .
\end{equation}
This will allow us to identify two of the main examples of dynamical potential
for quintessence:
\begin{itemize}
\item the exponential potential: 
\begin{equation}
\label{5-59b}
V(\phi) = V_0 e^{-\lambda \phi} \ ;
\end{equation} 
\item the Ratra-Peebles potential~\cite{RP88,PR88}, 
\begin{equation}
\label{5-58a}
V(\phi) = {M^{4+\alpha} \over \phi^\alpha} \ , \quad \alpha > 0 \ .
\end{equation}
\end{itemize}
As we will see, the interest of such solutions is that they correspond to 
attractors in the cosmological evolution of the scalar field. 

Since it
follows from (\ref{5-46g}) that $\dot \rho_\phi / \rho_\phi = - n_\phi H$,
we deduce from (\ref{5-53}) a relation between $n_\phi$ and the equation of 
state parameter $w_\phi$:
\begin{equation}
\label{5-56}
w_\phi = {n_\phi \over 3} - 1 \ . 
\end{equation}
Hence the scaling solutions that we look for, exist only in epochs of the 
cosmological evolution where the equation of state parameter may be considered 
as constant (it could still be constant piecewise). 

Since the dark energy (quintessence) density is expected to emerge from the 
background (radiation or matter) energy density, we consider the evolution of 
a scalar field $\phi$ with constant parameter $w_\phi$, during a phase 
dominated by a background fluid with equation of state parameter
\begin{equation}
\label{5-51}
w_{_B} = {n_{_B} \over 3} - 1
\end{equation}
Following (\ref{2-11a}), we have $a(t) \sim t^{2/n_{_B}}$ ($n_{_B} = 4$ for 
radiation, $3$ for non-relativistic matter,...).

From (\ref{5-45b}) and (\ref{5-45a}), we obtain 
\begin{equation}
\label{5-57}
\dot \phi^2 = {n_\phi \over 3} \rho_\phi \ , \quad V(\phi) = \left( 1 -
{n_\phi \over 6} \right) \rho_\phi \ .
\end{equation}
Hence $\dot \phi^2 \sim a^{-n_\phi}$ and $\dot \phi \sim t^{-n_\phi/n_{_B}}$.
We thus distinguish two cases:
\vskip .3cm
i) $n_\phi = n_{_B}$
\vskip .3cm
{\em Clearly this implies $w_\phi = w_{_B} >0$ and the quintessence field $\phi$ 
cannot be interpreted as the dark energy component. 
We have
\begin{equation}
\label{5-59}
\phi = \phi_0 + {2 \over \lambda}  \ \ln (t/t_0) \ , 
\end{equation}
with $\lambda$ constant. Then
\begin{equation}
\label{5-59a}
V(\phi) \sim \rho_\phi \sim a^{-n_\phi} \sim t^{-2} \sim e^{-\lambda \phi} \ .
\end{equation}
Hence, we find a scaling behavior for the exponential potential (\ref{5-59b}) 
in a background such that $n_{_B} = n_\phi$ ($w_{_B} = w_\phi$). The solution
of the equation of motion (\ref{5-52}) then reads
\begin{equation}
\label{5-59c}
\phi = {1 \over \lambda} \ln \left( {V_0 \lambda^2 \over 2} 
{n_{_B}  \over 6-n_{_B}} t^2 \right) \ ,
\end{equation}
and the energy density (\ref{5-45a})
\begin{equation}
\label{5-59cc}
\rho_\phi = {12 \over \lambda^2 n_{_B} t^2} \ .
\end{equation}
Since $H^2 = (\rho_{_B} + \rho_\phi)/3 \sim [2/(n_{_B}t)]^2$,
\begin{equation}
\label{5-59ce}
{\rho_\phi \over \rho_{_B} + \rho_\phi} \sim {n_{_B} \over \lambda^2}  \ .
\end{equation}
Hence $\rho_\phi/\rho_{_B}$ tends to be constant in this scenario. One calls 
this 
property ``tracking''. This is obviously compatible with our initial 
assumptions only if $\lambda^2 > n_{_B}$.

What happens if $\lambda^2 \le n_{_B}$?

It turns out that the scaling solution corresponds to a totally different 
regime: the scalar field is the dominant contribution to the energy density.
We do not have to redo the calculation: it is identical to the previous one 
with the only changes $w_{_B} \rightarrow w_\phi$ or $n_{_B} \rightarrow 
n_\phi$ (for example, $H^2 = (\rho_{_B} + \rho_\phi)/3 \sim [2/(n_\phi t)]^2$: 
the scalar energy density determines the evolution of the Universe). But then 
(\ref{5-59ce}) reads $1 \sim n_\phi / \lambda^2$, i.e. 
\begin{equation}
\label{5-59cd}
w_\phi = - 1 + {\lambda^2 \over 3} \ .
\end{equation}
Thus, if $\lambda^2 < 2$, the scalar field $\phi$ may provide the dark energy 
component. }

\begin{figure}[h]
\begin{center}
\vskip .5cm
\includegraphics[scale=0.5]{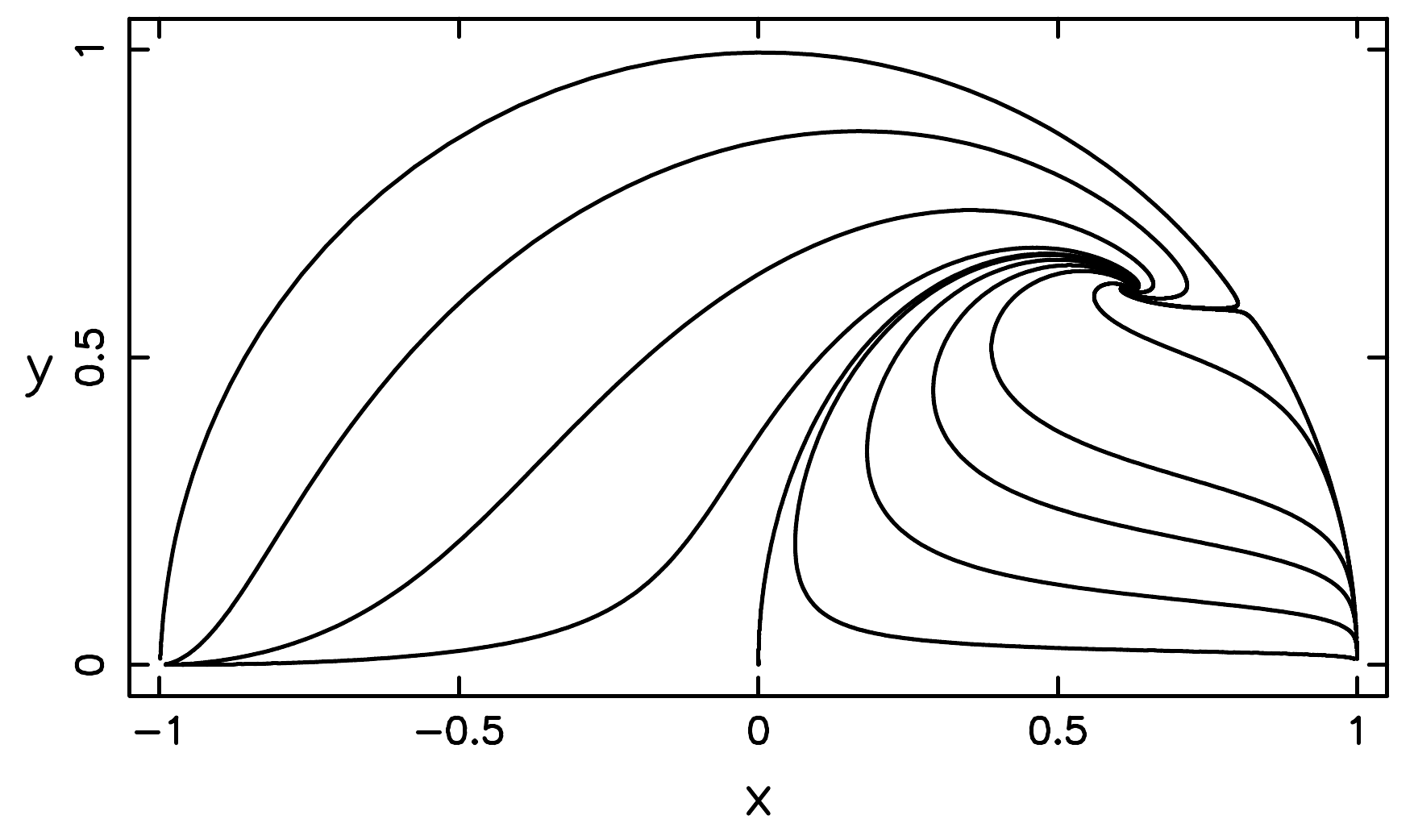}
\end{center}
\vskip -.5cm
\caption{Phase space diagram for the exponential potential (\ref{5-59b}) with
$\lambda = 2$ and $n_{_B} = 3$ (matter domination)~\cite{CLW98}}
\label{figN3-1}
\end{figure}

To summarize the two regimes that we have obtained for the exponential 
potential (\ref{5-59b}):
\begin{itemize}
\item if $\lambda^2 \le n_{_B}$, the scaling solution has $w_\phi = -1 
+ \lambda^2/3$ and $\rho_\phi/(\rho_\phi + \rho_{_B}) \sim 1$ ($\phi$ is the 
dominant species),
\item if $\lambda^2 > n_{_B}$, the scaling solution has $w_\phi = w_{_B}$
and $\rho_\phi / (\rho_{_B} + \rho_\phi) \sim n_{_B} / \lambda^2$ (the
background energy density dominates; the scalar field energy density tracks 
it).
\end{itemize}

ii) $n_\phi \not= n_{_B}$
\vskip .3cm
Then $\phi \sim t^{-{n_\phi \over n_{_B}}+1}$ and we now have
\begin{equation}
\label{5-58}
V(\phi) \sim \rho_\phi \sim a^{-n_\phi} \sim t^{-2n_\phi/n_{_B}}
\sim \phi^{-2{n_\phi \over n_{_B} - n_\phi}} \ .
\end{equation}
Hence, we find a scaling behaviour for the Ratra-Peebles potential 
(\ref{5-58a}) in a background characterized by $n_{_B} \not = n_\phi$ 
(or $w_{_B} \not = w_\phi$). We have
\begin{equation}
\label{5-58b}
n_\phi = {\alpha n_{_B} \over \alpha+2} \quad \hbox{or} \quad
w_\phi = {\alpha w_{_B} -2 \over \alpha +2} \ .
\end{equation}
 The complete solution of the equation of motion 
(\ref{5-52}) is 
\begin{equation}
\label{5-58c}
\phi = \left( {\alpha (\alpha+2)^2 n_{_B} \ M^{4+\alpha}t^2
\over 2 \left[6(\alpha +2) - n_{_B} \alpha \right]}\right)^{{1 \over 
\alpha +2}} \ .
\end{equation} 
\vskip .3cm
As we advertised in the beginning, these scaling solutions 
correspond to attractors in 
the cosmological evolution of the scalar field. 
We show in Fig.~\ref{figN3-1} the full phase diagram for the exponential 
potential with $\lambda= 2$ during matter domination ($n_{_B} =3$). The 
coordinates are $x \equiv \dot \phi/(\sqrt{6}H m_{_P}^2)$ and $y \equiv 
\sqrt{V}/(\sqrt{3}H m_{_P}^2)$. The stable (spiral) attractor is found at 
$x=y=\sqrt{3/8}$.

\vskip .5cm
{\em \underline{Exercise 5-2}~: We show in this exercise that the solution 
(\ref{5-58c}) of the Ratra-Peebles potential (\ref{5-58a}) is also an 
attractor. 

a) Show that  a small perturbation $\delta \phi$  satisfies the equation
\begin{equation}
\label{5-58d}
\delta \ddot \phi + {6 \over n_{_B} t} \delta \dot \phi + {2(\alpha +1) \over 
n_{_B} (\alpha +2)^2 t^2} \left[6(\alpha +2) - n_B \alpha \right]\delta \phi 
= 0 \ .
\end{equation}

b) Look for a solution of the form $\delta \phi \sim t^\gamma$ and express 
$\gamma$ in terms of $n_{_B}$ and $\alpha$.

c) Assume  $\alpha >0$ and standard values of $n_{_B}$. Show that the two 
solutions obtained in b) decay as $\delta \phi \sim 
t^{-(6-n_{_B})/2n_{_B}}$: the solution (\ref{5-58c}) is an attractor.

\vskip .3cm \noindent
Hints: 
b) 
\begin{equation}
\label{5-58e}
\gamma = -{6-n_{_B} \over 2 n_{_B}} \pm  
{\sqrt{ -\left[ 3\alpha^2 (3 n_{_B}-2) (6-n_{_B}) -12 \alpha (n_{_B}^2 
-16 n_{_B} +12) - 4 (n_{_B}^2 -36 n_{_B} +36) \right]}\over 2 n_{_B}
(\alpha +2)}  \ .
\end{equation}

c) The reduced discriminant of the second order polynomial in $\alpha$ which
is under the square root
is simply $288 n_{_B}^3 >0$. The corresponding roots are then negative for 
$8- \sqrt{52} \sim 0.79 < n_{_B} < 6$ and the term is thus negative for 
$\alpha >0$ and $n_{_B}$ in this range. The square root contributes as an 
oscillating term and the two 
solutions corresponding to (\ref{5-58e}) decay as $\delta \phi \sim 
t^{-(6-n_{_B})/2n_{_B}}$.}

\vskip .3cm

The reader may have noticed that the type of field solutions 
that we have found is very similar to those encountered at the onset of 
inflation. This is not surprising since we are introducing dynamics through 
the 
slow roll of a field down its potential. Let us look a little closer at the 
field evolution in the case of the Ratra-Peebles  potential (\ref{5-58a}).

We have found the attractor scaling solution~\cite{RP88,PR88} 
$\phi \propto a^{n_B/(2+\alpha)}$, $\rho_\phi \propto 
a^{-\alpha n_B/(2+\alpha)}$ in the case where the background density dominates.
Thus $\rho_\phi$ decreases at a
slower rate than the background density ($\rho_B \propto a^{-n_B}$) and 
tracks it until it becomes of the same order, at a given
value $a_{_Q}$. We thus have:
\begin{eqnarray}
{\phi \over m_{_P}} & \sim &  \left({a \over a_{_Q}} \right)^{n_B/(2+\alpha)}
, \label{5-64} \\
{\rho_\phi \over \rho_B} &\sim &  \left( {a \over a_{_Q}} 
\right)^{2n_B/(2+\alpha)}.
\label{5-65}
\end{eqnarray}
\vskip .5cm
{\em\underline{Exercise 5-3}~: Compute the time $t_{_Q}$ at which $\rho_\phi 
\sim \rho_{_M}$ in terms of $M$ and $m_{_P}$. Check that $\phi$ at $t_{_Q}$ does 
not depend on $M$.
\vskip .3cm \noindent
Hints: $\rho_M \sim m_{_P}^2/t^2$ and $\rho_\phi \sim 
M^{{2(\alpha+4) \over \alpha+2}} t^{-{2\alpha \over \alpha+2}}$ give 
$t_{_Q} \sim m_{_P}^{{\alpha +2 \over 2}} M^{-{\alpha+4 \over 2}}$.}
\vskip .3cm
The corresponding value for the equation of state parameter is given by 
(\ref{5-58b}):
\begin{equation}
\label{5-66}
w_\phi = -1 + {\alpha (1+ w_B) \over 2+ \alpha}. 
\end{equation}
Shortly after  $\phi$ has reached for $a=a_{_Q}$ a value of order $m_{_P}$, it 
satisfies the usual slow roll conditions (using the notations (\ref{5-17} and 
(\ref{5-18}) introduced in the context of inflation)
\begin{equation}
\label{5-66a}
\epsilon \equiv {1 \over 2} \left(  {m_{_P} V' \over V} \right)^2 
= (\alpha/2)(m_{_P}/\phi)^2 \ll 1 \ , \quad \eta \equiv {m_{_P}^2 V'' \over V}
= \alpha (\alpha+1)(m_{_P}/\phi)^2 \ll 1 \ .
\end{equation}
Therefore (\ref{5-66}) provides a good approximation to the
present value of $w_\phi$. 
Thus, at the end of the matter-dominated era, this field may provide
the quintessence component that we are looking for. 

Two features are  interesting in this respect. One is that
this scaling solution is reached for rather general initial conditions,
{\em i.e.} whether $\rho_\phi$ starts of the same order or much smaller
than  the background energy density~\cite{ZWS99}. 

The second is the present value of $\rho$. Typically, since  in this scenario
$\phi$ is of order $m_{_P}$ when the quintessence component emerges, we must 
choose the scale $M$ in such a way that $V(m_{_P}) \sim \rho_c$.
The constraint reads:
\begin{equation}
\label{5-67}
M \sim \left( H^2_0 m_{_P}^{2+\alpha} \right)^{1/(4+\alpha)} .  
\end{equation}
We may note that this gives for $\alpha = 2$, $M \sim 10$ MeV,
not such an atypical scale for high energy physics.

\vskip .5cm
{\em \underline{Exercise 5-4}~: In the case of slow roll, the equation of 
motion (\ref{5-52}) simply reads $3H \dot \phi = -V'(\phi)$. 

a) Under this assumption, show that 
\begin{equation}
\label{5-67aa}
\ddot \phi = - {4\pi G_{_N}\over 3}{V' \over H^2} \sum_i (p_i + \rho_i) 
\end{equation}
where the summation is over all components of the Universe.

b) Deduce that, in the case where only matter and dark energy are 
nonnegligible at present time $t_0$ ($\Omega_M + \Omega_\phi = 1$),
\begin{equation}
\label{5-67ab}
\left. {\ddot \phi \over V'} \right|_{t_0} \sim -{1 \over 2} (1-\Omega_\phi)
\ .
\end{equation}
Hence slow roll requires that $\Omega_\phi \sim 1$. 
\vskip .3cm \noindent
Hints: a) Use $\dot H = -4\pi G_{_N} \sum_i (p_i + \rho_i)$.}

\vskip .5cm

However appealing, the quintessence idea is difficult to implement in
the context of realistic models~\cite{Ca98,KL99}. The main problem lies in
the fact that the quintessence field must be extremely weakly coupled
to ordinary matter. This problem can take several forms:

\vskip .3cm $\bullet$ 
The quintessence field must be very light. If we
return to our example of the Ratra-Peebles potential (\ref{5-58a}), 
$V''(m_{_P})$ provides an order of magnitude for the mass-squared  of the 
quintessence component:
\begin{equation}
\label{5-68}
m_\phi \sim M \left( { M \over m_{_P} } \right)^{1+\alpha/2}
\sim H_0 \sim 10^{-33} \ {\rm eV}.
\end{equation}
using (\ref{5-67}). The exchange of such a field
leads to a long-range force: the range is typically $\ell_{H_0}$, the size of 
the presently observable Universe. Since this force has not been observed yet, 
this means that the field $\phi$ must be very weakly coupled to matter in 
order to comply with the constraints imposed on gravitational-type forces by 
the very stringent tests of the equivalence principle.

\vskip .3cm $\bullet$ 
The quintessence field is presently evolving with time. This may generate 
a time dependence of what we call the constants of Nature. Indeed, it turns 
out that, in modern particle theories, many of the constants of nature have a 
dynamical origin: they are expressed in terms of quantum fields which have 
settled down to their vacuum values in our present Universe. For example, in
the models discussed above, it is difficult to find a symmetry that would 
prevent any coupling of the form 
\begin{equation}
\label{phiFF}
\beta {\phi\over m_{_P}} F^{\mu \nu} F_{\mu\nu}
\end{equation} 
to the gauge field kinetic term (or any given power of $\phi$). Since the 
quintessence behaviour is
associated with time-dependent values of the field of order $m_{_P}$, this
would generate, in the absence of fine tuning, corrections of order one
to the gauge coupling\footnote{For example, the quantum electrodynamics 
Lagrangian reads, with our conventions ${\cal L} = -(1/4e^2) F^{\mu\nu} 
F_{\mu\nu}$. The extra term would lead to an effective electromagnetic coupling
coupling $ (1/4e_{{\rm eff}}^2) =  (1/4e^2) + \beta (\phi/m_{_P})$, hence a 
time-dependent fine structure constant $\alpha = e_{{\rm eff}}^2/(\hbar c)$.
\label{gaugecoupling}}.
But the time dependence of the fine structure
constant for example is very strongly constrained~\cite{Uz03}: $|\dot \alpha /
\alpha| < 5 \times 10^{-17} {\rm yr}^{-1}$. This yields a limit~\cite{Ca98}:
\begin{equation}
\label{5-69}
|\beta | \le 10^{-8} {m_{_P} H_0 \over \langle \dot \phi \rangle},
\end{equation}
where $\langle \dot \phi \rangle$ is the average over the last $2 \times 10^9$ 
years. Let us
recall~\cite{Wi06} that the non-constancy of constants is not compatible 
with the principle of local position invariance (i.e. independence on the 
the location in {\em time} and space where a non-gravitational experiment is 
performed) which forms part of the Einstein's equivalence principle. This in 
turn leads to violations of the universality of the weak equivalence 
principle~\cite{No02}.


\vskip .3cm $\bullet$
We have seen that, in the simplest models, the regime 
of interest is reached when the quintessence field $\phi$ value becomes larger
than $m_{_P}$. In this instance, as well as in the context of (single field)
chaotic inflation, there has been discussions 
whether this remains in the sub-Planck domain: strictly speaking, the answer is
yes since what characterizes the Planck domain are energy densities of
order $m_{_P}^4$, whereas here $\rho_\phi$ remains much smaller because of the 
specific form of the potential (see e.g. (\ref{5-59b}) or (\ref{5-58a})).  
It remains that, in such a context, one must take into account all 
non-renormalisable interactions of order $(\phi / m_{_P})^n$ compatible with 
the symmetries. 


\vskip .5cm
All the preceding shows that there is extreme fine tuning in the
couplings of the quintessence field to ordinary matter, unless they are
forbidden by some symmetry. This is somewhat reminiscent of the fine
tuning associated with the cosmological constant. Let us stress however that   
the quintessence solution does not claim to solve the cosmological
constant (vacuum energy) problem described above. Quintessence may explain 
the acceleration of the expansion of the Universe but has nothing to say about 
the cancellation of  the bulk of the vacuum energy arising from quantum 
fluctuations. 

\subsection{Back to the cosmological constant}

Let us conclude this Section by reviewing some of the attempts to address the 
problem of the cosmological constant.

\subsubsection{Relaxation mechanisms} \label{subsect:5-1}

In the days where it was believed that vacuum energy was vanishing, one had to 
look for a mechanism to fully cancel the contribution of order $m_{_P}^4$. One 
naturally advocated mechanisms that relaxed the cosmological constant to zero through 
the equations of motions of some dynamical fields.

For example, in the context of 
string models, any dimensionful
parameter is expressed in terms of the fundamental string scale $M_{_S}$
and of vacuum expectation values of scalar fields. The physics of the
cosmological constant and of its relaxation to a vanishing value would then 
be associated with the dynamics of the corresponding  scalar
fields.  

However, Steven  Weinberg~\cite{We89} has constrained the possible 
mechanisms for the relaxation of the cosmological constant by proving the 
following ``no-go'' theorem: it is not possible to obtain a 
vanishing cosmological constant as a consequence of the equations of motion 
of a finite number of fields. Weinberg's no-go theorem relies on a series 
of assumptions: 
Lorentz invariance, {\em finite} number of {\em constant} fields, 
possibility of globally redefining these fields...  All attempts to propose a 
relaxation mechanism have tried to avoid the conclusions of the theorem by 
relaxing one of these assumptions. 

\subsubsection{Anthropic considerations} 
\label{subsect:5-4}

The anthropic principle approach can be sketched as follows.
We consider regions of spacetime with different values of $t_G$ (time of 
galaxy formation)  and $t_\Lambda$, the time when the cosmological constant 
starts to dominate i.e. when the Universe enters a de Sitter phase of 
exponential expansion. Clearly galaxy formation must precede this phase 
otherwise no observer (similar to us) would be able to witness it. Thus
$t_G \le t_\Lambda$. On the other hand,  regions with $t_\Lambda \gg t_G$ 
have not yet undergone any de Sitter phase of re-acceleration and are thus 
``phase-space suppressed'' compared with regions with $t_\Lambda \sim t_G$. 
Hence the regions favoured have $t_\Lambda \stackrel{>}{\sim} t_G$ and 
thus $\rho_\Lambda \sim  \rho_M$.

This was quantified by S. Weinberg~\cite{We87,MSW97}, who obtained the 
following bound:
\begin{equation}
\label{7-22}
\rho_\Lambda < {\pi^2 \over 3} \rho_0 (1+z_G)^3 \ ,
\end{equation}
where $\rho_0$ is the present energy density and $z_G$ the redshift 
corresponding to galaxy formation. Using $z_G= 4.5$ as originally chosen by
Weinberg~\cite{We87}, one finds $0 < \rho_\Lambda/\rho_{_M} < 550$. 
More recent observations of a galaxy at $z=8.6$~\cite{Le10} or the existence 
of dwarf galaxies at 
$z \sim 10$~\cite{Lo06} give a larger anthropic range:
\begin{equation}
\label{7-23}
0 < \rho_\Lambda/\rho_{_M} < 4000 \ .
\end{equation}

\subsubsection{Emergent gravity} \label{subsect:5-2}

The alternative approach is to return to the origin of the vacuum energy 
problem. We stressed in Section~\ref{sect:5-2} that this problem arises in the 
context of a quantum treatment of gravity (both $\hbar$ and $G_{_N}$ are 
involved). At present we do not have a fully valid theory of quantum gravity.
Presumably, it involves as well a quantum version of spacetime. It is probable
that, just as our notion of continuous and elastic matter is only valid in 
a large distance approximation, our notion of continuous space and time is also
only valid at large distance. Of course, it remains to be seen by which 
"objects" one should replace continuous space and time, for distances smaller 
than the Planck length, and what is the corresponding theory. In any case, 
space and time would be emergent notions, and probably also gravity.
For what concerns us here, it could be that the solution to the vacuum problem 
should be searched in this deeper context. And maybe dark energy is telling us 
something about this underlying theory. Let us also note that, if spacetime is 
an emergent notion, then its symmetries are also emergent: one may expect at 
some level violations of Lorentz invariance for example, which lead to 
violations of Einstein's equivalence principle.  
    
\subsubsection{Holography} \label{subsect:5-5}

Until now we have considered gravity as a fundamental force which is on the 
same footing as the other three. However, one aspect of gravity is strikingly 
different from what we encounter with other interactions: it is the phenomenon 
of gravitational collapse. As we have seen in Section~\ref{2s6}, if a quantity $E$ of (gravitating) energy is
localized in a region of spacetime of size $R$ smaller than the Schwarzschild 
radius defined as:
\begin{equation}
\label{7-25}
R_S \equiv 2 {G_{_N} E \over c^2} \,
\end{equation}
it undergoes gravitational collapse. This has been used by some (e.g. 
\cite{DG10}) to 
consider that the high-energy (ultraviolet) regime of gravity is classical:
before reaching Planckian energies, regions of spacetime undergo gravitational 
collapse and turn into black holes, which are classical objects. This may have 
some far reaching consequences for the issues we are dealing with here, 
especially vacuum energy.  

Indeed, let us return to the considerations that led to the estimate 
$\rho \sim m_{_P}/\ell_{_P}^3$ (see (\ref{120})) for the vacuum energy density 
in the context of quantum field 
theory.  Consider a spherical region of radius $R$ and energy density given by
(\ref{120}): $\rho = m_{_P}^4$. Then the total energy reads:
\begin{equation}
\label{7-26}
E = {4\pi \over 3} R^3 \rho =  {4\pi \over 3} m_{_P} \left( R m_{_P} \right)^3
\ .
\end{equation}
But the system will undergo gravitational collapse when $R<R_S$ that is, using 
(\ref{7-25}) $R < \left( R m_{_P} \right)^3 /3m_{_P}$ i.e. $R > 1/m_{_P} = 
\ell_{_P}$. Hence, for any volume larger than the elementary cell, on cannot 
concentrate a vacuum energy density $\rho = m_{_P}^4$, at least in the case (that
we consider here) that vacuum energy is gravitating: the system is unstable 
and undergoes gravitational collapse. The maximal energy density for a 
macroscopic region of size $R$ is ($E < R/(2G_{_N})$)
\begin{equation}
\label{7-27}
\rho_{max} = E/(4 \pi R^3/3)= {3 \over 8\pi G_{_N} R^2}
\end{equation}
Let us extend these considerations to the 
whole observable Universe of size $R\sim H_0^{-1}$: we can only store a vacuum 
energy density
\begin{equation}
\label{7-28}
\rho < {3 H_0^2 \over 8 \pi G_N} = \rho_c \ .
\end{equation}
Taken at face value, this would mean that the vacuum energy density has the 
value it has because our observable Universe is very large. This cannot be true 
at all times: otherwise, one can easily check that the presence of dark energy 
can be absorbed in a redefinition of Newton's constant; up to this redefinition,
the Universe would behave as if there is no dark energy and thus the recent 
phase
of acceleration of the expansion would remain unexplained. If pushed to its full
consequences, this leads to a new way of considering the quantum evolution of 
the Universe~\cite{Bi12}. 

\subsection{Concluding remarks} \label{conclusion}

The most fascinating aspect of the dark energy problem is the number of 
fundamental questions it connects with: how did the Universe emerge from a
quantum state to become so large and so old? why does dark energy emerge so 
late in the evolution of the Universe? does it relate to the nature of space 
and time as we know them? has spacetime emerged from something else? is general 
relativity the ultimate theory of gravity? if not, are its basic principles 
violated at some scale? what is a quantum state of the Universe? what is the 
status of an observer in such a Universe? are there multiple universes?
are there more than four dimensions?...

One of the reasons is that dark energy appears to be connected with vacuum 
energy, which is the most fundamental issue faced by theorists in fundamental 
physics, an issue that illustrates the difficulties encountered at the 
interface between general relativity, the present theory of gravity, and the 
quantum theory. In some sense, the situation is reminiscent of the one 
encountered at the end of the XIX$^{\rm th}$ century, where one had two very 
successful theories, Newtonian gravity and electromagnetism (summarized into 
the Maxwell equations). The Michelson-Morley experiment in 1887 was the 
experimental observation that led Einstein and others to reconsider the 
foundations.
Similarly, both general relativity and the quantum theory, the latter described
at the level of (non-gravitational) fundamental interactions 
by the Standard Model,
are extremely successful theories. Moreover, the recent successes of cosmology 
have shown that our picture of the early Universe based on these two pillars
is not simply qualitative but is supported quantitatively by increasingly 
precise observations. Is dark energy the signal of a new era? It remains to be 
seen its exact connection with the issue of vacuum energy. But more 
importantly, 
it is at present a conceptual difficulty, rather than a clear experimental 
sign of the inconsistency of the overall picture. We are still lacking our 
Michelson-Morley experiment. 

From this perspective, it is reassuring that we have in front of us in the next 
decade or so a very substantial experimental programme, which includes not 
only increasingly 
precise and complete observational data, but also experiments of many types, 
that might help us identify the road to follow in order to reconsider the 
foundations of physics.

\appendix
\section{Astrophysical constants and scales}
\label{app:A}
\noindent
Constants
\vskip .3cm
The tradition in astrophysics is to use the CGS system. Whereas there are 
in some specific cases useful quantities to be defined (such as the parsec),
centimeter and gram seem hardly relevant. We thus use here the international 
system. Note that $1$ kg.m$^{-3} = 10^{-3}$ g.cm$^{-3}$ , $1$ J $= 10^{7}$ 
erg, $1$ W  $= 10^{7}$ erg.s$^{-1}$. 
\vskip .3cm
Speed of light: $c = 299 \ 792 \ 458$ m s$^{-1}$

Newtonian gravitational constant: $G_{_N} = 6.6742 \times 10^{-11}$ m$^3$ 
kg$^{-1}$ s$^{-2}$

$\alpha_G \equiv G_{_N} m_p^2/(\hbar c) = 5.906 \times 10^{-39}$

Fine structure constant: $\alpha \equiv e^2/(4 \pi \epsilon_0 \hbar c) = 7.297 
\times 10^{-3} = 1/137$

Thomson cross section: $\sigma_T = 8 \pi r_e^2/3 = 0.665$ barn $ = 0.665 \times
10^{-28}$ m$^2$ 

Boltzmann constant: $k_B = 1.380 \times 10^{-23}$ J.K$^{-1}= 8.617 \times 
10^{-5}$ eV.K$^{-1}$

Planck constant: $\hbar = 1.054 \times 10^{-34}$ J.s 

\vskip .5cm
\noindent
Typical length scales 
\vskip .3cm

Planck length: $\ell_{_P} = \sqrt{8\pi G_{_N}\hbar/c^3} = 8.1 \times 10^{-35}$ 
m 

Classical electron radius: $r_e = e^2/(4\pi \epsilon_0 m_e c^2) = 2.817 \times 
10^{-15}$ m

Solar radius: $R_\odot = 6.9598 \times 10^8$ m

Astronomical unit (au) = Sun-Earth distance = $1.4960\times 10^{11}$ m 

Parsec (au/arc sec): $1$ pc $= 3.262$ light-year $= 3.086 \times 10^{16}$ m

Sun-galactic center distance: $10$ kpc

Milky way galaxy disk radius (luminous matter): $15$ kpc 

Presently visible universe: $6$ Gpc

\vskip .5cm
\noindent
Typical mass scales 
\vskip .3cm

Reduced Planck mass: $m_{_P} = \sqrt{\hbar c/(8\pi G_{_N})} = 2.14 \times 
   10^{18}$ GeV$/c^2  = 3.81 \times 10^{-9}$ kg

Solar mass: $M_\odot  = 1.989 \times 10^{30}$ kg

Milky Way galaxy mass: $4$ to $10\times 10^{11} \ M_\odot$

\vskip .5cm
\noindent
Typical luminosities 
\vskip .3cm

Solar luminosity: $L_\odot = 3.85 \times 10^{33}$ erg/s $
= 3.85 \times 10^{26}$ W

\vskip .5cm
\noindent
Typical densities 
\vskip .3cm

Present mean density of the universe: $\rho_0 \sim \rho_c = 10^{-26}$ 
kg.m$^{-3}$

Interstellar medium: $10^{-22}$ kg.m$^{-3}$

Sun: $\rho_\odot = 1408$ kg/m$^{-3}$

Neutron star: $10^{18}$ kg.m$^{-3}$

\section{General relativity}
\label{app:B}

{\em In the context of general 
relativity, one defines the Christoffel symbol or affine connection 
$\Gamma^\rho{}_{\mu\nu}$ which is the analogue of the gauge field (it 
appears in covariant derivatives). It is defined in terms 
of the metric as:
\begin{equation}
\label{2-1a}
\Gamma^\rho{}_{\mu\nu} = {1 \over 2} g^{\rho\sigma} \left[ 
\partial_\mu g_{\nu\sigma}+ \partial_\nu g_{\mu\sigma} - \partial_\sigma 
g_{\mu\nu}\right] \quad ,
\end{equation}
where $g^{\rho\sigma}$ is the inverse metric tensor: $g^{\rho\sigma}
g_{\sigma\tau} = \delta^\rho_\tau$.

In the same way that one defines the field strength by differentiating the 
gauge field, one introduces the Riemann curvature tensor:
\begin{equation}
\label{2-1b}
R^\mu{}_{\nu\alpha\beta} = \partial_\alpha \Gamma^\mu{}_{\nu\beta} 
- \partial_\beta \Gamma^\mu{}_{\nu\alpha}+  \Gamma^\mu{}_{\alpha\sigma}
\Gamma^\sigma{}_{\nu\beta} -  \Gamma^\mu{}_{\beta\sigma}
\Gamma^\sigma{}_{\nu\alpha} \quad .
\end{equation}
By contracting indices, one then defines the Ricci tensor $R_{\mu\nu}$
and the curvature scalar $R$
\begin{equation}
\label{2-1c}
R_{\mu\nu} \equiv  R^\alpha{}_{\mu\alpha\nu}, \quad , \quad 
R \equiv g^{\mu\nu} R_{\mu\nu} \quad .
\end{equation}
One also uses the Christoffel symbols to define the covariant derivatives:
\begin{eqnarray}
V_{\mu;\nu} &=& \nabla_\nu V_\mu \equiv \partial_\nu V_\mu - \Gamma^\rho{}_{\mu\nu} 
V_\rho \ , \nonumber \\
V^\mu{}_{;\nu} &=& \nabla_\nu V^\mu \equiv \partial_\nu V^\mu + 
\Gamma^\mu{}_{\nu\rho} V^\rho \ .
\label{covder}
\end{eqnarray}

\vskip .5cm
\underline{Exercise B-1}~: In the case of the Robertson-Walker metric 
(\ref{2-2a}),

a) compute the non-vanishing Christoffel symbols (\ref{2-1a}),

b) using the fact that the Ricci tensor associated with the 3-dimensional 
metric $\gamma_{ij}$ is simply $R_{ij}(\gamma) = 2k \gamma_{ij}$, compute 
the components of the Ricci tensor and the scalar curvature (\ref{2-1c}), 

c) deduce the components of the Einstein tensor $G_{\mu\nu}$ defined in 
(\ref{2-Einstein}): 
The components of the Einstein tensor now read (see Exercise 2-1):
\begin{eqnarray}
G_{tt} &=& 3  \left( {{\dot a}^2   \over a^2 } + {k \over a^2} \right),  
\label{2-3a}\\
G_{ij} &=& - \gamma_{ij} \left( {\dot a}^2 + 2 a {\ddot a} + k \right), 
\label{2-3b}
\end{eqnarray}

\vskip .3cm 
Hints: a) $\Gamma^i{}_{jt} = \delta^i_j \dot a / a$, $\Gamma^t{}_{ij} = 
a \dot a \gamma_{ij}$, $\Gamma^i{}_{jk} = \Gamma^i{}_{jk}(\gamma)$.

b)
$R_{tt} = -3 \ddot a/a$, $R_{ij}= \left( 2k + \ddot a a + 2 \dot a^2 \right)
\gamma_{ij}$, $R= -6 \left( k + \ddot a a +  \dot a^2 \right)/a^2$.
}

\section{Measure of distances}
\label{app:C}

{\em
Measuring cosmological distances allows to study the geometry of spacetime. 
Depending on the type of observation, one may define several distances.

First consider a photon travelling in an expanding or contracting 
Friedmann universe. Its equation of motion is fixed by the condition 
$ds^2=0$ (as in Eq.~(\ref{2-13}) of Chapter~\ref{chap:1}). One then defines the 
proper distance as 
\begin{equation}
\label{4-30}     
d(t) \equiv  a(t)\int_0^r {dr \over \sqrt{1-kr^2}} = a(t) \int_t^{t_0} 
{cdt' \over a(t')} \ .
\end{equation}

Using 
$$\int_{t}^{t_0} {cdt \over a(t)} = \int_{a(t)}^{a_0} {cda \over a \dot a}
= \int_{a(t)}^{a_0} {cda \over a^2 H} = \int_0^z {cdz \over H(z)}\, $$
we may extract from (\ref{4-30}) the proper distance at time $t_0$:
\begin{eqnarray}
\label{4-32}
d(t_0) = a_0  \int_0^{r} {dr \over \sqrt{1-kr^2}} &=& a_0 \left\{ 
\begin{array}{ll} \sin^{-1} r & k=+1 \\ r & k=0 \\ \sinh^{-1} r & k=-1 
\end{array} \right.\\
&=&\ell_{H_0} \int_0^z {dz \over \left[\Omega_{_M} (1+z)^3 
+  \Omega_{_R} (1+z)^4 + \Omega_{k} (1+z)^2 + \Omega_\Lambda \right]^{1/2}} 
\nonumber
\end{eqnarray}
where $\ell_{H_0} = c H_0^{-1}$.

\vskip .3cm
If a photon source of luminosity $L$ (energy per unit time) is placed at 
a distance $r$ from the observer, then the energy flux $\phi$ (energy per unit 
time and unit area) received by the observer is given by 
\begin{equation}
\label{4-34}
\phi = {L \over 4\pi a_0^2 r^2 (1+z)^2} \equiv {L \over 4\pi d_L^2} \quad .
\end{equation}
The two powers of $1+z$ account for the photon energy redshift and the 
time dilatation between emission and observation. The quantity
$d_L \equiv a_0 r (1+z)$ is called luminosity distance.

If the source is at a redshift $z$ of order one or smaller, the effect of 
spatial curvature is unimportant and we can 
approximate the integral $\int_0^{r} dr / \sqrt{1-kr^2}$ in (\ref{4-30})
by simply $r$ (i.e. the value for $k=0$). This equation gives
\begin{equation}
\label{4-35}
a_0 r \sim \int_{t}^{t_0} {a_0 cdt \over a(t)} = \int_{a}^{a_0}{a_0 c da \over 
a \dot a} \sim \ell_{H_0} \int_a^{a_0} 
{da \over a \left[1-q_0H_0(t-t_0)\right]}
\end{equation}
where we have used the development (\ref{4-9}) with $ t_{H_0} =  \ell_{H_0}/c
= H_0^{-1}$. Using $H_0(t-t_0) \sim (a-a_0)/a_0 \ll 1$ and $a=a_0/(1+z)$, 
we obtain for $z\ll 1$
\begin{equation}
\label{4-36}
a_0 r = \ell_{H_0} z \left( 1 - {1+q_0 \over 2} z + \cdots \right)\quad .
\end{equation}
Thus, the luminosity distance reads, for $z \ll 1$,
\begin{equation}
\label{4-37}
d_L = \ell_{H_0} z \left( 1 - {1+q_0 \over 2} z + \cdots \right)(1+z)
= \ell_{H_0} z \left( 1 + {1-q_0 \over 2} z + \cdots \right) \quad .
\end{equation}
Hence measurement of deviations to the Hubble law ($d_L = \ell_{H_0} z$) at 
moderate redshift allow  to measure the combination $\Omega_{_M}/2 - 
\Omega_\Lambda$ (see (\ref{4-8})).

Another distance is defined in cases where one measures the angular diameter 
$\delta$ of a source in the sky. If $D$ is the diameter of the source, then 
$D/\delta$ would be the distance of the source in Euclidean geometry. In a 
universe with a Robertson-Walker metric, it turns out to be $a(t) r
= a_0r/(1+z)$. This defines the angular diameter distance $d_A$
\begin{equation}
\label{4-38}
d_A = {d_L \over (1+z)^2} \quad .
\end{equation}

Several distance measurements tend to point towards an evolution of the present
universe dominated by the cosmological constant contribution\footnote{at least 
when analyzed in the framework  of the model discussed in this section, i.e.
including non-relativistic matter, radiation and a cosmological constant.} 
and thus a late acceleration of its expansion, as we will now see. 

\vskip .3cm
\underline{Exercise C-1}~: We compute exactly the luminosity distance
$d_L = a_0 r (1+z)$ or angular distance $d_A = a_0 r/(1+z)$ in the case 
of a matter-dominated universe. Defining 
\begin{equation}
\label{D-4ex1}
\zeta_k(r) \equiv \left\{ 
\begin{array}{ll} \sin^{-1} r & k=+1 \\ r & k=0 \\ \sinh^{-1} r & k=-1 
\end{array} \right. \ ,
\end{equation}
use (\ref{4-32}) which reads, in the case of a matter-dominated universe,
\begin{equation}
\label{D-4ex2}
a_0 \zeta_k (r) = \ell_{H_0} \int_0^z {dz \over \left[\Omega_{_M} (1+z)^3  
+ (1-\Omega_{_M}) (1+z)^2  \right]^{1/2}} 
\end{equation}
to prove Mattig's formula~\cite{Ma58}:
\begin{equation}
\label{D-4ex3}
a_0 r = 2  \ell_{H_0} {\Omega_{_M} z + \left(\Omega_{_M}-2\right) \left[
\sqrt{1 + \Omega_{_M} z} - 1\right] \over \Omega_{_M}^2 (1+z)} \ .
\end{equation} 

\vskip .3cm 
Hints: For $k \not = 0$, change to the coordinate $u^2 = k(\Omega -1)/\left[ 
\Omega (1+z)\right]$ in order to compute the integral (\ref{D-4ex2}). Using
the last of equations (\ref{4-4}), which reads $\ell_{H_0}^2/a_0^2 = k
(\Omega -1)$, one obtains
$$\zeta_k(r) = 2 \left( \zeta_k \left[ \sqrt{{k(\Omega-1) \over \Omega}} 
\right] - \zeta_k \left[ \sqrt{{k(\Omega-1) \over(1+z) \Omega}} \right] 
\right) \ ,$$
from which (\ref{D-4ex3}) can be inferred.
 
}

\section{Perturbations with scalar fields}
\label{app:E}

{\em We study in this Appendix the perturbations of a scalar field coupled to 
gravity, following Ref.~\cite{GM99}. This has obvious implications for the study of 
inflation or dark energy models. 

We consider the most general local action for a scalar field 
coupled to Einstein gravity:
\begin{equation}
\label{E-1}
{\cal S} = -{m_{_P}^2 \over 2} \int d^4x  \sqrt{g} R + \int d^4x
\sqrt{g} \ p(X,\phi) \ ,
\end{equation}
where we have defined
\begin{equation}
\label{E-2}
X \equiv {1 \over 2} g^{\mu\nu} \partial_\mu \phi \partial_\nu \phi \ .
\end{equation}
One may describe this system as a perfect fluid, with the standard 
energy-momentum tensor (\ref{2-4}): 
\begin{equation}
\label{E-3}
T_{\mu\nu} = -  p g_{\mu\nu} + (p+\rho ) U_\mu U_\nu \quad ,
\end{equation}
Indeed, varying with respect to the metric, we find
\begin{eqnarray}
\label{E-4}
\delta {\cal S}  &=& \int \sqrt{g} \left[ Xp_{,X} U_\mu U_\nu - {1 \over 2} 
p g_{\mu\nu} \right] \delta g^{\mu\nu}  \nonumber \\
& \equiv & {1 \over 2} \int T_{\mu\nu} \delta g^{\mu\nu} \ , 
\end{eqnarray}
with $U_\mu \equiv \partial_\mu \phi /(2X)^{1/2}$. Thus, the energy-momentum 
tensor has the form (\ref{E-3}) with the function $p(X,\phi)$, i.e. the 
scalar Lagrangian, as the pressure (hence the notation) and the energy
density:
\begin{equation}
\label{E-5}
\rho = 2 X p_{,X} - p \ .
\end{equation}
In the case where $p=X -V(\phi)$, one recovers (\ref{5-45a},\ref{5-45b}).

In the following, a quantity will play an important role; it is the speed of 
sound:
\begin{equation}
\label{E-6}
c_s^2 \equiv {\delta p \over \delta \rho} = {p_{,X} \over \rho_{,X}}
= {p + \rho  \over 2 X \rho_{,X}} \ .
\end{equation} 

\vskip .3cm
We start with a background metric described by (\ref{2-2a}) (for simplicity,
we assume that space is flat: $k=0$; for the general case, see Ref.~\cite{GM99})
and with a background scalar configuration $\varphi (t)$ which satisfies
(\ref{2-11}) (or equivalently the scalar field equation of motion).

Perturbing this background, we write in the longitudinal gauge 
\cite{MFB92}\footnote{{\em We use the fact that the spatial part of the 
energy-momentum tensor is diagonal. Otherwise, two different functions $\Phi$ 
wand $\Psi$ would appear respectively as $(1+2\Phi)$ in the time component
and $(1-2\Psi)$ in the space component~\cite{MFB92}.}}  
\begin{equation}
\label{E-7}
ds^2 = (1 + 2 \Phi) dt^2 - (1- 2 \Phi) a^2(t) \gamma_{ij} dx^i dx^j \ ,
\end{equation}
where $\Phi$ is the Newtonian potential, and we take for the scalar field
\begin{equation}
\label{E-8}
\phi(t,x) = \varphi (t) + \delta \varphi(t,x) \ .
\end{equation}
Then
\begin{eqnarray}
\delta G^0{}_0 &=& 2 \left[ {1 \over a^2}\Delta \Phi - 3H \dot \Phi - 3 H^2 
\Phi \right] \ , \label{E-9} \\
\delta G^0{}_i &=& 2 \left[ \dot \Phi + H \Phi \right]_{,i}  \ , \label{E-10}
\end{eqnarray}
where, as usual, $H^2 = 8\pi G_{_N} \rho/3$.
As for the variation of the energy-momentum tensor, we have
\begin{equation}
\label{E-11}
\delta T^0{}_0 = \delta \rho = \rho_{,X} \delta X + \rho_{,\phi} \delta 
\varphi \ , \quad  \delta T^0{}_i = (p + \rho) \delta U_i \ .
\end{equation} 
Note for the latter that $U_0 =1$, $U_i = 0$ but $\delta U_i  =
(\delta \varphi / \dot \varphi)_{,i} \not = 0$. For the former, we use
$\dot \rho = -3H (p+\rho) = \rho_{,X} \dot X + \rho_{,\phi} \dot 
\phi$. One finds
\begin{eqnarray}
\delta T^0{}_0 &=& -3H (p+\rho) {\delta \varphi \over \dot \varphi}
+ {p + \rho \over c_s^2} \left[ \left( {\delta \varphi \over \dot \varphi}
\right)^\cdot - \Phi \right] \ , \label{E-12} \\
\delta T^0{}_i &=& (p + \rho) \left( {\delta \varphi \over \dot \varphi}
\right)_{,i} \ . \label{E-13}
\end{eqnarray}
We thus obtain from $\delta G_{\mu\nu} = 8 \pi G_{_N} \delta T_{\mu\nu}$
\begin{eqnarray}
 \left( {\delta \varphi \over \dot \varphi} \right)^\cdot &=& 
\left( 1 + {c_s^2 \over 4 \pi G_{_N}a^2 (p + \rho)}\Delta \right) \Phi 
\ , \label{E-14} \\
\left( a \Phi \right)^\cdot &=& 4 \pi G_{_N} a (p + \rho) \left( 
{\delta \varphi \over \dot \varphi}
\right) \ . \label{E-15}
\end{eqnarray}
The other Einstein's equations are redundant. We may now define the new 
variables $\xi$ and $\zeta$:
\begin{equation}
\label{E-16}
a\Phi = 4\pi G_{_N} H\xi \ , \quad {\delta \varphi \over \dot \varphi}
= {\zeta \over H} - {4 \pi G_{_N} \over a} \xi \ ,
\end{equation}
which satisfy the equations of motion
\begin{eqnarray}
\dot \xi &=& {a(p+\rho) \over H^2} \ \zeta\ , \label{E-17} \\
\dot \zeta &=& {c_s^2 H^2 \over a^3 (p+\rho)} \ \Delta \xi \ . \label{E-18}
\end{eqnarray}
Defining
\begin{equation}
\label{E-19}
z \equiv {a (p+ \rho)^{1/2} \over c_s H}
\end{equation}
and differentiating with respect to conformal time ($\xi' \equiv\  d\xi / 
d\eta = a \dot \xi$ and so on), 
we may write the system of differential equations simply as
\begin{equation}
\label{E-20}
\xi' = c_s^2 z \zeta \ , \quad \zeta' = {1 \over z^2} \Delta \xi \ ,
\end{equation}
which can be turned into a single differential equation for $\zeta$. Indeed,
defining $v \equiv z\zeta$, we find
\begin{equation}
\label{E-21} 
v'' - c_s^2 \Delta v - {z'' \over z} v = 0 \ .
\end{equation}
This can be derived from the following action:
\begin{equation}
\label{E-22}
{\cal S} = {1 \over 2} \int \left[ v'^2 + c_s^2 v \Delta v + {z'' \over z}
v^2 \right] d\eta d^3x \ .
\end{equation}
If we look for plane wave solutions of (\ref{E-21}) i.e. $v=v_{\bf k} e^{-i
{\bf k}.{\bf x}}$, we find two regimes:
\begin{itemize}
\item at long wavelength (${\bf k}^2 c_s^2 \ll |z''/z|$), a non-decaying 
solution $v_{\bf k} \propto z$. 
\item at short wavelength (${\bf k}^2 c_s^2 \ll |z''/z|$), an oscillating 
solution $v_{\bf k} \propto \exp (ikc_s\eta)$.
\end{itemize}

\vskip .5cm
{\bf Quantization of scalar field in curved spacetime}
\vskip .5cm

We now turn to the quantization of the scalar degrees of freedom. The fact 
that we are in a non-trivial background gravitational field brings some new 
features but, since the background is only time dependent, the  
quantization procedure may be broadly inspired by the flat spacetime case.

Let us consider a generic scalar field (this could be for example the 
gravitational potential). The standard commutation relations
\begin{equation}
\label{E-23}
\left[ \Phi(\eta,{\bf x}),\Phi(\eta,{\bf x'})\right] =
\left[ \Pi(\eta,{\bf x}),\Pi(\eta,{\bf x'})\right]= 0 \ , \quad
\left[ \Phi(\eta,{\bf x}),\Pi(\eta,{\bf x'})\right] = i \delta^3({\bf x}
- {\bf x'}) 
\end{equation}
involves the canonical momentum $\Pi = \delta {\cal L} / \delta \partial_\eta
\Phi$ (we are using here the conformal time $\eta$). 

One may decompose the operator $\Phi$ over the complete orthonormal basis of 
the eigenfunctions of the Laplace operator. In the spatially flat case that we 
are considering here, these are simply the plane waves: $\chi_{\bf k}(\eta) 
e^{-i{\bf k}.{\bf x}}$ (we note that we are making full use of spatial 
translation invariance, which remains a symmetry). We thus write
\begin{equation}
\label{E-24}
\Phi(\eta,{\bf x}) = {1 \over \sqrt{2}} \int {d^3 k \over (2\pi)^{3/2}}
\left[ e^{-i {\bf k}.{\bf x}} \chi_{\bf k}(\eta) a^\dagger_{\bf k} +
e^{i {\bf k}.{\bf x}} \chi_{\bf k}^*(\eta) a_{\bf k} \right] \ ,
\end{equation}
where the operators $a_{\bf k}$ and $a^\dagger_{\bf k}$ satisfy the 
commutation rules
\begin{equation}
\label{E-25}
\left[ a_{\bf k},a_{\bf k'} \right] = \left[ a^\dagger_{\bf k},
a^\dagger_{\bf k'} \right] = 0 \ , \quad \left[ a_{\bf k},a^\dagger_{\bf k'} 
\right] = \delta^3 ({\bf k} - {\bf k'} ) \ .
\end{equation}
This is consistent with (\ref{E-23}) under the condition
\begin{equation}
\label{E-26}
\chi'_{\bf k} (\eta) \chi_{\bf k}^* (\eta) - \chi'^*_{\bf k} (\eta) 
\chi_{\bf k} (\eta) = 2i \ .
\end{equation}
The $\chi_k(\eta)$ modes satisfy an equation of the type
\begin{equation}
\label{E-27}
\chi''_{\bf k} (\eta) + E_{\bf k}^2 \chi_{\bf k}(\eta) =0 \ ,
\end{equation}
where $E_{\bf k}^2$ includes a mass-squared term and possibly other 
contributions such as the one that would arise from a non-minimal coupling of 
the scalar field to gravity.

In Minkowski spacetime, on constructs a Fock space of states obtained by 
applying a product of creation (negative frequency) operators on the  vacuum 
state $|0\rangle$, defined as the state annihilated by all positive frequency 
operators  $a_{\bf k}$:  $a_{\bf k}|0\rangle = 0$. This relies on the 
invariance under the Poincar\'e group which gives an absolute meaning to these 
notions. More precisely, in Minkowski spacetime, the operator $\partial / 
\partial t$ is a Killing vector orthogonal to the spacelike hypersurfaces $t =$
constant and the plane wave modes $e^{-ik.x}$ are eigenfunctions of this 
Killing vector with eigenvalues $-ik_0 = - i \omega$ of a given sign.

In curved spacetime (see for example the book by Birrell and Davies
\cite{Birrell}), the Poincar\'e group is no longer a symmetry group of 
spacetime and correspondingly 
there is no time-invariant notion of positive or negative frequency. There is 
thus no possibility of agreeing on a specific vacuum state for all inertial 
measuring devices.  One may still rely, in some cases, 
on specific symmetries such as translation invariance, conformal symmetry or 
the de Sitter group to constrain the description of vacuum states.

[See the review by Mukhanov, Feldman and Brandenberger~\cite{MFB92}]

We thus choose a given time $\eta_0$ in order to define a vacuum state
$| 0 \rangle_{\eta_0}$ such that, for all ${\bf k}$, $a_{\bf k}
| 0 \rangle_{\eta_0} = 0$. These annihilation operators are the operator 
factors of the positive frequency modes $\chi^+_{\bf k} \equiv 
\chi_{\bf k}^*(\eta)$ in the expansion
(\ref{E-24}) (similarly we define the negative frequency modes
$\chi^-_{\bf k} \equiv  \chi_{\bf k}(\eta)$). 
If all $E_{\bf k}$ are positive, it turns out that one find such 
modes:  they are the solutions of (\ref{E-27}) with the following initial 
conditions at time $\eta_0$:
\begin{equation}
\label{E-28}
\chi_{\bf k}(\eta_0) = E_{\bf k}^{-1/2}(\eta_0) \ , \quad
\chi'_{\bf k}(\eta_0) = i E_{\bf k}^{1/2}(\eta_0) \ ,
\end{equation}
consistent with the consistency condition (\ref{E-26}). Since these solutions 
obviously depend on $\eta_0$, we will affect them a superscript $(0)$ in 
what follows.

At a later time $\eta_1$, we define along the same lines a new vacuum
$| 0 \rangle_{\eta_1}$, which is annihilated by all operators $b_{\bf k}$. 
These operators appear in an expansion of the type (\ref{E-24}) but with 
new positive frequency modes $\chi^{+(1)}_{\bf k}$. Since equation (\ref{E-27})
is linear, there is a linear relation between the positive and negative 
frequency modes at $\eta_0$ and $\eta_1$:
\begin{eqnarray}
\chi_{\bf k}^{(1)+} &=& \alpha_{\bf k} \chi_{\bf k}^{(0)+} 
+ \beta_{\bf k} \chi_{\bf k}^{(0)-} \ , \quad \left| \alpha_{\bf k} \right|^2
- \left| \beta_{\bf k} \right|^2 = 1 \ , \nonumber \\
\chi_{\bf k}^{(1)-} &=& \beta^*_{\bf k} \chi_{\bf k}^{(0)+} 
+ \alpha^*_{\bf k} \chi_{\bf k}^{(0)-} \ , \label{E-29}
\end{eqnarray}
where we have used (\ref{E-26}).

This defines the Bogoliubov coefficients $\alpha_{\bf k}$ and
$\beta_{\bf k}$. Obviously we have in parallel for the operators
\begin{eqnarray}
b_{\bf k} &=& +\alpha^*_{\bf k} a_{\bf k} - \beta^*_{\bf k} a^\dagger_{\bf k}
\ , \nonumber \\
b^\dagger_{\bf k} &=& - \beta_{\bf k} a_{\bf k} + \beta_{\bf k} 
a^\dagger_{\bf k} \ . \label{E-30}
\end{eqnarray}

Let us give an example to illustrate the physical meaning of the Bogoliubov 
coefficients. We start with the vacuum $| 0 \rangle_{\eta_0}$ at time
$\eta_0$ and compute at $\eta_1$ the number of particles 
$b^\dagger_{\bf k} b_{\bf k}$. It is given by
\begin{equation}
\label{E-31}
{}_{\eta_0} \langle 0\left| b^\dagger_{\bf k} b_{\bf k} \right| 0
\rangle _{\eta_0}  = \left| \beta_{\bf k} \right|^2  \ ,
\end{equation}
where we have used (\ref{E-30}). Thus, even though we have prepared the system 
in the vacuum state at time $\eta_0$, the number of particles is non-vanishing
at time $\eta_1$. Fluctuations can be produced quantum mechanically from the 
vacuum through the coupling of the scalar field to gravity. 
 
If not all energies are positive, then we cannot define a set of modes through 
the boundary conditions (\ref{E-28}). This is in particular the situation 
encountered in the case of inflation. There, the symmetries of de Sitter space 
help to define the so-called de Sitter invariant vacuum through the
conditions:
\begin{equation}
\label{E-32}
\chi_{\bf k}(\eta_0) = {1 \over k^{3/2}} \left( {\cal H}_0 +ik \right)
 \ , \quad \chi'_{\bf k}(\eta_0) = {i \over k^{1/2}} \left( {\cal H}_0 +ik 
- i {\cal H}'_0 / k \right) \ ,
\end{equation}  
where ${\cal H}_0 = a'(\eta_0)/a(\eta_0)= a(\eta_0) H_0$. We recover 
(\ref{E-28}) at small wavelength i.e. for $k \ll {\cal H}_0$. We note that, 
whereas the small wavelength behavior is universal, the large wavelength 
behaviour strongly depends on the choice of vacuum.

\vskip .5cm
{\bf Perturbations}
\vskip .5cm

Let us now apply this formalism to the quantum generation of perturbations.
We are interested in fluctuations of the Newtonian gravitational potential.
Using (\ref{E-15}) and (\ref{E-16}), we have 
\begin{equation}
\label{E-33}
\zeta = \Phi \left[ 1+ {2 \over 3} {\rho \over p+ \rho} \right]
+ {2 \over 3} {\rho \over p+ \rho} {\dot \Phi \over H} \ .
\end{equation}
Since $\Phi$ is constant in any phase where $p/\rho$ is constant (say matter
or radiation domination), then in such a phase, $\zeta$ is simply 
proportional to the gravitational potential. We thus consider the scalar
variable $\zeta$ in what follows and write: 
\begin{equation}
\label{E-34}
\zeta(\eta,{\bf x}) = {1 \over \sqrt{2}} \int {d^3 k \over (2\pi)^{3/2}}
\left[ e^{-i {\bf k}.{\bf x}} \zeta_{\bf k}(\eta) a^\dagger_{\bf k} +
e^{i {\bf k}.{\bf x}} \zeta_{\bf k}^*(\eta) a_{\bf k} \right] \ ,
\end{equation}
We have $\zeta_{\bf k} = v_{\bf k}/z$, where $v_{\bf k}$ satisfies, according
to (\ref{E-21}),
\begin{equation}
\label{E-35}
v''_{\bf k} + \left( c_s^2 {\bf k}^2 - {z'' \over z} \right) v_{\bf k} = 0 \ .
\end{equation}
One often characterizes the fluctuations through the power spectrum 
${\cal P}^\zeta_{\bf k}(\eta)$ which is 
defined from the $2$-point correlation function: 
\begin{equation}
\label{E-36}
\langle 0 \left| \zeta({\bf x}, \eta) \zeta({\bf x} + {\bf r}, \eta) \right| 
0 \rangle = \int_{k=0}^{k=+ \infty} {dk \over k} {\sin kr \over kr} 
{\cal P}^\zeta_{\bf k}(\eta) \ .
\end{equation}
Using the decomposition (\ref{E-34}), one easily obtains
\begin{equation}
\label{E-37}
{\cal P}^\zeta_{\bf k}(\eta) = {k^3 \over 2 \pi^2} \left|\zeta_{\bf k}
\right|^2 = {k^3 \over 2 \pi^2} {\left|v_{\bf k}\right|^2 \over |z|^2} \ .
\end{equation}
As we have seen above, Eq. (\ref{E-35}) has two distinct regimes depending of 
the relative magnitude of $c_s k$ and $z''/z$. In the case of slow roll 
inflation, the main dependence with time in $z$, as given in (\ref{E-19}), 
comes from the scale factor. Hence $z''/z \sim a''/a \sim (aH)^2$. Hence
we have to compare $c_s k$ with $aH$ i.e. the comoving wavelength $a/k$
with the sound horizon length $c_s/H$.

In the case of short wavelength (smaller than the sound horizon), the 
normalized solution is
\begin{equation}
\label{E-38}
v_{\bf k} = {1 \over (2kc_s)^{1/2}} e^{ikc_s \eta} \ .
\end{equation}
For long wavelengths (larger than the sound horizon), it is
\begin{equation}
\label{E-39}
v_{\bf k} = C_{\bf k} z \ ,
\end{equation}
where the constant $C_{\bf k}$ may be obtained by continuity between the two 
approximate solutions at the scale of sound horizon: $|C_{\bf k}|^2 =
1/(2kc_sz_s)$, with $z_s$ the value of $z$ at horizon crossing.

We thus have the behavior indicated on Fig.~\ref{fig5-1}: the quantum 
fluctuations are created at small wavelength and grow until they cross the 
(sound) horizon. From then on they grow mechanically until they reneter the 
horizon and are observed in the matter dominated epoch. We thus obtain the
power spectrum:
\begin{equation}
\label{E-40}
{\cal P}^\zeta_{\bf k} = \left. {k^3 \over 2\pi^2} |C_{\bf k}|^2 
\right|_{k=aH/c_s} = \left. {1 \over c_s} {H^2 \over p+\rho} \left({H \over 2 
\pi} \right)^2  \right|_{k=aH/c_s} \ .
\end{equation}
The spectral index is defined as
\begin{equation}
\label{E-40a}
n_S(k) - 1 = {d \ln {\cal P}^\zeta_{\bf k} \over d \ln k} \ .
\end{equation}
Thus, inflation predicts a departure from a scale invariant spectrum 
($n_S=1$).

\vskip .5cm
\underline{Exercise D-1}~: Show that, in the case of slow roll inflation
(Section~\ref{5s1}), the spectral index is simply given by
\begin{equation}
\label{E-41}
n_S = 1 - 6 \epsilon + 2 \eta \ .
\end{equation}

\vskip .3cm 
Hints: In this case, $c_s = 1$. Moreover, 
$$n_S(k) - 1 \sim {1 \over H} {d \ln {\cal P}^\zeta_{\bf k} \over dt}
= -6\left(1+{p \over \rho}\right) - 2 {\ddot \phi \over H \dot \phi}.$$
Use then (\ref{5-18a}).
 
}

\bibliography{08_Binetruy}
\bibliographystyle{unsrt}

\end{document}